# The Design Reference Asteroid for the OSIRIS-REx Mission Target (101955) Bennu

Carl W. Hergenrother[1], Maria Antonietta Barucci[2], Olivier Barnouin[3], Beau Bierhaus[4], Richard P. Binzel[5], William F. Bottke[6], Steve Chesley[7], Ben C. Clark[4], Beth E. Clark[8], Ed Cloutis[9], Christian Drouet d'Aubigny[1], Marco Delbo[10], Josh Emery[11], Bob Gaskell[12], Ellen Howell[13], Lindsay Keller[14], Michael Kelley[15], John Marshall[16], Patrick Michel[10], Michael Nolan[13], Bashar Rizk[1], Dan Scheeres[17], Driss Takir[8], David D. Vokrouhlický[18], Ed Beshore[1], Dante S. Lauretta[1]

E-mail contact: chergen@lpl.arizona.edu

[1]Lunar and Planetary Laboratory, University of Arizona, Tucson, Arizona
[2]LESIA-Observatoire de Paris, CNRS, Univ. Paris-Diderot, Meudon, France
[3]Applied Physics Laboratory, John Hopkins University, Laurel, Maryland
[4]Space Systems Company, Lockheed Martin, Denver, Colorado
[5]Dept. of Earth, Atmospheric and Planetary Sciences, Massachusetts Institute of Technology, Cambridge, Massachusetts
[6]Southwest Research Institute, Boulder, Colorado
[7]Jet Propulsion Laboratory, California Institute of Technology, Pasadena, California
[8]Dept. of Physics, Ithaca College, Ithaca, New York
[9]Dept. of Geography, University of Winnipeg, Winnipeg, Canada
[10]Laboratoire Lagrange, UNS-CNRS, Observatoire de la Cote d'Azur, Nice, France
[11]Earth and Planetary Science Dept., University of Tennessee, Knoxville, Tennessee
[12]Planetary Science Institute, Tucson, Arizona
[13]Arecibo Observatory, Arecibo, Puerto Rico
[14]NASA Johnson Space Center, Houston, Texas
[15]Dept. of Astronomy, University of Maryland, College Park, Maryland
[16]SETI Institute, Mountain View, California
[17]Dept. of Aerospace Engineering Sciences, University of Colorado-Boulder, Boulder, Colorado
[18]Astronomical Institute, Charles University, Prague, Czech Republic

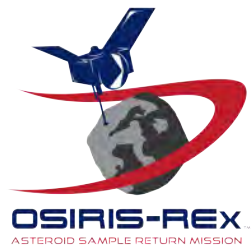

# Table of Contents

# I. Background on Bennu

The Apollo near-Earth asteroid (101955) Bennu (formerly 1999 RQ36) is the target of the OSIRIS-REx sample return mission. The selection of Bennu as the primary target was based on scientific, risk assessment and mission feasibility factors. As a B-type carbonaceous asteroid, Bennu represents an important source of volatiles and organic matter to Earth as well as being a direct remnant of the original building blocks of the terrestrial planets. With the exception of the few asteroids that have been visited by spacecraft, Bennu is the best-characterized near-Earth asteroid having been extensively studied in the visible, infrared and with radar. Its relatively Earth-like, low delta-V orbit is conducive for a low energy New Frontiers-level mission.

Bennu was discovered on September 11, 1999 by the Lincoln Laboratory Near Earth Research (LINEAR) survey with a 1.0-meter telescope located near Socorro, New Mexico (Williams, 1999). At the time of discovery it was a bright V magnitude 15.8 and located within 0.05 AU of Earth. After discovery Bennu approached to within 0.015 AU of Earth (~6 lunar distances) and brightened to V magnitude 14.4. The 1999/2000 apparition was the best opportunity for discovery and characterization of Bennu since 1970.

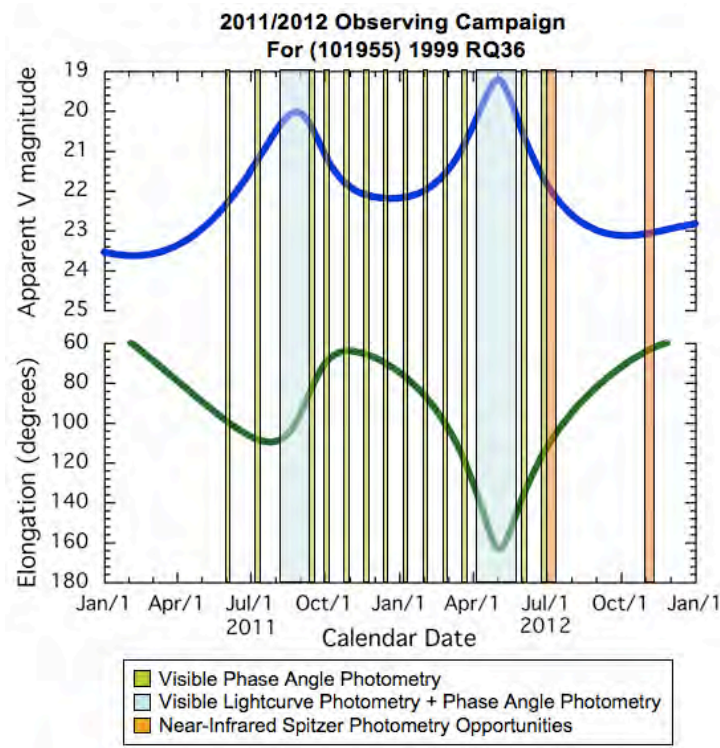

**Figure 1 - The 2011/2012 observing campaign for Bennu starts in the summer of 2011 and extends through the summer of 2012.**

The orbital period of Bennu is 1.20 years, which corresponds to a synodic period of 6 years resulting in relatively close approaches to Earth every 6 years. After its discovery apparition the next opportunity for easy observation of Bennu was 2005/2006 when Bennu approached within 0.033 AU of Earth and brightened to V = 16.1. The 1999/2000 and 2005/2006 apparitions are the best for Earth-based observation of Bennu until 2060.

Though a relatively poor apparition, 2011/2012 provided the last opportunity to characterize Bennu before OSIRIS-REx's launch in 2016. The 2011/2012 observing campaign consisted of 3 phases (maximum brightness in Sept 2011 at V = 20.0, maximum brightness in April/May 2012 at V = 19.2, and all other months when it was much fainter)(Figure 1). The 2011-2012 campaign consisted of observations from many telescopic assets including Arecibo, Herschel, HST, Magellan 6.5-m, SOAR 4-m, Spitzer, TNG 3.6-m, VATT 1.8-m, VLT 8.4-m, WHT 4.2-m and Kuiper 1.5-m. A complete list of all Bennu observations made by the OSIRIS-REx team and collaborators is shown in Table 1 (Lauretta et al. 2014).

Table 1 – Summary of Bennu observations from Lauretta et al. (2014).

| Dates | Telescope + Instrument | Observation Type |
|---|---|---|
| **1999** | | |
| 1999 Sep 15-20 | McDonald Obs. 2.1-m | VIS spectroscopy |
| 1999 Sep 23-25 | Goldstone 70-m | radar imaging and ranging |
| 1999 Sep 23,25, Oct 1 | Arecibo 305-m | radar imaging and ranging |
| **2005-2007** | | |
| 2005 Sep 4 | IRTF + SpeX | NIR spectroscopy |
| 2005 Sep 14-17 | Kuiper 1.5-m + CCD | ECAS color & lightcurve photometry |
| 2005 Sep 16,20,28, Oct 2 | Arecibo 305-m | radar imaging and ranging |
| 2005 Sep 18-19 | Goldstone 70-m | radar imaging and ranging |
| 2005 Sep-2006 May (8 nights) | Kuiper 1.5-m + CCD | phase function photometry |
| 2006 Jun 9 | VATT 1.8-m + CCD | phase function photometry |
| 2007 May 3,4,8 | Spitzer + IRS PUI & IRAC | thermal spectroscopy & photometry |
| **2011-2012** | | |
| 2011 Jul 26 | Magellan 6.5-m + FIRE | NIR spectroscopy |
| 2011 Aug 13,29, Sep 14 | WHT 4.2-m | lightcurve photometry |
| 2011 Sep 9 | Herschel Space Obs. | Far IR photometry |
| 2011 Sep 27-29 | Arecibo 305-m | radar ranging |
| 2011 Sep-2012 May (13 nights) | Kuiper 1.5-m + Mont4K | phase function photometry |
| 2012 May 2 | Magellan 6.5-m + FIRE | NIR spectroscopy |
| 2012 May 15 | VATT 1.8-m + CCD | phase function photometry |
| 2012 May 19 | SOAR 4-m + SOI | lightcurve & BVR color photometry |
| 2012 Aug 21 | Spitzer + IRAC | thermal photometry |
| 2012 Sep 17-18, Dec 10 | HST + WFPC3 | lightcurve photometry |

FIRE - Folded-port InfraRed Echellette; HST - Hubble Space Telescope; IRAC - Infrared Array Camera; IRS PUI - Infrared Spectrograph 'peak-up' imaging channels; IRTF – NASA Infrared Telescope Facility; Mont4K – Montreal 4K CCD imager; SOAR - Southern Astrophysical Research Telescope; SOI - SOAR Opitcal Imager; SpeX – 0.8-5.5 micron medium-resolution spectrograph; VATT - Vatican Advanced Technology Telescope; WFPC3 - Wide-Field Planetary Camera 3; WHT - William Heschel Telescope.

# II. Design Reference Asteroid

*The 'Design Reference Asteroid', or 'DRA', is a compilation of all that is known about the OSIRIS-REx mission target, asteroid (101955) Bennu. It contains our best knowledge of the properties of Bennu based on an extensive observational campaign that began shortly after its discovery, and has been used to inform mission plan development and flight system design. The DRA will also be compared with post-encounter science results to determine the accuracy of our Earth-based characterization efforts. The extensive observations of Bennu in 1999 has made it one of the best-characterized near-Earth asteroids. Many physical parameters are well determined, and span a number of categories: Orbital, Bulk, Rotational, Radar, Photometric, Spectroscopic, Thermal, Surface Analog, and Environment Properties. Some results described in the DRA have been published in peer-reviewed journals while others have been reviewed by OSIRIS-REx Science Team members and/or external reviewers. Some data, such as Surface Analog Properties, are based on our best knowledge of asteroid surfaces, in particular those of asteroids Eros and Itokawa.*

*This public release of the OSIRIS-REx Design Reference Asteroid is a annotated version of the internal OSIRIS-REx document "OREX-DOCS-04.00-00002, Rev 9" (accepted by the OSIRIS-REx project on 2014-April-14). The supplemental data products that accompany the official OSIRIS-REx version of the DRA are not included in this release. We are making this document available as a service to future mission planners in the hope that it will inform their efforts.*

# 1 Orbital Properties

*Orbital properties include all parameters pertaining to the orbit of Bennu including orbital elements, ephemerides, positional uncertainties and Yarkovsky accelerations.*

Bennu is defined as a potentially hazardous Apollo-type near-Earth asteroid. Any object with a perihelion distance (q) (closest distance between the object's orbit and the Sun) ≤ 1.30 au is defined as a near-Earth asteroid. Bennu is an Apollo asteroid die to its Earth crossing orbit with a perihelion distance within Earth's orbit (q = 0.89 au) and an orbital period longer than one year (P = 1.2 y). Its classification as a potentially hazardous object is based on its diameter being larger than 150 meters and a Minimum Orbit Intercept Distance (MOID, or the closest distance between the orbits of Bennu and Earth) less than 0.05 AU. Bennu's MOID is 0.0032 au (epoch 2011-Jan-1.0).

Relevant Observations:

- Ground-based optical astrometry: Between September 11, 1999 and January 20, 2013 569 CCD astrometric have been reported to the Minor Planet Center. These observations were conducted by a number of professional and amateur astronomers from around the world.
- Arecibo and Goldstone radar astrometry: Mike Nolan and Lance Benner led teams that used the Arecibo and Goldstone radio observatories to make radar observations of Bennu on 5 dates in Sept/Oct 1999, 5 dates in Sept/Oct 2005 and 3 dates in September 2011. The data resulted in accurate line-of-sight velocities and distances for Bennu.

The current orbital solution for Bennu ephemerides is JPL solution 76, which is based on 478 ground-based RA/DEC observations, as well as 22 radar delay and 7 Doppler measurements from the Arecibo and Goldstone radar observatories. This orbital solution and the associated ephemeris files are documented in JPL IOM 343R-13-001, available at ftp://ssd.jpl.nasa.gov/pub/eph/small_bodies/orex/.

Many of the Orbital Properties have been published in:

"Chesley, S.R., Farnocchia, D., Nolan, M.C., Vokrouhlický, D., Chodas, P.W., Milani, A., Spoto, F., Benner, L.A.M., Busch, M.W., Emery, J., Howell, E.S., Lauretta, D., Margot, J.-L., Rozitis, B., Taylor, P.A., 2014. Orbit and bulk density of the OSIRIS-REx target asteroid (101955) Bennu. Icarus 235, 10-22.".

## 1.1 Orbital Elements

Orbital elements consist of 6 elements that describe the orbit's shape (semi-major axis, eccentricity) and orientation (inclination, longitude of the ascending node, argument of perihelion) and the location of an object in the orbit (time of perihelion or mean anomaly).

Due to precise radar measurements and a large number of ground-based astrometry, the orbit of Bennu is very well characterized with a perihelion distance 1-sigma error of only ~4 km. Figure 2 is a diagram of the orbit of Bennu.

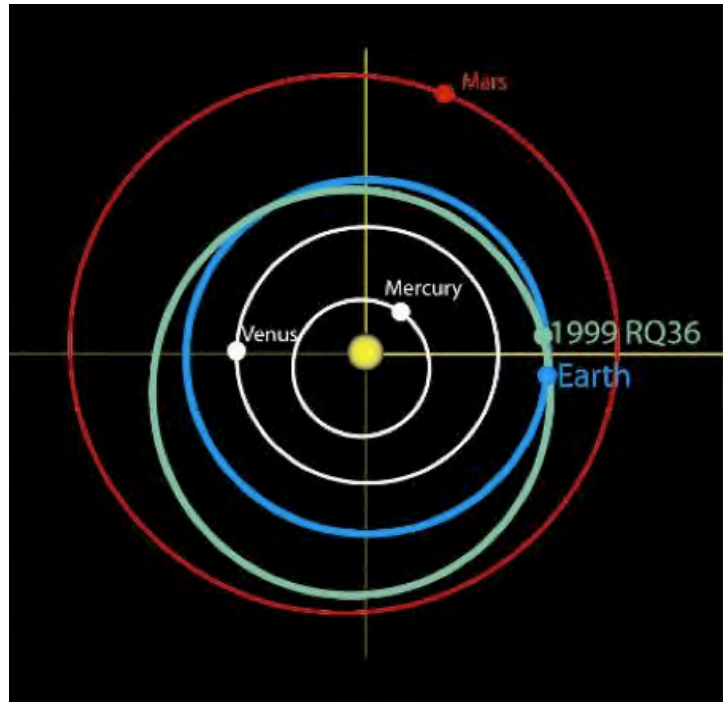

**Figure 2 - Bennu has an Earth-like low delta-V orbit that is conducive to low-cost missions.**

*Dynamical Model*
The Bennu orbit is estimated based on an extraordinarily refined dynamical model. The dominant forces on the asteroid come from the point-mass gravitational perturbations of the Sun, Moon, eight planets, and Pluto. The positions of these bodies are taken from DE424 (Folkner 2011), which also defines the reference frame of the Bennu ephemeris.

Relativistic perturbations are applied from the Sun, Moon and eight planets based on the EIH formulation (e.g., Moyer 1971). While the relativistic perturbation of the Sun is often crucial in fitting asteroid orbits, this is the first case where the relativistic perturbation of the Earth affects the ephemeris in a statistically significant way. Other small perturbations in the force model include point-mass perturbations from 25 main-belt asteroids, the $J_2$ spherical harmonic term in the geopotential, and direct solar radiation pressure (assumed area-to-mass ratio $3.07 \times 10^{-6}$ m$^2$/kg). Finally, the Yarkovsky effect is perhaps the most interesting and informative perturbation in the force model.

There are two primary numerical models at hand for the Yarkovsky effect. The simplest model applies a transverse acceleration of the form $\mathbf{a}_T = \mathbf{A}_2 r^{-d}$ where $\mathbf{A}_2$ is an estimated parameter, $\mathbf{r}$ is the heliocentric distance, and $\mathbf{d}$ is strictly in the range 0.5–3.5, but for most NEAs the value is in the range 2-3 (Farnocchia et al. 2012). For the known physical characteristics of Bennu, $\mathbf{d} = 2.25$. This model neglects out-of-plane and radial accelerations, as well as some finer details in the transverse acceleration such as hysteresis, but it captures the key aspects of the Yarkovsky effect, is computationally fast and requires no information on the physical properties of the asteroid and thus is the most generally applicable.

Our highest fidelity Yarkovsky model is produced in collaboration with D. Vokrouhlický of Charles University in Prague. This model uses the shape model of Nolan et al. (2013) in a fully nonlinear, finite-element heat transfer model, with shadowing and re-absorption. The technique computes the thermal evolution over several orbital periods until it has stabilized, and then produces a look- up table of thermal recoil acceleration as a function of orbital anomaly. This look-up table is interpolated in the orbit propagation and estimation process to derive the acceleration at any point in the orbit.

Note that the precision of the values listed below is greater than that of the uncertainties. It is customary to provide the full numerical precision when stating orbital initial conditions. This facilitates comparisons between analysts and software sets. Without the full precision it becomes difficult to disentangle force model disparities from the initial conditions. The stated uncertainties allow one to discern how many digits are significant, even while the non-significant digits do have some utility.

The formal uncertainties stated here do not capture the possible effects of mismodeling, either in the force model or the observational error model, either of which can in some cases increase the dispersions by perhaps a factor of two. While this problem is dramatically more severe for comets, in the present case we are pushing the force model fidelity and observational accuracy to unprecedented levels, which means that possible mismodeling of small perturbations still leaves some potential for statistically significant effects. For this reason any OSIRIS-REx mission operations that rely critically on the Bennu ephemeris accuracy should apply safety factors of 2-3 to the ephemeris uncertainties stated here.

### 1.1.1 Reference Frame

- Defines the reference frame that the orbital elements are measured with respect to.
- Source: Steve Chesley, JPL IOM 343R-13-001 – Chesley et al. (2014)

> Sun-centered, Earth ecliptic & equinox of J2000.0

### 1.1.2 Epoch Date (JD/ET)

- Defines the date (Julian Date/Ephemeris Time) the orbital elements are valid for.
- Source: Steve Chesley, JPL IOM 343R-13-001 – Chesley et al. (2014)

> Epoch (JD/ET) = 2455562.5

### 1.1.3 Epoch Date (Calendar Date/ET)

- Defines the date (Calendar Date/Ephemeris Time) the orbital elements are valid for.
- Source: Steve Chesley, JPL IOM 343R-13-001 – Chesley et al. (2014)

> Epoch (ET) = 2011 Jan 1.00000

### 1.1.4 Semi-major Axis

- Defines the average distance of Bennu from the Sun.
- Source: Steve Chesley, JPL IOM 343R-13-001 – Chesley et al. (2014)

$$a = 1.126391025571644 \pm 4.111300\text{E-}11 \text{ au (1-sigma uncertainty)}$$

Note that this value of semi-major axis depends critically on the assumed solar radiation pressure model. For this solution the assumed area-to-mass ratio is $3.07 \times 10^{-6}$ m$^2$/kg.

### 1.1.5 Perihelion

- Defines the minimum distance between Bennu and the Sun.
- Source: Steve Chesley, JPL IOM 343R-13-001 – Chesley et al. (2014)

$$q = 0.8968943569669300 \pm 2.390158\text{E-}08 \text{ au (1-sigma uncertainty)}$$

### 1.1.6 Aphelion

- Defines the maximum distance between Bennu and the Sun.
- Source: Steve Chesley, JPL IOM 343R-13-001 – Chesley et al. (2014)

$$Q = 1.3558876919756 \text{ au} \pm 2.5\text{E-}08 \text{ au (1-sigma uncertainty)}$$

### 1.1.7 Eccentricity

- Defines the shape of the orbit ellipse (0 for circle, 1 for parabola).
- Source: Steve Chesley, JPL IOM 343R-13-001 – Chesley et al. (2014)

$$e = 0.2037451146135014 \pm 2.123081\text{E-}08 \text{ (1-sigma uncertainty)}$$

### 1.1.8 Inclination

- Defines the vertical tilt of the orbit with respect to the reference plane.
- Source: Steve Chesley, JPL IOM 343R-13-001 – Chesley et al. (2014)

$$i = 6.03493867976381 \pm 4.721372\text{E-}08° \text{ (1-sigma uncertainty)}$$

### 1.1.9 Longitude of Ascending Node

- Defines the horizontal orientation of the ascending node (point where the orbit passes up through the reference plane) with respect to the reference frame's vernal point.

- Source: Steve Chesley, JPL IOM 343R-13-001 – Chesley et al. (2014)

$$\Omega = 2.06086819910204 \pm 6.531279\text{E-}08°\ (1\text{-sigma uncertainty})$$

### 1.1.10 Argument of Perihelion

- Defines the horizontal orientation of the point of perihelion (closest distance to Sun) as an angle measured from the ascending node to the semi-major axis.
- Source: Steve Chesley, JPL IOM 343R-13-001 – Chesley et al. (2014)

$$\omega = 66.22306886293201 \pm 9.685025\text{E-}08°\ (1\text{-sigma uncertainty})$$

### 1.1.11 Mean Anomaly

- Defines the position of Bennu on its orbit at the time of epoch.
- Source: Steve Chesley, JPL IOM 343R-13-001 – Chesley et al. (2014)

$$M = 101.70394725713800 \pm 2.6\text{E-}06°\ (1\text{-sigma uncertainty})$$

### 1.1.12 Time of Perihelion (JD/ET)

- Defines the time (Julian Date/Ephemeris Time) when Bennu reaches perihelion (closest distance to the Sun).
- Source: Steve Chesley, JPL IOM 343R-13-001 – Chesley et al. (2014)

$$TP(JD) = 2455439.1419464422 \pm 3.025935\text{E-}06\ \text{days}\ (1\text{-sigma uncertainty})$$

### 1.1.13 Time of Perihelion (Calendar Date/ET)

- Defines the Ephemeris Time Date when Bennu reaches perihelion.
- Source: Steve Chesley, JPL IOM 343R-13-001 – Chesley et al. (2014)

$$TP(ET) = 2010\ \text{Aug.}\ 30.64195$$

### 1.1.14 Orbital Period

- Defines the length of time for Bennu to orbit the Sun.
- Source: Steve Chesley, JPL IOM 343R-13-001 – Chesley et al. (2014)

$$P = 436.6487279241047 \pm 2.390600\text{E-}08\ \text{days}\ (1\text{-sigma uncertainty})$$

### 1.1.15 Minimum Orbit Intercept Distance (MOID)

- Defines the minimum distance between the orbit of Bennu and Earth
- Source: Steve Chesley – Chesley et al. (2014)

$$MOID = 0.003223 \text{ AU}$$

## 1.2 Ephemerides

*Ephemerides are the predicted positions of Bennu at future dates. These positions are derived from orbital elements valid for the epoch of interest.*

### 1.2.1 Ephemerides

- Defines the position of Bennu during the OSIRIS-REx mission
- Source: Steve Chesley, JPL IOM 343R-13-001

> The Bennu ephemeris files are described in JPL IOM 343R-13-001. The files extend from 2015-Jan-01 to 2023-May-31 and are available in both SPK and NAVIO formats. They can be downloaded from
> ftp://ssd.jpl.nasa.gov/pub/eph/small_bodies/orex/

## 1.3 Positional Uncertainties

*These are the uncertainties in the expected position of Bennu at the time of encounter. Uncertainties are due to the propagation small errors in our understanding of the current position and orbit of Bennu. All errors on the uncertainty measurements are 1-sigma.*

### 1.3.1 Positional Uncertainty Epoch (Ephemeris Time)

- Defines the Ephemeris Time (ET) when the uncertainty calculations are valid.
- Source: Steve Chesley, JPL IOM 343R-13-001 – Chesley et al. (2014)

2018 Sep 10.00000 ET

### 1.3.2 Positional Uncertainty Epoch (Julian Date)

- Defines the Julian Date (JD) when the uncertainty calculations are valid.
- Source: Steve Chesley, JPL IOM 343R-13-001 – Chesley et al. (2014)

JD 2458371.5

### 1.3.3 Radial Position Uncertainty

- Defines the uncertainty in the radial position (distance from the Sun)
- Source: Steve Chesley, JPL solution 76, JPL IOM 343R-13-001 – Chesley et al. (2014)

$$R = 3.3 \text{ km (1-sigma uncertainty)}$$

### 1.3.4 Transverse Position Uncertainty

- Defines the uncertainty in the transverse position (along the orbit or direction of travel)
- Source: Steve Chesley, JPL solution 76, JPL IOM 343R-13-001 – Chesley et al. (2014)

$$T = 3.8 \text{ km (1-sigma uncertainty)}$$

### 1.3.5 Normal Position Uncertainty

- Defines the uncertainty in the normal position (normal to the plane of the orbit of Bennu)
- Source: Steve Chesley, JPL solution 76, JPL IOM 343R-13-001 – Chesley et al. (2014)

$$N = 6.9 \text{ km (1-sigma uncertainty)}$$

### 1.3.6 Radial Velocity Uncertainty

- Defines the uncertainty in the radial velocity (velocity directly towards or away from the Sun)
- Source: Steve Chesley, JPL solution 76 – Chesley et al. (2014)

$$V_R = 8.3 \text{ mm/s (1-sigma uncertainty)}$$

### 1.3.7 Transverse Velocity Uncertainty

- Defines the uncertainty in the transverse position (along the orbit or direction of travel)
- Source: Steve Chesley, JPL solution 76 – Chesley et al. (2014)

$$V_T = 6.8 \text{ mm/s (1-sigma uncertainty)}$$

### 1.3.8 Normal Velocity Uncertainty

- Defines the uncertainty in the normal position (normal to the plane of the orbit of Bennu)
- Source: Steve Chesley, JPL solution 76 – Chesley et al. (2014)

$$V_N = 9.2 \text{ mm/s (1-sigma uncertainty)}$$

## 1.4 Yarkovsky Properties

These properties include direct measurements of the Yarkovsky force on the orbit of Bennu.

### 1.4.1 Semi-major Axis Drift Rate

- Defines the change in the semi-major axis of Bennu due to Yarkovsky forces.
- Source: Steve Chesley – Chesley et al. (2014)

$$da/dt = -19.0 \pm 0.1 \times 10^{-4} \text{ au/Myr (1-sigma uncertainty)}$$

### 1.4.2 Transverse Non-gravitational Acceleration Parameter

- Defines the acceleration of Bennu in the transverse direction (normal to radial unit vector **r**/r in the orbit plane, usually in the general direction of motion of the asteroid). In the below formula, A_transverse = A2 * (r/1 au)$^{-2.25}$ and r is the Bennu-Sun distance in au.
- Source: Steve Chesley, JPL Solution 87 – Chesley et al. (2014)

$$A2 = (-4.618 \pm 0.024) \times 10^{-14} \text{ au/day}^2 \text{ (1-sigma uncertainty)}$$

# 2 Bulk Properties

*These properties are related to the shape, volume and mass of Bennu as derived by Doppler imaging with Arecibo and Goldstone.*

Relevant Observations:

- Arecibo and Goldstone Radar Ranging and Resolved Imaging: Mike Nolan and Lance Benner led teams that used the Arecibo and Goldstone radio observatories to make radar observations on 3 dates in Sept/Oct 1999 and 6 dates in Sept/Oct 2005. The data resulted in accurate line-of-sight velocities and distances for Bennu as well as 7.5-m resolution images.
- Hubble Space Telescope Lightcurve Photometry: Mike Nolan was awarded Director's Discretionary Time on the Hubble Space Telescope in order to obtain lightcurve photometry in September and December of 2012. These observations were used to tie the rotation phase with Spitzer near-IR observations made in August/September 2012. The observations were also used to refine the rotation period and shape model. As of this DRA revision, the September observations have been collected and are consistent with the shape model.

The radar observations at Arecibo Observatory were carried out from 23 to 25 September, 1999, and again from 16 Sep to 2 Oct in 2005. Further radar observations were obtained in September, 2011, but were too low signal-to-noise ratio (SNR) to be useful for characterizing the shape of the asteroid. Observations at Goldstone were obtained on additional days. They have much lower SNR than the Arecibo images, but were obtained at times when the object was not visible from Arecibo. No further ground-based radar observations will be possible until 2037. Some additional optical observations may be possible in 2017, though it will be quite faint, $M_V = 20$, making detailed studies difficult.

Bennu is in a pseudo-resonance with the Earth, and makes close passes about every six years, though at increasingly larger distances. Data were obtained and processed using standard techniques (Magri et al., 2007). A monochromatic circularly-polarized signal is transmitted, and the reflected echo is received in both circular polarizations, though for some of these observations, data-rate limitations prevented us from recording both polarizations. The total energy received depends upon the reflectivity of the surface material and its orientation with respect to the observer. The distribution of energy in the two orthogonal polarizations gives an estimate of the surface roughness, or the amount of light multiply scattered compared to that returned by specular reflectance. Other compositional factors may also influence the polarization ratio (Benner et al. 2008; Magri et al., 2001).

By visually examining the raw images, it is clear that Bennu is roughly spheroidal with some large-scale but fairly subtle features (Figure 3). In the data from 1999, the terminator is clearly discernible due to the very high SNR of those data. The visible extent of those images is about 15-20 pixels, or 225-300 meters, and the illuminated area is roughly hemispherical. Thus an initial rough estimate of the diameter is 450-600m. The Doppler bandwidth is 4 Hz, which for this diameter lets us estimate that the rotation period is about 4 h which compares well with the final derived period of ~4.28 hours.

The Bulk Property quantities are all computed across the shape model of Bennu, which was estimated from range-Doppler imaging. Mike Nolan produced the shape model. It has 2,292 surface facets uniformly distributed across the shape. The total surface area of the shape is 0.7786 km$^2$, leading to an average area of 340 m$^2$ for each facet, and an average dimension of each facet of 18 meters. For the current shape model the circular target area of 25-meter radius corresponds to approximately 5 surface facets of the asteroid model.

*Some of the models presented in this section are based on a preliminary density. Though Chesley et al. (2014) has produced a better density estimate, the preliminary one is used in the determination of the 2.3.1 Global Shape Model, 2.3.2 Global Gravity Field Model, 2.3.3 Global Surface Slope Distribution Model, 2.3.4 Global Surface Acceleration Model, 2.3.5 Global Tilt Model and 2.3.6 Global Surface Escape Velocity Model. Updated maps are being produced.*

Steve Chesley recently determined the density of the asteroid by measuring the Yarkovsky force (see italicized note above). The nominal estimate was 0.96 g/cm$^3$ with ~14% uncertainty, leading to a range of densities from 0.83 to 1.09 g/cm$^3$. The mean radius of the asteroid is 246 meters, leading to a total volume of 0.0623 km$^3$. The corresponding mass of the asteroid for a density of 0.96 g/cm$^3$ is 5.97 x 10$^{10}$ kg, and ranges between 5.06 – 6.88 x 10$^{10}$ kg across the range of densities. The associated nominal gravitational parameter $\mu$ equals 3.98 x 10$^{-9}$ km$^3$ s$^{-2}$, with a range of 3.37 to 4.59 x 10$^{-9}$ km$^3$ s$^{-2}$ across the density range.

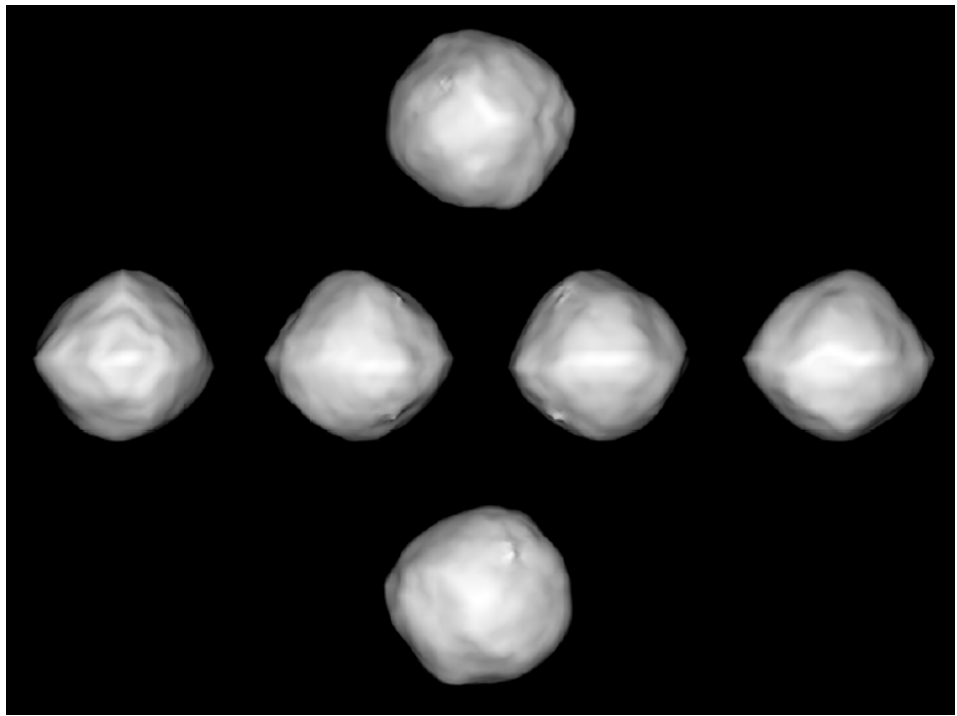

**Figure 3 - The shape of Bennu showing the views along the principal axes. Some of the apparent north-south symmetry is a model artifact from the near-equatorial geometry. The "boulder" feature is actually in the southern hemisphere.**

Dan Scheeres has developed models for the surface gravity field, surface slope distribution and surface accelerations. The slopes have been computed across the surface of Bennu for the range of different densities. Given the rapid rotation of the body and its peculiar shape, we find that the slope distribution changes significantly as a function of density. Figure 4 shows the global slope distributions for the different densities, where it is significant to note that the mean slope changes from 24° at the low density of 0.85 g/cm$^3$ to 15° at the high density of 1.15 g/cm$^3$. The surface distribution of the slopes can be seen in Figure 5, which has the slope vectors superimposed on it. These maps are for the nominal density of 1.0 g/c m$^3$, and we note that the highest slope regions are at the mid-latitudes with the lowest slopes occurring at the equatorial ridge. Although the absolute value of the slopes changes with different densities, the locations of the extreme values do not shift appreciably.

A final measure of the surface environment is the net acceleration that a particle would experience on the surface of the asteroid has also been developed by Dan Scheeres. For the nominal density these accelerations range from 0.018 to 0.068 mm/s$^2$ across the body. For the low density range they go from 0.008 to 0.058 mm/s$^2$ while for the high density it ranges from 0.027 to 0.078 mm/s$^2$ (Figure 6). Thus across all of the plausible densities the surface acceleration field strength ranges between 1 and 10 micro-Gs.

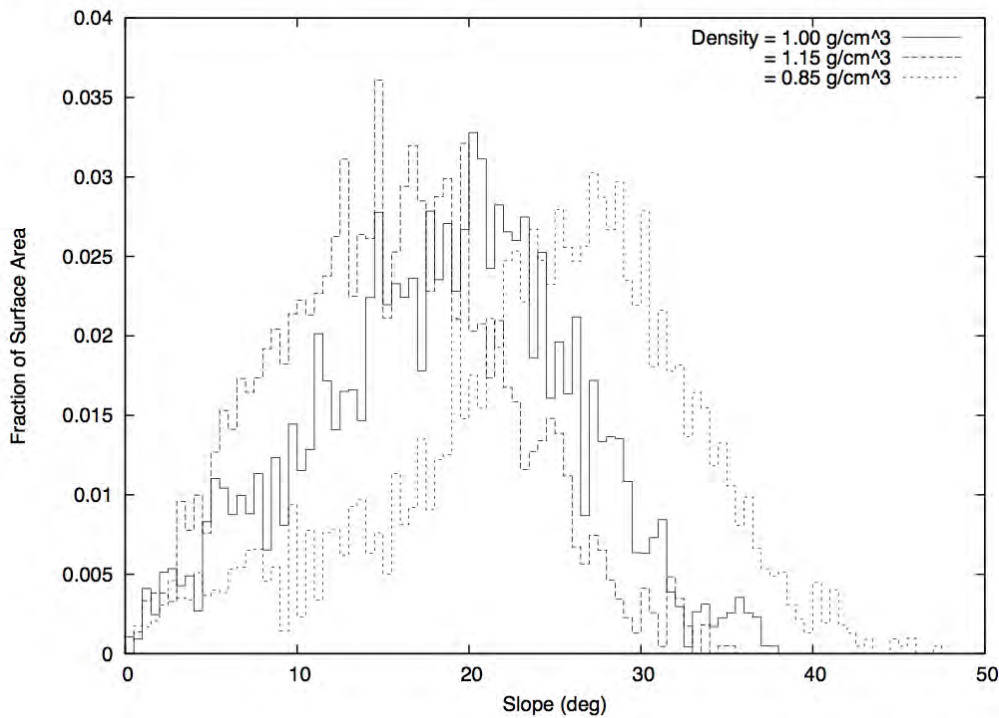

Figure 4 - Slope distributions at 0.5 degrees steps for the extreme and mean densities.

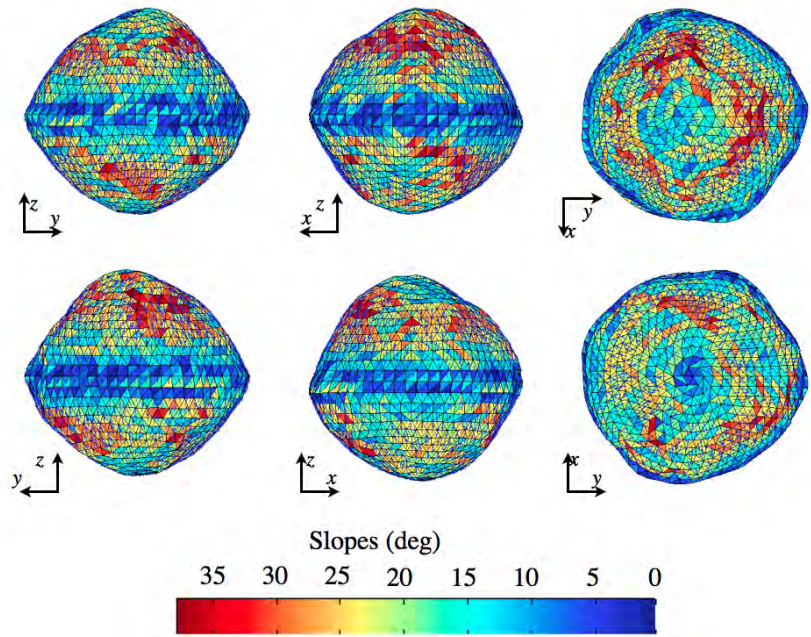

**Figure 5** - Slope distributions across the Bennu surface for the nominal density of 1 g/cm³.

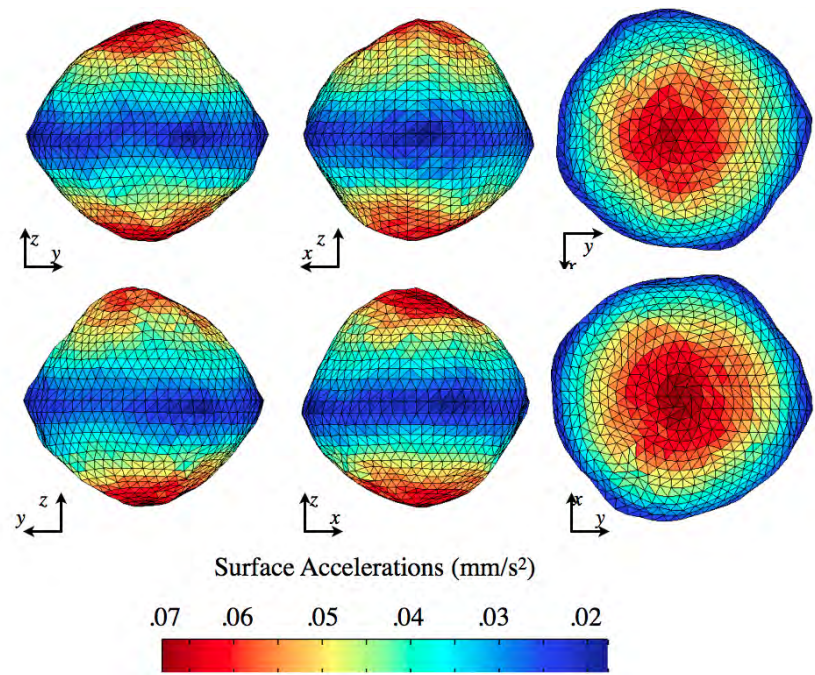

**Figure 6** - Surface accelerations across the shape model for the nominal density of 1 g/cm³.

## 2.1 Size

Size parameters have been published in:

"Nolan, M.C., Magri, M., Howell, E.S., Benner, L.A.M., Giorgini, J.D., Hergenrother, C.W., Hudson, R.S., Lauretta, D.S., Margot, J-.L., 2013. Shape model and surface properties of the OSIRIS-REx target asteroid (101955) Bennu from radar and lightcurve observations. Icarus 226, 629-640.".

### 2.1.1 Mean Diameter

- Defined as the diameter of a sphere with the equivalent volume of Bennu
- Source: Mike Nolan – Nolan et al. (2013)

$$D_{mean} = 492 \pm 20 \text{ meters (1-sigma uncertainty)}$$

### 2.1.2 Polar Dimension

- Defined as the dimension of the principal moment of inertia through the poles
- Source: Mike Nolan – Nolan et al. (2013)

$$D_{polar} = 508 \pm 52 \text{ meters (1-sigma uncertainty)}$$

### 2.1.3 Equatorial Dimensions

- Defined as the dimensions of the principal moment of inertia on the equator
- Source: Mike Nolan – Nolan et al. (2013)

$$D_{equator} = 565 \text{ x } 535 \pm (10,10) \text{ meters (1-sigma uncertainty)}$$

### 2.1.4 Dynamically Equivalent Equal Volume Ellipsoid (DEEVE) Dimensions

- Defined as the dimensions of an ellipsoid with the same volume and inertia tensor as the model shape, so that it will respond the same way to torques. Error bars have still not been derived for the DEEVE dimensions.
- Source: Mike Nolan – Nolan et al. (2013)

$$D_{DEEVE} = 517 \text{ x } 500 \text{ x } 460 \pm \text{ TBD meters}$$

### 2.1.5 Volume

- Defined as the three-dimensional space enclosed by the shape model of Bennu
- Source: Mike Nolan – Nolan et al. (2013)

$$Vol = 0.0623 \pm 0.006 \text{ km}^3 \text{ (1-sigma uncertainty)}$$

### 2.1.6 Surface Area

- Defined as the area of the exposed surface of Bennu
- Source: Mike Nolan – Nolan et al. (2013)

$$A_{surface} = 0.786 \pm 0.004 \text{ km}^2 \text{ (1-sigma uncertainty)}$$

### 2.1.7 Projected Area at Equatorial Viewing Aspect

- Defined as the projected area of the surface of Bennu as viewed from above the equator of Bennu
- Source: Mike Nolan – Nolan et al. (2013)

$$A_{proj\_equ} = 0.191 \pm 0.006 \text{ km}^2 \text{ (1-sigma uncertainty)},$$
$$493 \text{ m equivalent diameter}$$

## 2.2 Mass and Density

Mass and Density parameters have been published in:

Chesley, S.R., Farnocchia, D., Nolan, M.C., Vokrouhlický, D., Chodas, P.W., Milani, A., Spoto, F., Benner, L.A.M., Busch, M.W., Emery, J., Howell, E.S., Lauretta, D., Margot, J.-L., Rozitis, B., Taylor, P.A., 2014. Orbit and bulk density of the OSIRIS-REx target asteroid (101955) Bennu. Icarus 235, 10-22.

### 2.2.1 Bulk Density

- Defined as the mass of Bennu divided by the total volume of Bennu
- Source: Steve Chesley – Chesley et al. (2014)

$$\rho = 1260 \pm 70 \text{ kg/m}^3 \text{ (1-sigma uncertainty)}$$

### 2.2.2 Mass

- Defined as a measure of the magnitude of the gravitational force of Bennu
- Source: Steve Chesley – Chesley et al. (2014)

$$M = (7.8 \pm 0.9) \times 10^{10} \text{ kg (1-sigma uncertainty)}$$

### 2.2.3 Gravitational Parameter

- Defined as the product of the gravitational constant and the mass of Bennu
- Source: Steve Chesley – Chesley et al. (2014)

$$GM = \mu = 5.2 \pm 0.6 \text{ m}^3/\text{s}^2 \text{ (1-sigma uncertainty)}$$

### 2.2.4 Area-to-Mass Ratio

- Defined as the ratio between the cross-sectional area exposed to solar radiation and the mass of Bennu
- Source: Steve Chesley – Chesley et al. (2014)

$$(2.4 \pm 0.1) \times 10^{-6} \text{ m}^2/\text{kg (1-sigma uncertainty)}$$

## 2.3 Global Models

Global Shape Model has been published in

Nolan, M.C., Magri, M., Howell, E.S., Benner, L.A.M., Giorgini, J.D., Hergenrother, C.W., Hudson, R.S., Lauretta, D.S., Margot, J-.L., 2013. Shape model and surface properties of the OSIRIS-REx target asteroid (101955) Bennu from radar and lightcurve observations. Icarus 226, 629-640, and at the Small Bodies Node of the Planetary Data System [Asteroid (101955) Bennu Shape Model V1.0, EAR-A-I0037-5-BENNUSHAPE-V1.0].

Other global models have been internally reviewed.

### 2.3.1 Global Shape Model

- Defined as the shape model of Bennu as derived from Arecibo and Goldstone radar imaging data obtained in 1999 and 2005. The current version of the radar-derived global shape model is the October 2012 (oct12) version. It is provided in vector (x,y,z coordinates) and Lat/Long versions.
- Source: Mike Nolan – Nolan et al. (2013)

> The vector (x,y,z) model can be found at "Small Bodies Node of the Planetary Data System [Asteroid (101955) Bennu Shape Model V1.0, EAR-A-I0037-5-BENNUSHAPE-V1.0]".

### 2.3.2 Global Gravity Field Model

- Defined as a spatial model of the gravity experience by a particle on the surface of Bennu.

- Source: Dan Scheeres

> The data files for this model are not part of this public release of the OSIRIS-REx Bennu Design Reference Asteroid.

### 2.3.3 Global Surface Slope Distribution Model

- Defined as the complement of the angle between the surface normal vector and the total gravitational plus centripetal acceleration acting on the center of the facet.
- Source: Dan Scheeres

> The data files for this model are not part of this public release of the OSIRIS-REx Bennu Design Reference Asteroid.

### 2.3.4 Global (Tangent/Normal/Total) Surface Acceleration Model

- Defined as the component of the acceleration at the center of a modeled surface facet that is normal to the modeled surface facet of Bennu.
- Source: Dan Scheeres

> The data files for this model are not part of this public release of the OSIRIS-REx Bennu Design Reference Asteroid.

### 2.3.5 Global Surface Tilt Model

- Defined as the angle between the surface normal and the radius vector of the surface facet. Equals "0" for a sphere.
- Source: Dan Scheeres

> The data files for this model are not part of this public release of the OSIRIS-REx Bennu Design Reference Asteroid.

### 2.3.6 Global Surface Escape Velocity Model

- Defined as the minimum speed that a particle can be launched normal to the surface of the rotating Bennu and will ideally escape.
- Source: Dan Scheeres

> The data files for this model are not part of this public release of the OSIRIS-REx Bennu Design Reference Asteroid.

# 3 Rotational Properties

*Rotational properties are those related to the rotation state of Bennu about its axis(axes).*

Relevant Observations

- Arecibo and Goldstone Radar Ranging and Resolved Imaging: Mike Nolan and Lance Benner led teams that used the Arecibo and Goldstone radio observatories to make radar observations on 3 dates in Sept/Oct 1999 and 6 dates in Sept/Oct 2005. The data resulted in accurate line-of-sight velocities and distances for Bennu as well as 7.5-m resolution images.
- Krugly rotational lightcurve photometry – Yu. Krugly conducted lightcurve photometry over the course of 3 nights in September 1999 with 0.7-m and 1.0-m telescopes in the Ukraine (Krugly et al. 2002). Due to their short observational arc on each night, their rotation period was estimated to be one-half of the true value.
- Kuiper rotational lightcurve photometry – Carl Hergenrother obtained lightcurve photometry on Sept. 14-17, 2005 with the Kuiper 1.5-m telescope. Photometry was conducted in the ECAS *w* band.
- WHT rotational lightcurve photometry – Javier Licandro acquired lightcurve observations with the 4.2-m WHT in August and September of 2011. These observations have been reduced. Due to poor observing conditions the data was not of a sufficient quality to add to our rotation state knowledge of Bennu.
- HST rotational lightcurve photometry – Mike Nolan obtained lightcurve photometry with the Hubble Space Telescope in Sept. and Dec. 2012.

## 3.1 Rotation State

Time-series photometry was obtained over the course of 4 consecutive nights in September 2005. Observations were obtained using multiple Eight-Color Asteroid Survey (ECAS) filters though the *w* (0.70 μm) filter was primarily used. A $10^{th}$ order Fourier series fit finds a rotation period of 4.297812 with amplitude of 0.17 magnitudes (Figure 7). The low amplitude and trimodal (three maxima and three minima) lightcurve is consistent with the rotation of a nearly spherical body observed at high phase angles.

Yu. Krugly conducted lightcurve photometry over the course of 3 nights in September 1999 with 0.7-m and 1.0-m telescopes in the Ukraine. Krugly et al. (2002) found a period that is half of our result (2.146 hr vs 4.2986 hr). The discrepancy is due to not observing a total rotation period during any of their nights. As a result, they incorrectly arrived at a sub-multiple of the actual period. By comparison, three of our four nights covered a complete rotation giving more credence to the 4.2986 hr period. The longer period is also consistent with the radar observations.

The pole orientation of Bennu was measured from the radar and thermal observations to be (ʃ,β) = 45°,-88° ± 2° which denotes retrograde rotation and a pole nearly perpendicular to the ecliptic plane (Nolan et al. 2013). The lightcurves from Krugly et al. (2002) in 1999, the lightcurves from 2005, and the radar images from both apparitions were compared against the expected lightcurves and radar images based on the shape model. The rotation rate was fit while holding the shape model constant. Due to the uncertainty in the pole position, the rotation rate was fit against shape models with the following pole positions (in ecliptic longitude, ecliptic latitude): 0°, -90°; 0°, -85°; 60°, -85°; 120°, -85°; 180°, -85°; 240°, -85°; and 300°, -85°.

The best-fit sidereal rotation period is 4.297461 ± 0.002 h. The error is a 2-sigma error and corresponds to ~11 rotations over the 6-year observation interval. The result is one of many discrete solutions hence the reason the period has more significant digits than the error. This period matches the shape fit within the error bars though the lightcurve data were used to help constrain the shape and rotation fit. The low amplitude and trimodal (three maxima and three minima) lightcurve is consistent with the rotation of a nearly spheroidal body with some large but subtle features body observed at high phase angles.

Rotation State parameters have been published in

"Nolan, M.C., Magri, M., Howell, E.S., Benner, L.A.M., Giorgini, J.D., Hergenrother, C.W., Hudson, R.S., Lauretta, D.S., Margot, J.-L., 2013. Shape model and surface properties of the OSIRIS-REx target asteroid (101955) Bennu from radar and lightcurve observations. Icarus 226, 629-640."

and

"Hergenrother, C.H., M. Nolan, R. Binzel, E. Cloutis, M.A. Barucci, P. Michel, D. Scheeres, C.D. d'Aubigny, D. Lazzaro, N. Pinilla-Alonso, H. Campins, J. Licandro, B.E. Clark, B. Rizk, E. Beshore, D. Lauretta 2013. Lightcurve, Color and Phase Function Photometry of the OSIRIS-REx Target Asteroid (101955) Bennu. Icarus 226, 663-670.".

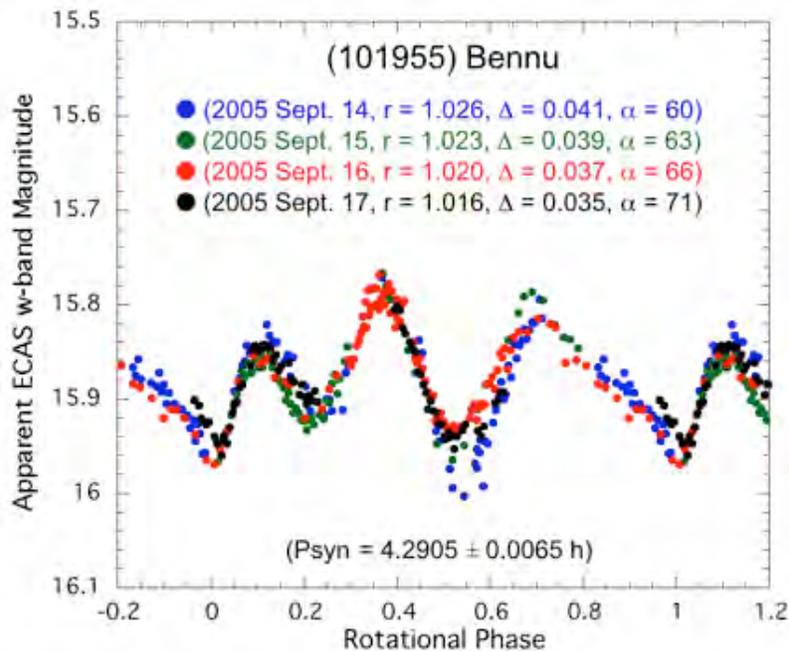

**Figure 7 - Composite rotational lightcurve of Bennu (formerly 1999 RQ36) from observations made during 4 consecutive nights in September 2005. This figure is a version of Figure 1 from Hergenrother et al. (2013).**

### 3.1.1 Sidereal Rotation Period

- Defined as the period of time for Bennu to rotate about its axis relative to the stellar reference frame.
- Source: Mike Nolan and Carl Hergenrother – Nolan et al. (2013) and Hergenrother et al. (2013)

$$P = 4.297461 \pm 0.002 \text{ hours (1-sigma uncertainty)}$$

### 3.1.2 Direction of Rotation

- Defined as the direction of rotation about the rotation axis of Bennu.
- Source: Mike Nolan – Nolan et al. (2013)

$$\text{Dir of Rot} = \text{Retrograde}$$

### 3.1.3 Obliquity (with respect to the orbit plane of Bennu)

- Defined as the angle of the rotation axis of Bennu with respect to the orbit plane of Bennu. The angle with respect to the Ecliptic is 178°.
- Source: Mike Nolan – Nolan et al. (2013)

$$\text{Obl} = 176 \pm 2° \text{ (1-sigma uncertainty)}$$

### 3.1.4 Pole Position

- Defined as the ecliptic coordinates of Bennu's north pole.
- Source: Mike Nolan – Nolan et al. (2013)

$$\text{Ecliptic Long, Lat} = (45°, -88°) \pm 3° \text{ (1-sigma uncertainty)}$$

### 3.1.5 Lightcurve Amplitude

- Defined as the amplitude of the rotational lightcurve.
- Source: Mike Nolan and Carl Hergenrother – Hergenrother et al. (2013)

$$\text{Rot Amp} = 0.15 \pm 0.15 \text{ magnitudes (3-sigma uncertainty)}$$

### 3.1.6 Non-principal Axis (NPA) Rotation

- Defined as the existence of rotation about two or more axes.
- Source: Carl Hergenrother

No NPA rotation detected

# 4 Radar Properties

*Properties listed in section 4 are related to surface properties derived from radar observations.*

For specifics on the radar observations and methods, see Nolan et al. (2013) and DRA section 2. Radar Properties have been published in
"Nolan, M.C., Magri, M., Howell, E.S., Benner, L.A.M., Giorgini, J.D., Hergenrother, C.W., Hudson, R.S., Lauretta, D.S., Margot, J-.L., 2013. Shape model and surface properties of the OSIRIS-REx target asteroid (101955) Bennu from radar and lightcurve observations. Icarus 226, 629-640.".

## 4.1 Radar Characteristics

Relevant Observations

- <u>Arecibo and Goldstone radar ranging and imaging</u> - Mike Nolan and Lance Benner led teams that used the Arecibo and Goldstone radio observatories to make radar observations on 3 dates in Sept/Oct 1999, 6 dates in Sept/Oct 2005 and 3 dates in Sept 2011. The data resulted in accurate line-of-sight velocities and distances for Bennu as well as 7.5-m resolution images.

### 4.1.1 Opposite-circular (O-C) Radar Albedo (12.6-cm wavelength)

- Defined as the fraction of radar energy reflected back at the emitter compared to a perfect reflector. Measured at the 12.6-cm wavelength. Variations of ± 10% seen over the surface.
- Source: Mike Nolan – Nolan et al. (2013)

$$\text{O-C Radar Albedo (12.6-cm)} = 0.12 \pm 0.04$$

### 4.1.2 Radar Circular Polarization Ratio (12.6-cm wavelength)

- Defined as a polarization in which the electric field of the wave does not change strength but only changes direction in a rotary type manner. Measured at the 12.6-cm wavelength. Variations of ±10% seen over the surface.
- Source: Mike Nolan – Nolan et al. (2013)

$$\text{Radar polarization ratio (12.6-cm)} = 0.18 \pm 0.03 \text{ (1-sigma uncertainty)}$$

### 4.1.3 Radar Circular Polarization Ratio (3.5-cm wavelength)

- Defined as a polarization in which the electric field of the passing wave does not change strength but only changes direction in a rotary type manner. Measured at the 3.5-cm wavelength. Variations of ± 10% seen over the surface.
- Source: Mike Nolan – Nolan et al. (2013)

$$\text{Radar polarization ratio (3.5-cm)} = 0.19 \pm 0.03 \text{ (1-sigma uncertainty)}$$

# 5 Photometric Properties

Relevant Observations

- <u>Kuiper 1.5-m visible photometry</u> - Carl Hergenrother used *V, R* and ECAS filter photometry obtained between Sept 2005 and June 2012. The data taken with telescopes at the University of Arizona covered a range of phase angles from 15° to 100°. The absolute magnitude and slope of the phase function suggest a very low albedo asteroid consistent with being carbonaceous. ECAS filter photometry in September 2005 confirmed Ellen Howell's classification of Bennu as a B-type asteroid.
- <u>SOAR 4.2-m visible photometry</u> - The SOAR 4.2-m (May 2012) was used for lightcurve work to measure rotationally resolved colors of Bennu in the B, V, R and I bands. Due to poor viewing conditions only V and R colors were obtained. The poor conditions limited the usefulness of the SOAR photometry for constraining the rotation period and lightcurve of Bennu.

## 5.1 Phase Function

Determining the relationship between the brightness of an asteroid and the observation phase angle ($\alpha$) allows an estimate of the absolute magnitude (H) or brightness of the asteroid at zero degrees phase angle (where the phase angle is defined as the Sun-asteroid-Earth angle). The relationship between brightness and phase angle is determined by normalizing the observed apparent magnitude of Bennu to a heliocentric and geocentric distance of 1 AU. The phase function is modeled by a simple linear least squares fit and by the IAU H-G photometric system derived for airless bodies by Bowell et al. (1989).

Italicized text in this section is directly from pages 667-668 of

"Hergenrother, C.H., M. Nolan, R. Binzel, E. Cloutis, M.A. Barucci, P. Michel, D. Scheeres, C.D. d'Aubigny, D. Lazzaro, N. Pinilla-Alonso, H. Campins, J. Licandro, B.E. Clark, B. Rizk, E. Beshore, D. Lauretta 2013. Lightcurve, Color and Phase Function Photometry of the OSIRIS-REx Target Asteroid (101955) Bennu. Icarus 226, 663-670.".

*The dataset for Bennu consists of V-, R-, and ECAS w-band magnitude measurements made during the 2005/2006 apparition between 2005 Sept. 14 UT and 2006 Jun. 19 UT and during the 2011/2012 apparition between 2011 Sep. 26 UT and 2012 May 29 UT. The range of phase angles observed in 2005/2006 range from 15.0° to 80.2° and those observed in 2011/2012 range from 17.7° to 95.6°. R-band data were transformed to V using the measured V-R color index of +0.36 from this work. The ECAS w-band data were transformed to V using a V-w color index of +0.30 derived from Howell (1995). A linear least squares fit through the data yields an absolute magnitude of ($H_v$) = 20.61 ± 0.20 and phase slope (β) of 0.040 ± 0.003 magnitude per degree of phase angle (*Figure 8*). The linear fit does not include any opposition effect, which usually occurs at phase angles of much less than 15°. Low albedo carbonaceous asteroids show shallow opposition effects of < 0.3 magnitudes (Shevchenko et al. 2008, Muinonen et al. 2010). Shevchenko and Belskaya (2010) found a correlation between the albedo and the magnitude of the opposition effect. Their study of 33 low albedo asteroids found that the magnitude of the opposition effect decreased with decreasing albedo. Based on their results (and Figure 2 of*

*Shevchenko and Belskaya [2010] in particular), Bennu with an albedo of 0.045 ± 0.015 should have an opposition effect of 0.10 $^{+0.10}_{-0.10}$ magnitudes. Assuming such an opposition effect for Bennu results in an $H_V$ of 20.51 $^{+0.10}_{-0.10}$.*

*The Minor Planet Center (MPC) assumes a G value of +0.15 for most asteroids. A fit to the photometry reported here using G=+0.15 yields an $H_v$ of 20.73 which is consistent with the value determined by the MPC of 20.9. The discrepancy between the values for $H_v$ is because the MPC uses all photometry submitted to them. Most of the MPC photometric data are low-precision and low-accuracy due to systematic and random errors (Oszkiewicz et al. 2011). Our dataset is limited to the data collected by Carl Hergenrother and is presented in Hergenrother et al. (2013). Solving for both H and G produces values of $H_v$ = 19.97 ± 0.26 and G = –0.12 ± 0.06. The H-G system is known to have difficulty with low albedo objects like Bennu (Shevchenko et al. 2010). The $H_V$ value from this system is derived from an extrapolated opposition effect that is much larger than those observed for dark objects (Shevchenko and Belskaya 2010). As a result the values of $H_v$ = 19.97 ± 0.26 and G = –0.12 ± 0.06 are suspect and should not be used without low phase angle observations to constrain the opposition effect.*

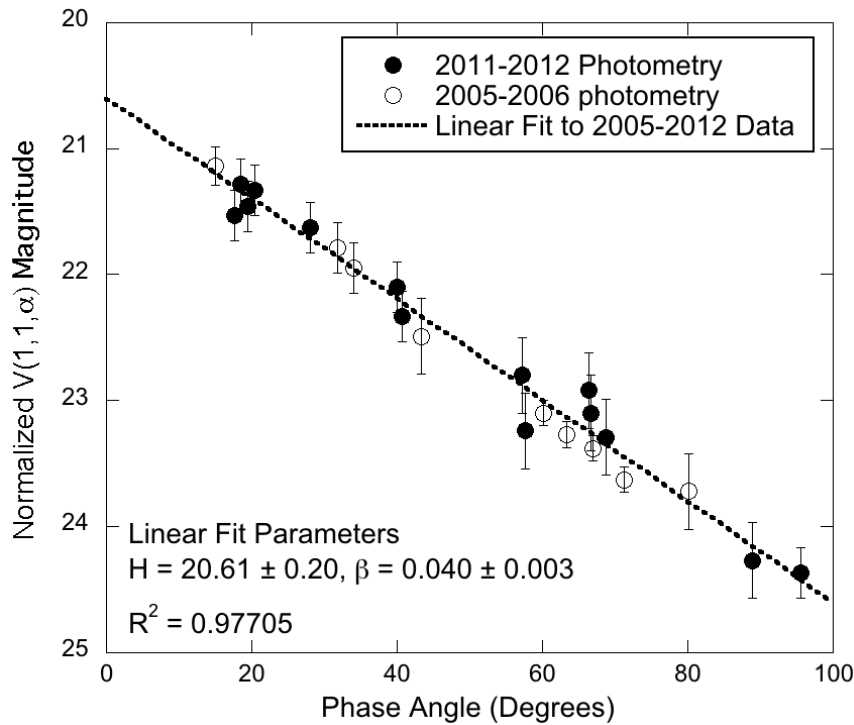

**Figure 8 - Linear phase function for Bennu phase photometry obtained between 2005 and 2012. Observed magnitudes were normalized to a distance of 1 AU from the Sun and Earth. The relation is known to deviate from linearity for other dark asteroids at phase angles < 3° and > 100°.**

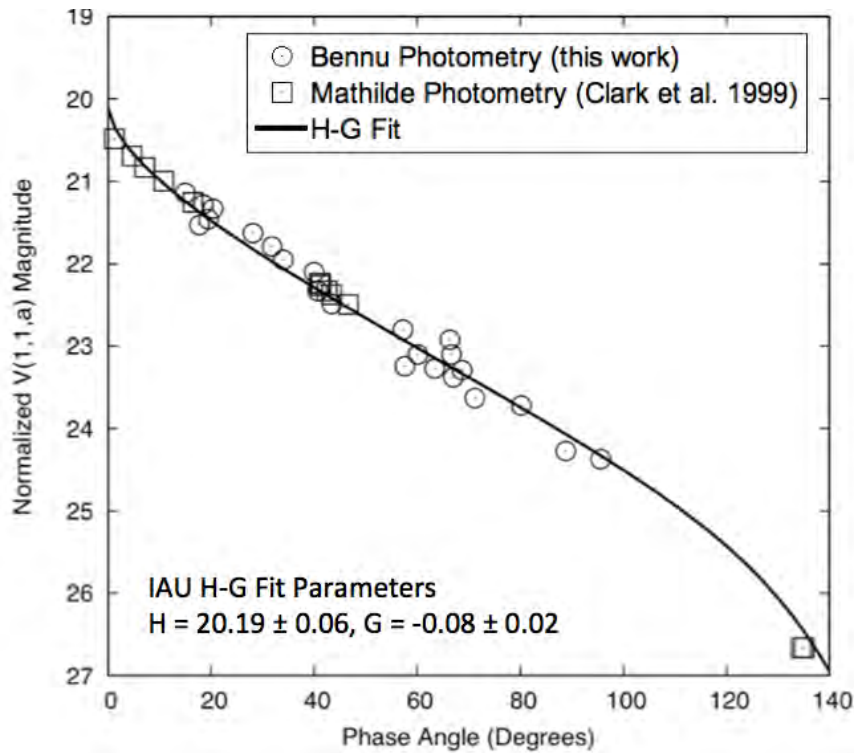

Figure 9 - H-G phase function fit to a combination of Bennu data and ground-based and NEAR photometry of (253) Mathilde from Clark et al. (1999). The Mathilde photometry has been shifted fainter by 10.02 magnitudes to match the Bennu data. Magnitudes are normalized to 1AU from the Sun and Earth as in Figure 8.

*Few carbonaceous asteroids have been precisely observed over a large range of phase angles. The NEAR spacecraft obtained measurements of the Main Belt asteroid (253) Mathilde over a wide range of phase angles during its 1997 flyby. Mathilde, with a mean diameter of 52.8 km, is a larger object than Bennu at 0.5 km (Veverka et al. 1999, Nolan et al. 2012). According to the analysis by Delbo and Tanga (2009), asteroids in the 50 km diameter range have lower thermal inertia values than those in the sub-km range, suggesting that larger asteroids have surfaces covered with finer regolith material than smaller asteroids like Bennu. Similarities between the two objects include carbonaceous taxonomies, low albedos (Mathilde with $0.047 \pm 0.005$ and Bennu with $0.03 \pm 0.01$) and low bulk densities (Mathilde with a density of $1.3 \pm 0.3$ g cm$^{-3}$ and Bennu with $0.98 \pm 0.15$ g cm$^{-3}$) (Yeomans et al. 1997, Veverka et al. 1999, Clark et al. 1999).*

Photometry obtained over a range of phase angles from 1.2° to 136.2° are presented in Clark et al. (1999). For our purposes, we merged the two distinct data sets by shifting the Mathilde photometry fainter by 10.02 magnitudes to match the Bennu photometry. Since the data overlie each other in the region of overlap, we suggest that Mathilde low and high-phase angle observations can be used as approximate expected measurements for Bennu. IAU H-G values for the combined data are $H_v = 20.19 \pm 0.06$ and $G = -0.08 \pm 0.02$ (Figure 9). Due to the measurements at very low phase angles the opposition effect is better constrained by the *H-G* system, however a close examination of the curve fit shows that the solution predicts an exaggerated opposition effect for Bennu (Figure 9). A visual inspection of the combined data at very low phase angles suggests, rather, a shallow opposition effect with a best fit $H_v$ of $20.40 \pm 0.05$ is more appropriate.

Phase Function parameters have been published in

"Hergenrother, C.H., M. Nolan, R. Binzel, E. Cloutis, M.A. Barucci, P. Michel, D. Scheeres, C.D. d'Aubigny, D. Lazzaro, N. Pinilla-Alonso, H. Campins, J. Licandro, B.E. Clark, B. Rizk, E. Beshore, D. Lauretta (2013). Lightcurve, Color and Phase Function Photometry of the OSIRIS-REx Target Asteroid (101955) Bennu. Icarus 226, 663-670.".

### 5.1.1 Absolute Magnitude without an Opposition Effect

- Defined as the brightness of an object at 0° phase angle and 1 AU from the Sun and observer. This absolute magnitude was derived from a linear fit to the Bennu V- and R-band phase function photometry (data spans from 15° to 100° phase angle). No provision for an opposition effect was included.
- Source: Carl Hergenrother – Hergenrother et al. (2013)

$$H = 20.61 \pm 0.20 \text{ (3-sigma uncertainty)}$$

### 5.1.2 Absolute Magnitude with an Opposition Effect

- Defined as the brightness of an object at 0° phase angle and 1 AU from the Sun and observer. This absolute magnitude was derived from a linear fit to the combined Bennu and (253) Mathilde (from Clark et al. 1999) V- and R-band phase function photometry (data spans from 15° to 100° phase angle).
- Source: Carl Hergenrother – Hergenrother et al. (2013)

$$H = 20.40 \pm 0.05 \text{ (3-sigma uncertainty)}$$

### 5.1.3 Magnitude of Opposition Effect

- Defined as the brightening of a surface at very small phase angles due to either shadow hiding or coherent backscattering. This parameter is based on values typical for dark asteroids (Shevchenko et al. 2008, Icarus 196, 601-611.).
- Source: Carl Hergenrother – Hergenrother et al. (2013)

$$\Delta H = 0.10 \pm \frac{+0.10}{-0.10} \text{ magnitudes (3-sigma uncertainty)}$$

### 5.1.4 Linear Phase Slope

- Defined as the change in astronomical brightness of an object at different phase angles. The brightness is normalized to 1 AU from the Sun and observer. The value for this parameter is based on data that spans from 15° to 100° phase angle.
- Source: Carl Hergenrother – Hergenrother et al. (2013)

$$\beta = 0.040 \pm 0.003 \text{ magnitudes per degree of phase angle (3-sigma uncertainty)}$$

### 5.1.5 IAU H-G Phase Function

- Defined as a phase function based on the IAU H-G system. This system is the standard for modeling the brightness behavior of asteroids at different phase angles. The combined Bennu and Mathilde data are modeled for this parameter. A close examination of the curve fit shows an exaggerated opposition effect solution resulting in an H value that is too high.
- Source: Carl Hergenrother – Hergenrother et al. (2013)

$$H = 20.19 \pm 0.06, G = -0.08 \pm 0.02 \text{ (1-sigma uncertainty)}$$

## 5.2 Bidirectional Reflectance Distribution Function

The Bidirectional Reflectance Distribution Function (BRDF) is a function that defines how light is reflected by an opaque surface. The function is relative to the incoming and outgoing light direction with respect to the surface normal.

The Photometric Correction Working Group (PCWG) was tasked with updating predictions for the disk-resolved brightness of Bennu. The PCWG effort was led by Driss Takir, Beth Clark, and Christian Drouet d'Aubigny, internally reviewed by the PCWG and externally reviewed by Jian-Yang Li (Planetary Science Institute) and Bonnie Buratti (Jet Propulsion Laboratory).

Christian d'Aubigny produced a preliminary set of OCAMS models for the BRDF of Bennu (d'Aubigny 2011). The PCWG first reproduced the d'Aubigny models using the same data and the same functions that d'Aubigny used in 2011. This was done in order to validate and verify our software. Inputs to Christian d'Aubigny's models were updated and the functions optimized to fit the new data for Bennu. This resulted in a new set of Bidirectional Reflectance Distribution Functions (BRDFs) for Bennu. The new inputs and their errors are shown in (Table 2):

- New size for Bennu: $0.492 \pm 0.020$ km (from Nolan et al. 2013).
- Newly calibrated ground-based photometric phase curve data (from DRA).
- New low phase-angle (proxy) data from asteroid 253 Mathilde from Clark et al. (1999).

Using the error estimates shown in Table 2, Minimum, Maximum, and Nominal BRDFs were produced for Bennu at 550 nm. The Minimum and Maximum models capture the uncertainties in the size and the low and high phase-angle behavior, and the scatter in the moderate phase angle ground-based observations of Bennu.

**Table 2 - Description of inputs used to reproduce nominal, maximum, and minimum models.**

|  | **Maximum Brightness** | **Nominal Brightness** | **Minimum Brightness** |
|---|---|---|---|
|  | Reduced Magnitude (Vmag-error)* | Reduced Magnitude (Vmag)* | Reduced Magnitude (Vmag+error)* |
| Diameter (km)** | 0.472 | 0.492 | 0.512 |

*The Vmag values are in Hergenrother et al. (2013):
 (https://sciwik.lpl.arizona.edu/wiki/pages/A7H8y8c/Bennu_and_the_Operational_Environment.html).

For our nominal model, the Reduced Vmag values also include NEAR spacecraft data of Mathilde (as the best available proxy data) at the lowest and highest phase angles (Clark et al. 1999). The Minimum and Maximum models capture the scatter in the moderate phase angle ground-based observations of Bennu, and the uncertainties in the size and the low and high phase-angle behavior.

**The Diameter of Bennu is from the OSIRIS-REx Design Reference Asteroid document (this document).

*Model BRDFs for Bennu*

Reflectance, $r$, is directly related to $[I/\mathcal{F}](i,e,\alpha)$ as described in the following Lommel- Seeliger Radiance Factor (*RADF*) function (D'Aubigny 2011, Hapke 1993, Buratti et al. 2012; Li et al 2004 and 2009). We will use this function to model the disk-resolved brightness when we get to asteroid Bennu (Takir and Clark 2013). For now, we can use it to constrain the average disk-resolved brightness across the surface of Bennu by fitting the disk-integrated ground-based phase curve of Bennu from Hergenrother et al. (2013), assuming the following form of the Lommel-Seeliger *RADF* function:

$$RADF(i,e,\alpha) = \pi r_{LS}(i,e,\alpha) = \frac{\varpi_o}{4} \frac{\mu_o}{\mu_o + \mu} f(\alpha) = [\frac{I}{\mathcal{F}}](i,e,\alpha) \ .$$

(Seeliger 1884, Helfenstein and Veverka 1989)

Solving for $r_{LS}$, this disk-resolved function can be integrated to give the disk-integrated magnitude of Bennu. We start with the disk averaged albedo, $A(\alpha)$, as a function of phase angle, assuming a spherical shape (Li 2004):

$$A(\alpha) = 2 \int_{\theta=0}^{\pi/2} \int_{\phi=\alpha-\pi/2}^{\pi/2} r_{LS}(i,e,\alpha) \, \mu \cos(\theta) \, d\phi d\theta \ ,$$

where $\mu_o = \cos(i)$, $\mu=\cos(e)$, $i$ is the incidence angle, $e$ is the emission angle, $\alpha$ is the phase angle, $\theta$ is the photometric latitude, and $\phi$ is the photometric longitude.

The disk-integrated absolute magnitude is then a function of the disk-averaged albedo (Pravec and Harris 2007):

$$H(\alpha) = -5 log_{10}(\frac{D\sqrt{A(\alpha)}}{1329})$$

where D is the diameter (in km), and 1329 is the value in (km) determined as a good empirical reference diameter in Fowler and Chillemi (1992).

Functions in the form of BRDFs are requested by the OSIRIS-REx Instrument Teams to be used to predict Bennu's brightness.

Therefore, we present our Modified Lommel-Seeliger model BRDF for use in predicting the $I/\mathcal{F}(i,e,\alpha)$(reflectance) of Bennu (Takir and Clark 2013):

$$BRDF = \frac{\varpi_o}{4\pi}\frac{1}{\mu_o + \mu}f(\alpha) = \frac{\left[\frac{I}{F}\right](i,e,\alpha)}{\mu_o\pi} = \frac{A_{LS}}{\mu_o + \mu}f(\alpha)$$

where $A_{LS} = \varpi_o / 4\pi$ is the Lommel-Seeliger "albedo" and $f(\alpha) = e^{\beta\alpha+\gamma\alpha^2+\delta\alpha^3}$. Table 3 shows the models for Minimum, Nominal, and Maximum predicted brightness of Bennu at 550 nm.

**Table 3 - Modified Lommel-Seeliger functions that predict $[I/\mathcal{F}](i,e,\alpha)$(reflectance) of Bennu at 550 nm. $A_{LS}$ is Lommel-Seeliger Albedo and $f(\alpha) = e^{\beta\alpha+\gamma\alpha^2+\delta\alpha^3}$.**

|  | $A_{LS}$ | β | γ | δ |
|---|---|---|---|---|
| **Nominal** | 0.030 | -4.36x10$^{-2}$ | 2.69x10$^{-4}$ | -9.90x10$^{-7}$ |
| **Maximum** | 0.033 | -2.51x10$^{-2}$ | 1.62x10$^{-4}$ | -18.77x10$^{-7}$ |
| **Minimum** | 0.021 | -4.17x10$^{-2}$ | 2.72x10$^{-4}$ | -11.96x10$^{-7}$ |

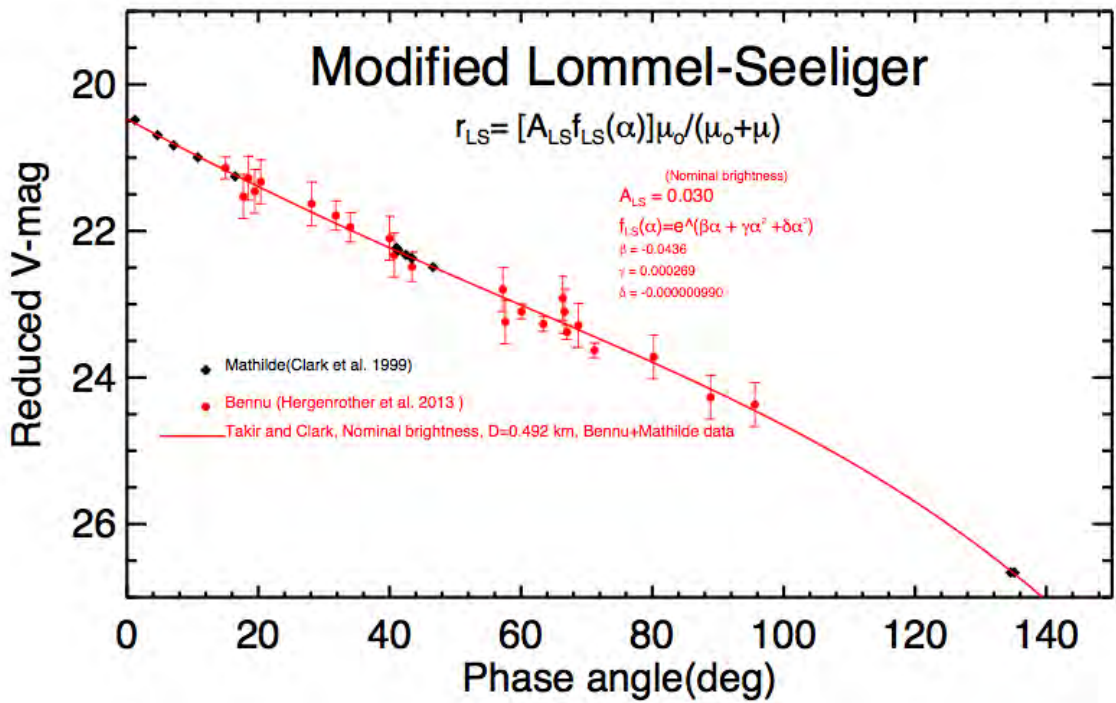

**Figure 10 -** The reduced V magnitude of Bennu as a function of phase angle predicted by the Lommel-Seeliger model is shown compared with the ground-based measurements. Shown is the Nominal model (red), which also includes Mathilde data (black diamonds) $r_{LS}$ is in units of sr$^{-1}$.

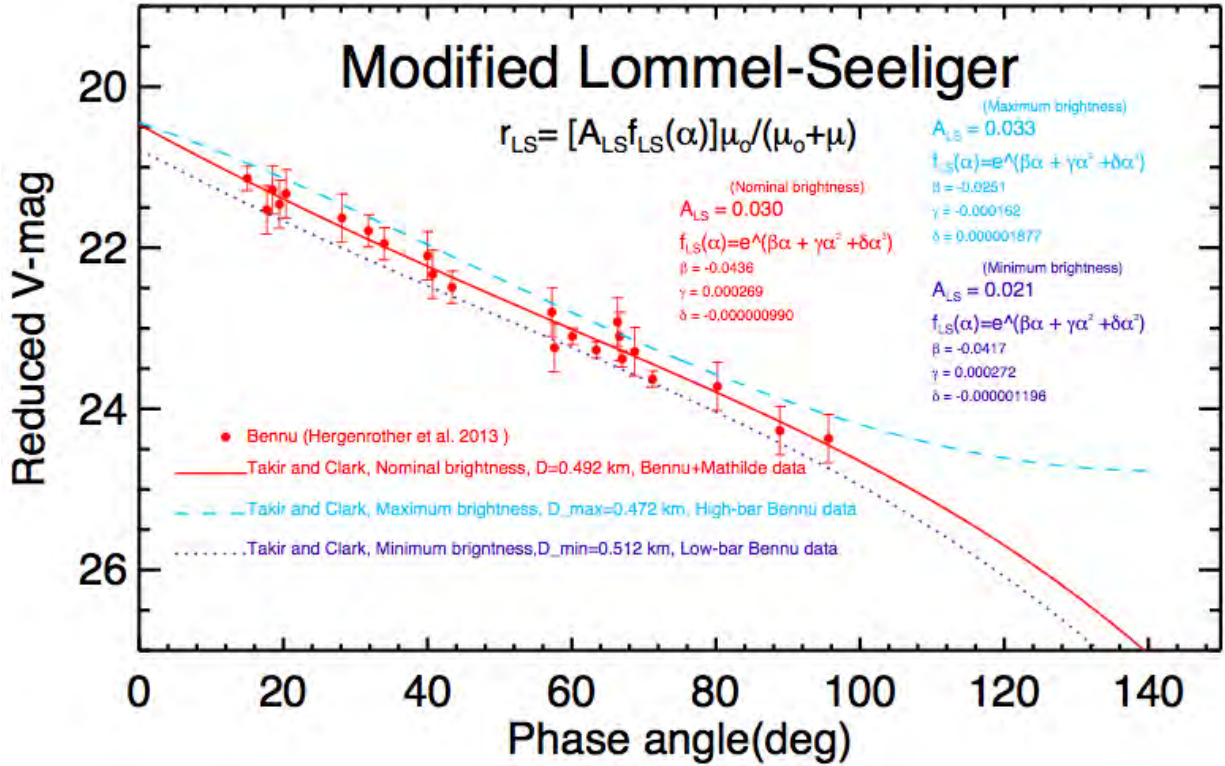

Figure 11 - The reduced V magnitude of Bennu as a function of phase angle predicted by the Lommel-Seeliger model is shown compared with ground-based measurements. Shown are the Minimum (purple dots), Maximum (light blue dashes), and Nominal (red solid lines) models. The minimum and maximum models do not include the Mathilde data, however the nominal model does. $r_{LS}$ is in units of sr$^{-1}$.

Figure 10 and Figure 11 show the data for Bennu as compared with our models. We note that the BRDF value at $i=0°$, $e=0°$, and $\alpha=0°$, calculated using the BRDF models is $0.015^{+0.002}_{-0.005}$ almost $1/\pi$ times the Geometric Albedo ($0.045\pm0.015$) as expected for Lommel-Seeliger spheres (Lester et al. 1979) and published in the Design Reference Asteroid Document (this document). As an example calculation relevant to viewing geometries from Detailed Survey, we find that our nominal BRDF value at $i=15°$, $e=15°$, and $\alpha=30°$ is $0.005^{+0.002}_{-0.001}$. As another example calculation relevant to viewing geometries from terminator orbits, we find that the nominal BRDF value at $i=30°$, $e=30°$, and $\alpha=60°$ is $0.003^{+0.001}_{-0.001}$. Also note that the Lommel-Seeliger "albedo" ($A_{LS}$) is twice the BRDF value at $i=0°$, $e=0°$, and $\alpha=0°$.

*Extrapolations to Other Wavelengths*

To predict the brightness of Bennu at wavelengths other than 550nm, we advise using spectral scaling factors, assuming the same phase functions at all wavelengths. So, our next step is to extrapolate from 550 nm to other wavelengths of interest to the OSIRIS-REx instrument team.

We used the Clark et al. 2011 spectrum to estimate Maximum and Minimum models (Figure 12). We used the scatter in the data to define a reasonable brighter model (Maximum) and a reasonable darker model (Minimum). Using the Minimum, Maximum, and Nominal models normalized to 1.0 at 550 nm, we extrapolated out along smooth fits to the ground-based spectral data to each of the requested wavelengths to get Min, Max, and Nominal scaling factors for each of the Min, Max and Nominal BRDFs. To make the fewest assumptions, we suggest use of the Minimum spectral factor for the Minimum BRDF, the Maximum spectral factor for the Maximum BRDF, and the Nominal spectral factor for the Nominal BRDF.

We determined spectral factors for 10 distinct wavelengths (Table 4). To predict the brightness of Bennu at these wavelengths, simply multiply the $A_{LS}$ values in Table 3 by these scaling factors (constants).

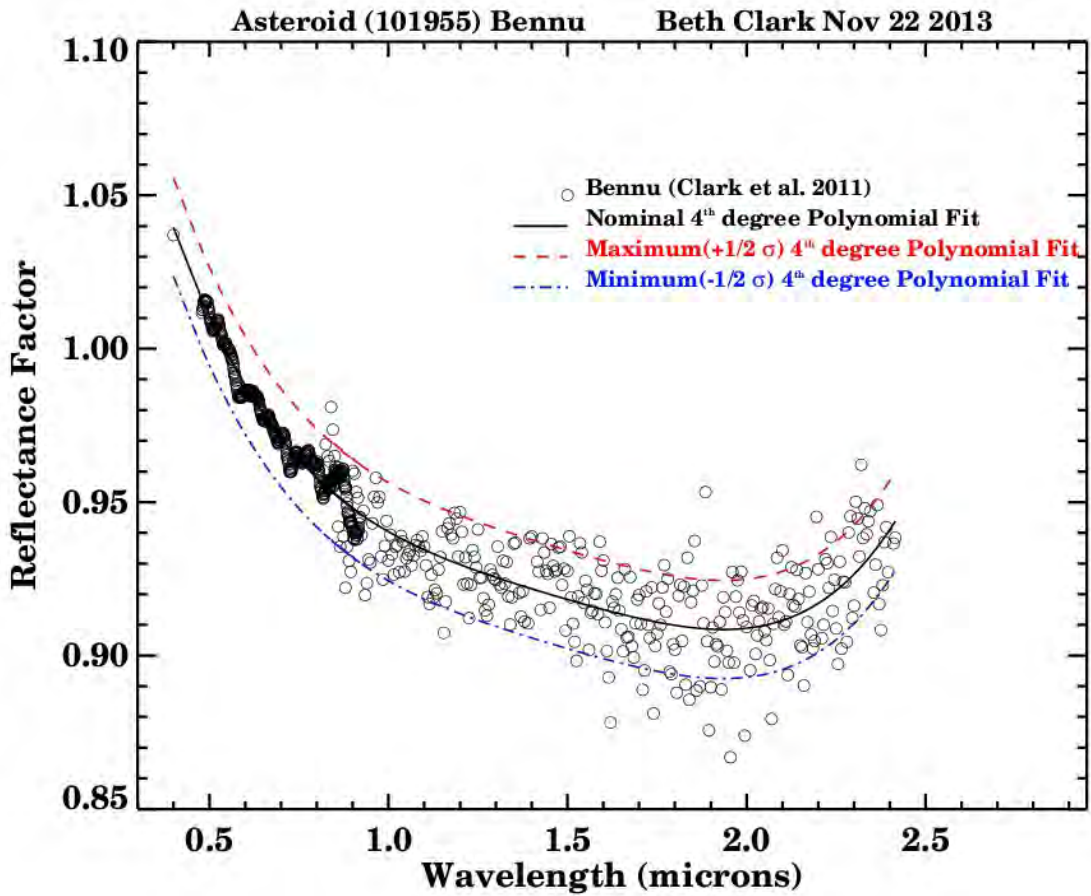

Figure 12 - Smooth spectral curves are fit to the available ground-based spectral data for Bennu from Clark et al. 2011. The spectral scaling factors shown in Table 4 are interpolations at specific wavelengths along these curves.

Table 4 – Spectral scaling factors for Bennu

| Wavelength (nm) | Minimum Model | Nominal Model | Maximum Model |
|---|---|---|---|
| 400 | 1.024 | 1.040 | 1.056 |
| 500 | 0.995 | 1.011 | 1.027 |
| 700 | 0.955 | 0.971 | 0.987 |
| 900 | 0.932 | 0.948 | 0.964 |
| 1000 | 0.924 | 0.940 | 0.956 |
| 1064 | 0.920 | 0.936 | 0.952 |
| 1100 | 0.918 | 0.934 | 0.950 |
| 1400 | 0.906 | 0.922 | 0.938 |
| 1800 | 0.894 | 0.910 | 0.926 |
| 2400 | 0.925 | 0.941 | 0.957 |

Bidirectional Reflectance Distribution Functions have been internally and externally reviewed. A peer-reviewed journal article is in preparation by Takir et al. (in prep).

### 5.2.1 Nominal Modified Lommel-Seeliger BRDF Model

- Defined as the nominal fit of the Modified Lommel-Seeliger model for Bidirectional Reflectance Distribution Function (BRDF).
- Source: Driss Takir and Beth Clark

$$r_{LS} = [A_{LS} f_{LS}(\alpha)] \mu_o / (\mu_o + \mu)$$

where $A_{LS} = 0.030$

$$f_{LS}(\alpha) = e^{(\beta \alpha + \gamma \alpha^2 + \delta \alpha^3)}$$

$$\beta = -0.0436$$

$$\gamma = 0.000269$$

$$\delta = -0.000000990$$

### 5.2.2 Minimum Modified Lommel-Seelinger BRDF Model

- Defined as the fit of the Modified Lommel-Seeliger model for BRDF to the minimum uncertainty in size and phase angle behavior.

- Source: Driss Takir and Beth Clark

$$r_{LS} = [A_{LS}f_{LS}(\alpha)]\mu_o/(\mu_o + \mu)$$

where $A_{LS} = 0.021$

$$f_{LS}(\alpha) = e^{(\beta\alpha + \gamma\alpha^2 + \delta\alpha^3)}$$

$$\beta = -0.0417$$

$$\gamma = 0.000272$$

$$\delta = -0.000001196$$

5.2.3 Maximum Modified Lommel-Seelinger BRDF Model

- Defined as the fit of the Modified Lommel-Seeliger model for BRDF to the maximum uncertainty in size and phase angle behavior.
- Source: Driss Takir and Beth Clark

$$r_{LS} = [A_{LS}f_{LS}(\alpha)]\mu_o/(\mu_o + \mu)$$

where $A_{LS} = 0.033$

$$f_{LS}(\alpha) = e^{(\beta\alpha + \gamma\alpha^2 + \delta\alpha^3)}$$

$$\beta = -0.0251$$

$$\gamma = -0.000162$$

$$\delta = -0.000001877$$

## 5.3 Broadband Colors

Concurrent with the rotation lightcurve data taken on 2005 September 14-17 UT, ECAS photometry was obtained in the *u* (0.320 μm), *b* (0.430 μm), *v* (0.545 μm), *x* (0.860 μm), and *p* (0.955 μm) filters in addition to the *w* (0.705 μm) filter. Photometric variability due to rotation and change in phase angle has been corrected.

Italicized text in this section is directly from pages 667-668 of

"Hergenrother, C.H., M. Nolan, R. Binzel, E. Cloutis, M.A. Barucci, P. Michel, D. Scheeres, C.D. d'Aubigny, D. Lazzaro, N. Pinilla-Alonso, H. Campins, J. Licandro, B.E. Clark, B. Rizk, E. Beshore, D. Lauretta 2013. Lightcurve, Color and Phase Function Photometry of the OSIRIS-REx Target Asteroid (101955) Bennu. Icarus 226, 663-670.".

*V- and R-band photometry conducted with the SOAR 4.2-m in May of 2012 yielded a V-R color of +0.37 ± 0.03 magnitudes. Transformations from the ECAS system to the Johnsons-Cousins*

*system in Howell (1995) give a V-R color of +0.35. The mean value of V-R = +0.36 ± 0.04 is from both the ECAS and SOAR VR photometry.*

*Bennu has been identified as a B-type asteroid with a bluish slope across visible wavelengths (Clark et al. 2011) (*Figure 12 *and* Figure 13*). The extension of measurements toward the ultra-violet allows some comparison with spectral properties characterized by the Tholen taxonomy (Tholen 1984, Tholen and Barucci 1989). Modern CCD spectroscopy typically does not provide sensitivity for measurements below 0.45 microns; the Eight-Color Asteroid Survey (Zellner et al. 1985) extends down to 0.34 microns. While Bennu maintains the general spectral qualities of the C-class and its various subtypes (denoted by Tholen as B, C, F), the UV turnover for Bennu is less pronounced.*

Broadband Color parameters have been published in

"Hergenrother, C.H., M. Nolan, R. Binzel, E. Cloutis, M.A. Barucci, P. Michel, D. Scheeres, C.D. d'Aubigny, D. Lazzaro, N. Pinilla-Alonso, H. Campins, J. Licandro, B.E. Clark, B. Rizk, E. Beshore, D. Lauretta (2013). Lightcurve, Color and Phase Function Photometry of the OSIRIS-REx Target Asteroid (101955) Bennu. Icarus 226, 663-670.".

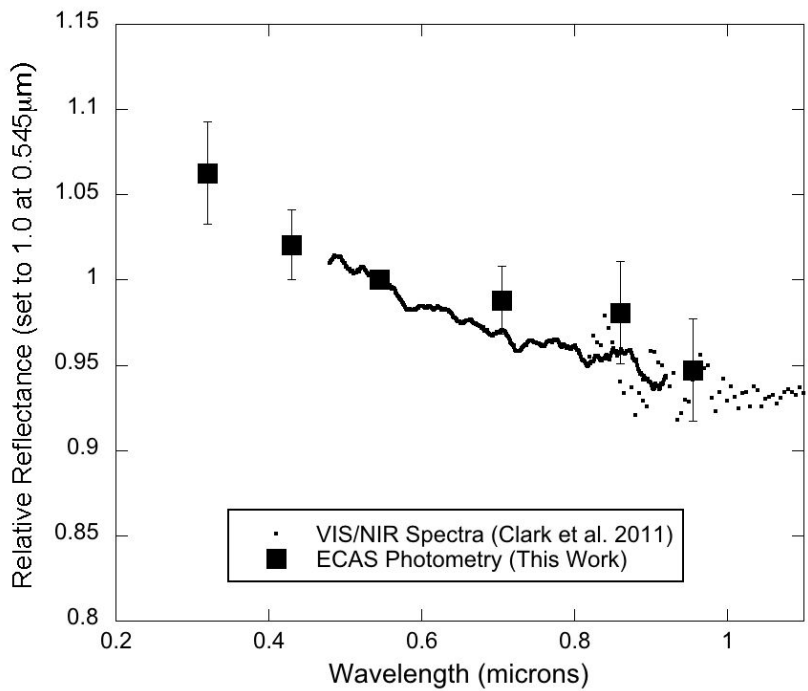

**Figure 13 - Comparison of ECAS spectrophotometry from Hergenrother et al. (2013) and visible to near-IR spectroscopy from Clark et al. (2011).**

5.3.1 ECAS u-v (0.32-0.54) μm Color Index

- Defined as the astronomical magnitude difference between the ECAS u (0.32-micron) and ECAS v (0.54-micron) broadband filters. Solar colors have been removed from this index.
- Source: Carl Hergenrother – Hergenrother et al. (2013)

ECAS u-v (0.32-0.54 microns) = -0.07 ± 0.03 (1-sigma uncertainty)

### 5.3.2 ECAS b-v (0.43-0.54) µm Color Index

- Defined as the astronomical magnitude difference between the ECAS b (0.43-micron) and ECAS v (0.54-micron) broadband filters. Solar colors have been removed from this index.
- Source: Carl Hergenrother – Hergenrother et al. (2013)

$$\text{ECAS b-v (0.43-0.54 micron)} = -0.03 \pm 0.02 \text{ (1-sigma uncertainty)}$$

### 5.3.3 ECAS v-w (0.54-0.71) µm Color Index

- Defined as the astronomical magnitude difference between the ECAS v (0.54-micron) and w (0.71-micron) broadband filters.
- Source: Carl Hergenrother – Hergenrother et al. (2013)

$$\text{ECAS v-w (0.54-0.71 micron)} = -0.01 \pm 0.02 \text{ (1-sigma uncertainty)}$$

### 5.3.4 ECAS v-x (0.54-0.86) µm Color Index

- Defined as the astronomical magnitude difference between the ECAS v (0.54-micron) and x (0.86-micron) broadband filters.
- Source: Carl Hergenrother – Hergenrother et al. (2013)

$$\text{ECAS v-x (0.54-0.86 micron)} = -0.01 \pm 0.03 \text{ (1-sigma uncertainty)}$$

### 5.3.5 ECAS v-p (0.54-0.96) µm Color Index

- Defined as the astronomical magnitude difference between the ECAS v (0.54-micron) and p (0.96-micron) broadband filters.
- Source: Carl Hergenrother – Hergenrother et al. (2013)

$$\text{ECAS v-p (0.54-0.96 micron)} = -0.05 \pm 0.03 \text{ (1-sigma uncertainty)}$$

### 5.3.6 ECAS 0.71-µm Feature [depression below (0.86-0.54) µm slope]

- Defined as the astronomical magnitude difference between the ECAS w (0.71-micron) filter and the average magnitude of the ECAS v (0.54 micron) and x (0.84 micron) filters.
- Source: Carl Hergenrother – Hergenrother et al. (2013)

$$\text{ECAS 0.7 micron depression} = -0.01 \pm 0.04 \text{ magnitudes or effectively 0 (1-sigma uncertainty)}$$

### 5.3.7 Johnson U-B (0.36-0.44) µm Color Index

- Defined as the astronomical magnitude difference between the Johnson U (0.36 micron) and B (0.44 micron) broadband filters. Solar colors are included in this index. U-B index derived from the ECAS u-b index via the following formula from Ellen Howell's PhD thesis (1995): $(U-B)_J = 1.070 * (u-b) + 0.203$.
- Source: Carl Hergenrother – Hergenrother et al. (2013)

Johnson U-B (0.36-0.44 micron) = +0.16 ± 0.05 (1-sigma uncertainty)

### 5.3.8 Johnson B-V (0.44-0.54) µm Color Index

- Defined as the astronomical magnitude difference between the Johnson B (0.44 micron) and V (0.54 micron) broadband filters. Solar colors are included in this index. B-V index derived from the ECAS b-v index via the following formula from Ellen Howell's PhD thesis (1995): $(B-V)_J = 0.968 * (b-v) + 0.674$.
- Source: Carl Hergenrother – Hergenrother et al. (2013)

Johnson B-V (0.44-0.54 micron) = +0.64 ± 0.04 (1-sigma uncertainty)

### 5.3.9 Cousins V-R (0.54-0.65) µm Color Index

- Defined as the astronomical magnitude difference between the Cousins V (0.54 micron) and R (0.65 micron) broadband filters. Solar colors are included in this index. V-R index derived from the ECAS v-w index via the following formula from Ellen Howell's PhD thesis (1995): $(V-R)_C = 0.836 * (v-w) + 0.355$.
- Source: Carl Hergenrother – Hergenrother et al. (2013)

Cousins V-R (0.54-0.65 micron) = +0.36 ± 0.04 (1-sigma uncertainty)

### 5.3.10 Cousins V-I (0.54-0.79) µm Color Index

- Defined as the astronomical magnitude difference between the Cousins V (0.54 micron) and I (0.79 micron) broadband filters. Solar colors are included in this index. V-I index derived from the ECAS v-x index via the following formula from Ellen Howell's PhD thesis (1995): $(V-I)_C = 0.875 * (v-x) + 0.695$.
- Source: Carl Hergenrother – Hergenrother et al. (2013)

Cousins V-I (0.54-0.79 micron) = +0.69 ± 0.04 (1-sigma uncertainty)

# 6 Spectroscopic Properties

Relevant Observations

- *McDonald 2.1-m visible spectroscopy* – Ellen Howell used the McDonald Observatory 2.1-m and visible spectrograph to obtain 0.47 to 0.92 μm spectra over 5 nights in September 1999. This data identified Bennu as a B-type asteroid. No statistically significant spectral variation was observed across different longitudes.
- 3.6-36 μm Spitzer Space Telescope Photometry and Spectroscopy: Josh Emery obtained director's Discretionary Time on Spitzer in May 2007. Spitzer observed Bennu during the time period 3 to 9 May 2007. Thermal spectra from 5.2 to 38 μm were measured with the Infrared Spectrograph (IRS) of opposite hemispheres of the body. Photometry at 16 and 22 μm was obtained with the IRS peak-up imaging (PUI) mode and at 3.6, 4.5, 5.8, and 8.0 μm with the Infrared Array Camera (IRAC). Because of greater sensitivity of the imaging modes, 10 equally distributed longitudes were targeted in order to search for rotational heterogeneities. IRAC was also used to observe Bennu in August of 2012.
- 70-160 μm Herschel Space Observatory Photometry: Antonella Barucci used the Herschel Space Observatory on September 9, 2011 to measure the thermal flux of Bennu at 70, 100 and 160 μm with the PACS instrument. The highest flux for Bennu during the entire Herschel mission occurred in September 2011 during closest approach with the maximum flux expected for PACS to be 35/20/9 mJy at 70/100/160 μm.
- 0.8-2.5 μm IRTF and Magellan Spectroscopy: Rick Binzel observed Bennu with the IRTF as part of the MIT-UH-IRTF Joint Campaign for NEO Spectral Reconnaissance. This near-Infrared spectral data (spanning from 0.82 to 2.5 μm) is consistent with Bennu being a dark carbonaceous object. Additional observations were obtained in August 2011 and May 2012 with the Magellan 6.5-m.

## 6.1 Taxonomy

The average visible wavelength spectrum (0.4 to 0.9 μm) was combined with the average near-infrared spectrum (0.82 to 2.49 μm) and ECAS filter photometry (0.32 to 0.96 μm) to form a composite spectrum that is normalized to unity at 0.55 μm (Figure 14). Notice that a long-wavelength "tail" due to the thermal emission of the object exists beyond about 2 μm. Fits to the thermal tail used a range of geometric albedos ($P_v$ = 0.035 to 0.046) and beaming parameters (n = 1.26 to 1.76) to yield a range of model surface temperatures ($T_{ss}$ = 340 to 370 K). The nominal model used a 4% albedo and a beaming parameter of 1.48 and yielded a surface temperature of 350 K.

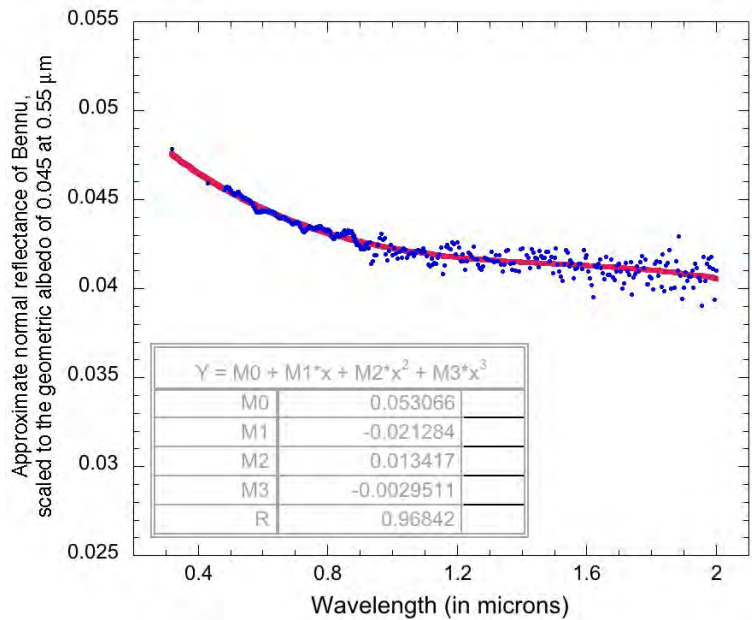

Figure 14 - Approximate normal reflectance of Bennu from 0.32 to 2.0 µm. Reflectance is scaled to a geometric albedo of 0.045 at a wavelength of 0.55 µm. A 3rd-degree polynomial is fit to the data. Measurements are from Hergenrother et al. (2013) (ECAS filter photometry from 0.32 to 0.96 µm), Ellen Howell (visual spectroscopy from 0.48 to 0.90 um) and Rick Binzel (NIR spectroscopy from 0.82 to 2.00 µm).

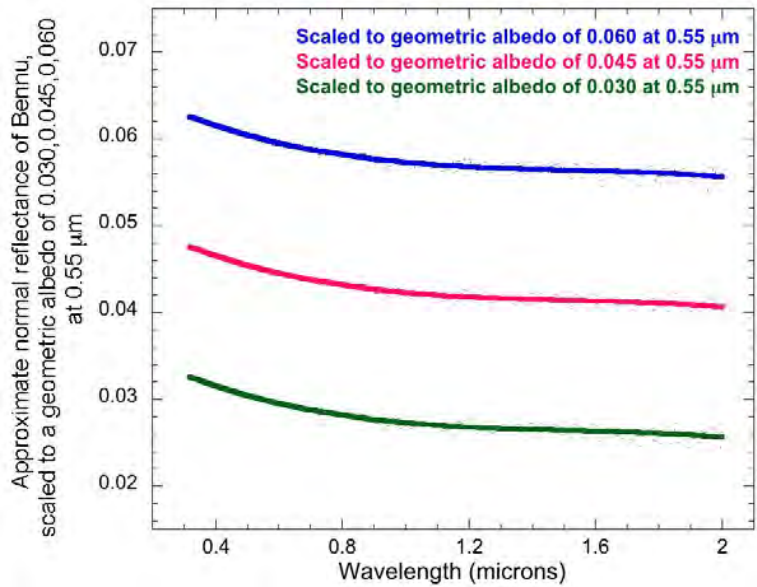

Figure 15 - The geometric albedo measurements have a 3-sigma uncertainty of ±0.015. Third-order polynomials were fit to nominal, minimum and maximum approximate normal reflectances normalized to 0.55 µm albedos of 0.030, 0.045, and 0.060. Note the fits should not be extrapolated to wavelengths outside of the 0.32-2.0 µm range since they are not valid outside of this range.

Using our composite spectrum, we can summarize the spectral properties of asteroid Bennu. We find that Bennu is a B-type asteroid (DeMeo et al. 2009), which means that it exhibits a generally negatively sloped spectral shape across the visible and near-infrared wavelength range.

Comparisons to other B-type asteroids show that Bennu is comparable to asteroid (24) Themis in terms of its albedo, visible spectrum, and near-infrared spectrum from 1.1 to 1.45 um. However, Bennu is more similar to asteroid (2) Pallas in terms of its near-infrared spectrum from 1.6 to 2.3 um. Thus it is possible that B-type asteroids form a spectral continuum and that Bennu is a transitional object, spectrally intermediate between the two end-members. This is particularly interesting because Asteroid (24) Themis was recently discovered to have $H_2O$ ice and organics on the surface (Rivkin and Emery 2010, Campins et al. 2010). It is thus conceivable that OSIRIS-REx will rendezvous with a very primitive object, and return the first samples of a water-rich asteroid to Earth.

The Taxonomy of Bennu has been published in

"Clark, B.E., Binzel, R.P., Howell, E.S., Cloutis, E.A, Ockert-Bell, M., Christensen, P., Barucci, M.A., DeMeo, F., Lauretta, D.S., Connolly, H., Soderberg, A., Hergenrother, C., Lim, L., Emery, J., Mueller, M., 2011. Asteroid (101955) 1999 RQ36: spectroscopy from 0.4 to 2.4 μm and meteorite analogs. Icarus 216, 462-475."

and

"Hergenrother, C.H., M. Nolan, R. Binzel, E. Cloutis, M.A. Barucci, P. Michel, D. Scheeres, C.D. d'Aubigny, D. Lazzaro, N. Pinilla-Alonso, H. Campins, J. Licandro, B.E. Clark, B. Rizk, E. Beshore, D. Lauretta (2013). Lightcurve, Color and Phase Function Photometry of the OSIRIS-REx Target Asteroid (101955) Bennu. Icarus 226, 663-670.".

### 6.1.1 Taxonomy

- Defined as the taxonomic classification of Bennu.
- Source: Rick Binzel

Carbonaceous B-type – Clark et al. (2011) and Hergenrother et al. (2013)

## 6.2 Relative Reflectance

Relative Reflectance parameters have been internally reviewed.

### 6.2.1 Visible/Near-Infrared wavelength (0.32-2.0 μm) Approximate Normal Reflectance

Defined as the approximate normal reflectance of Bennu at wavelengths between 0.32 and 2.00 μm scaled to a geometric albedo of 0.045 at 0.54 μm. Measurements are from Carl Hergenrother (ECAS filter photometry from 0.32 to 0.96 μm), Ellen Howell (visual spectroscopy from 0.48 to 0.92 μm) and Rick Binzel (NIR spectroscopy from 0.82 to 2.00 μm). The approximate normal reflectance is an approximation of the reflectance of Bennu at zero phase angle (angle of incidence and angle of emission of 0°). The input measurements consist of radiance scattered into the Earth's direction (at phase angle 63° to 71° for the ECAS filter photometry, phase angle 60° for the VIS spectroscopy, and phase angle 44° for the NIR spectroscopy) divided by the radiance of a solar standard star. The geometric albedo is the ratio of integrated brightness (extrapolated to zero phase angle) to an idealized flat, fully reflecting, diffusively scattering (Lambertian) disk with the same cross-section. The relationship between the average bidirectional reflectance of the surface material and the geometric albedo of a spherical surface depends on limb darkening and opposition surge. For comparison, the spacecraft measured quantity - I/F - equals the geometric albedo when the phase angle of the observation is zero. Because limb darkening is less severe on low-albedo surfaces, you could consider that the average I/F of Bennu at zero phase angle would be approximately equal to the geometric albedo. But in the lab, we measure bidirectional reflectance, not I/F. The 0.32 to 2.00 μm range is fit with a $3^{rd}$-degree polynomial, which is the lowest degree polynomial that reasonably matches the data. The geometric albedo of Bennu has a 3-sigma uncertainty of 0.015 as a result the approximate normal reflectance of Bennu also has a 3-sigma uncertainty of order ±0.015 over the 0.32 to 2.00 μm range. The approximate normal reflectance is expected to be within the range between the blue curve and the green curve on Figure 15. *[Note: the polynomial fit is not valid outside of the 0.32-2.0 μm range and should not be extrapolated to wavelengths outside of the 0.32-2.0 μm range.]*

- Source: Carl Hergenrother, Ellen Howell, Rick Binzel

$$- y(\lambda) = (0.053066 - 0.021284*\lambda + 0.013417*\lambda^2 - 0.0029511*\lambda^3) \pm 0.015$$

where λ is wavelength in microns

## 6.3 Meteorite Analogs

Italicized text is directly from the Discussion section (pages 472-474) of

"Clark, B.E., Binzel, R.P., Howell, E.S., Cloutis, E.A, Ockert-Bell, M., Christensen, P., Barucci, M.A., DeMeo, F., Lauretta, D.S., Connolly, H., Soderberg, A., Hergenrother, C., Lim, L., Emery, J., Mueller, M., 2011. Asteroid (101955) 1999 RQ36: spectroscopy from 0.4 to 2.4 μm and meteorite analogs. Icarus 216, 462-475.".

*Three key characteristics of Bennu are notable: (1) the lack of well-defined absorption bands; (2) the overall blue spectral slope; and (3) the lack of a reflectance dropoff below ~0.55 μm. The lack of well-defined absorption bands expected near 0.7, 0.9, and 1.1 μm in the case of Fe-bearing phyllosilicates, near 1.05 μm for olivine, and near 0.9 and 1.9 μm for orthopyroxenes suggests at least two possibilities – either these components are not present, or they are present and masked by opaque materials or the low signal-to-noise of our data. The low albedo of Bennu suggests that the latter may be the case. Mixtures of various silicates with different opaques, comparable to those found in carbonaceous chondrites (e.g., lampblack, magnetite) indicate that a few weight percent of these opaques, when finely dispersed, are very effective at dramatically reducing silicate band depths (e.g., [Cloutis et al., 1990], [Milliken and Mustard, 2006] and Yang, 2010).*

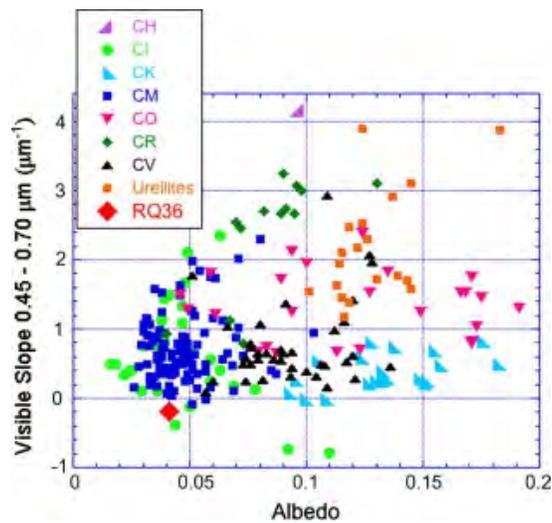

Figure 16 - Asteroid Bennu (red diamond) is compared parametrically with the carbonaceous meteorite classes (CH yellow circle, CI light green circles, CM blue squares, CR dark green diamonds, CV black triangles, CO pink upside-down triangles, CK light blue corner triangles, UR – ureilites orange squares). Albedo is the geometric albedo measured at 0.55 μm for Bennu, and is the reflected light value at 0.55 μm for the meteorites. Notice that Bennu is most similar to CI meteorites in terms of the low albedo and negative visible slope. The uncertainties in parameter values are smaller than the symbols.

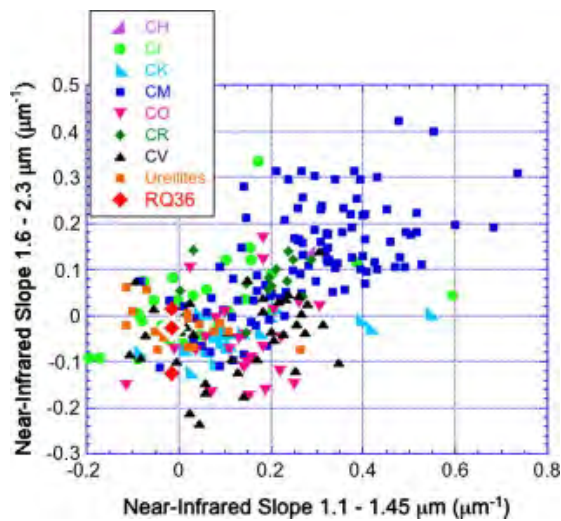

Figure 17 - Asteroid Bennu (red diamonds) is compared parametrically to the carbonaceous meteorite classes (CH yellow circle, CI light green circles, CM blue squares, CR dark green diamonds, CV black triangles, CO pink upside-down triangles, CK light blue corner triangles, UR – ureilites orange squares). Near-infrared slope is measured as rise over run for two spectral regions: NIR1: 1.10–1.45 μm and NIR2: 1.6–2.3 μm. Notice that Bennu is similar to most of the meteorites in terms of the near-infrared continuum slopes. The uncertainties in parameter values are smaller than the symbols. However the uncertainty in Bennu's NIR2 value is indicated with three red diamonds. The low temperature thermal model resulted in a positive NIR2 value (top red diamond), and the high temperature thermal model resulted in a negative NIR2 value (bottom red diamond), with nominal (red diamond outlined in white) being slightly negative. The spread in these points illustrates the uncertainty in the NIR2 parameter that is due to the thermal tail corrections to the observed spectrum.

*The pure mafic silicates and phyllosilicates that we have examined which are plausible candidates for primitive asteroids generally exhibit a flat to red-sloped spectrum ([Cloutis et al. 2010] and [Cloutis et al. 2011]). Three notable exceptions include a magnetite-bearing serpentinite, nanophase magnetite (Morris et al., 1985), and the insoluble organic matter in the Murchison CM2 and Allende CV3 chondrite (Johnson and Fanale 1973).*

*Both organic matter and magnetite are more abundant in CI than in other types of chondrites (Brearley and Jones, 1998) and either one could plausibly account for the blue slope seen in some CI spectra (e.g., Johnson and Fanale, 1973). The effect that magnetite can have on overall spectral slope can be seen by comparing two serpentinite reflectance spectra, one of which contains finely-dispersed magnetite at the few percent level (*Figure 18*). Physical mixtures (not modeled) of a coarser magnetite (<45 μm) with a red-sloped serpentinite (*Figure 19*) show how the presence of magnetite can convert a red-sloped spectrum to a blue-sloped spectrum. The insoluble organic matter in the Murchison is also blue-sloped, most markedly at small phase angles (*Figure 20*). Both magnetite and insoluble carbonaceous matter is present in the more aqueously altered carbonaceous chondrites (CI and CM), and magnetite abundance in the blue-sloped serpentinite, partially constrained from stoichiometry, is less than its abundance in CI chondrites such as Orgueil.*

*The fact that most carbonaceous chondrite spectra become bluer-sloped and darker with increasing grain size is well established (e.g., Johnson and Fanale, 1973). While our understanding of this bluing is incomplete, appreciable blue slopes in samples with constrained grain size are seen in most carbonaceous chondrites, and with the exception of a few CI chondrites, the size fractions that have blue-sloped spectra require the absence of fine-grained (generally <~75 μm) particles (e.g., Johnson and Fanale, 1973). In our analysis of spectra of the Orgueil CI chondrite, we find that there is a general trend of increasingly blue slope with decreasing reflectance, with blue overall slopes restricted to Orgueil spectra with reflectance near the 0.6 μm peak of <5%.*

*The third characteristic of Bennu is the lack of a reflectance downturn shortward of ~0.48 μm. CI and CM chondrite spectra invariably exhibit a reflectance downturn shortward of ~0.55 μm. Changes in grain size can cause a shift in the wavelength position of this downturn, with larger grain sizes often exhibiting the start of the reflectance downturn at shorter wavelengths than fine-grained sample spectra (Johnson and Fanale, 1973). Various mineral—opaque mixture spectra do show a systematic decrease in the wavelength position of the local reflectance peak in this region, as discussed below.*

*When we look at the best spectral matches to Bennu from the spectral data base search we find that each has some limitations or uncertainties.*

*The CM Mighei spectrum that provides a good fit to Bennu is the 100–200 μm spectrum of a matrix-rich fraction (Moroz and Pieters, 1991). The <40 μm matrix-rich fraction spectrum is more red-sloped, and the peak reflectance position (~0.56 μm) is the same for the two size fraction spectra. Peak reflectance of the 100–200 μm spectrum is 2.6%. This suggests that both matrix enrichment and removal of a fine grain size fraction from a CM chondrite like Mighei are required to match the blue slope, but overall reflectance may be too low, and the wavelength position of the reflectance downturn toward the UV is still not matched.*

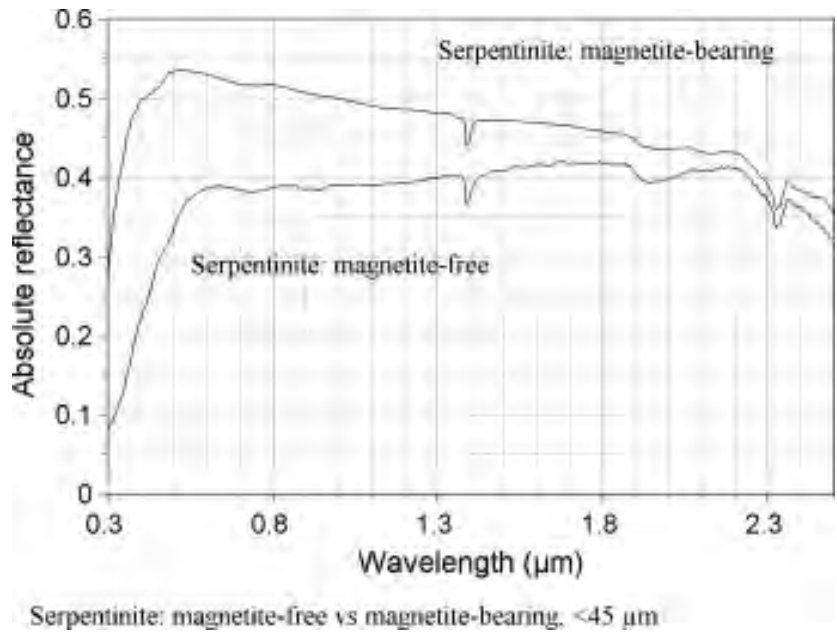

Figure 18 - Serpentinite, a common component of CI meteorites, is shown with and without finely dispersed magnetite.

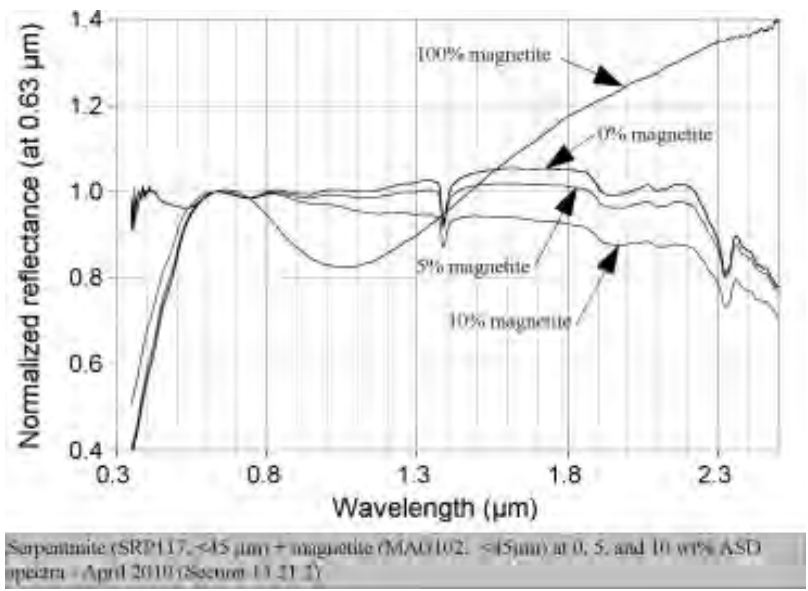

Figure 19 - Mixtures of a coarser magnetite (<45 μm) with a red-sloped serpentinite show how the presence of magnetite can convert a red-sloped spectrum to a blue-sloped spectrum

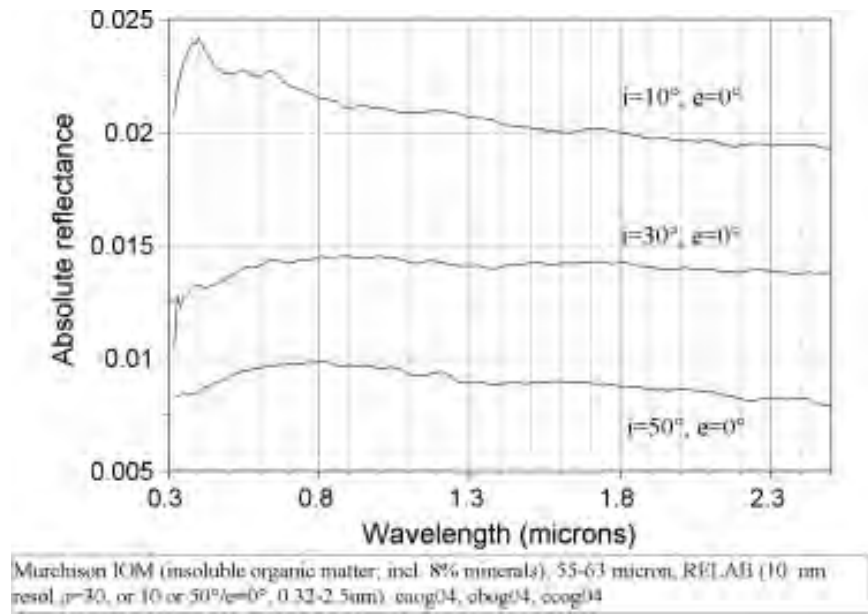

**Figure 20 - The insoluble organic matter in Murchison is also spectrally blue at several viewing geometries.** *i* is incidence angle, and *e* is emission angle.

*The CI Ivuna spectrum that matches Bennu is a <125 μm fraction that has been heated to 700 °C. Unheated samples of Ivuna and samples heated to lower temperatures (up to 600 °C) do not exhibit the overall blue slope and have their reflectance maximum near 0.65 μm. Only the Ivuna sample heated to 700 °C has its reflectance maximum below 0.65 μm and shows a sufficiently blue slope. Progressive heating does not lead to a gradual change in spectral slope, rather the strong blue slope appears only in the sample heated to 700 °C. Progressive heating also leads to a gradual decrease in overall reflectance to 500 °C, followed by an increase to 600 and 700 °C. The physical and compositional changes that accompany heating to this temperature are unknown. Thus, the spectrum of Ivuna heated to 700 °C is the only heated CI chondrite spectrum that provides a good match to Bennu.*

*Additional observations worth mentioning in the context of Bennu are laboratory spectra of serpentinite.* Figure 18 *shows reflectance spectra of two terrestrial serpentinites. Compositionally these two samples are nearly identical, and the only significant difference is that the presence of magnetite was detected in the upper spectrum by X-ray diffractometry. The abundance of magnetite is unknown, but cannot exceed a few percent on the basis of stoichiometry. The presence of this fine-grained magnetite seems to cause a progressively bluer spectral slope and also probably causes the local reflectance maximum to shift from ~0.6 to ~0.5 μm. Fine-grained magnetite is a component of CI chondrites, and many CI chondrites are blue-sloped, strongly suggesting that this phase is probably a required component for creating blue-sloped spectra in fine-grained carbonaceous chondrites. A similar effect is seen for some phyllosilicate + lampblack mixtures (e.g., [Clark, 1983] and [Milliken and Mustard, 2006]), where a few percent fine-grained and finely-dispersed lampblack can cause phyllosilicate spectra to become darker (reflectance <10%), more blue-sloped, and also cause the local reflectance maximum to shift to short wavelengths (<0.5 μm). It should be noted that not all lampblack samples can cause all of* *these spectral changes however. Therefore the spectral relevance of lampblack to the insoluble carbon-rich component of carbonaceous chondrites is unknown.*

*These laboratory spectra suggest that fine-grained and finely-dispersed opaque materials, such as insoluble organic matter/lampblack and magnetite can affect silicate spectra in ways that would match the salient features of Bennu: an overall blue slope, low reflectance, and a low wavelength (<0.5 μm) reflectance peak. As both of these materials are present in CI carbonaceous chondrites, CI chondrite-like assemblages, with only small adjustments in opaque abundances, can quite plausibly account for the spectral properties of Bennu.*

*In order to make a case for the most likely composition of Asteroid (101955) Bennu, this paper synthesizes results from spectral analyses. We present results from our RELAB spectral library search, and results from our parametric comparisons with meteorites. We conclude that the most likely meteorite analogs for Bennu are the CI and/or CM meteorites, and the least likely analogs are the CO and CH meteorites. Bennu has a spectrally "blue" continuum slope that is also observed in carbonaceous chondrites containing magnetite. When compared with other B-type asteroids, we find that Bennu is transitional between Pallas-like spectral properties and Themis-like spectral properties.*

### 6.3.1 Most Similar Meteorite Analogs

- Defined as the meteorite types that are most similar to Bennu based on Visible and Near-IR spectra.
- Source: Beth Clark – Clark et al. (2011)

> CI meteorites (specific best fit is a sample of CI Ivuna <125-μm fraction heated to 700°C and
>
> CM meteorites (specific best fit is a sample of CM Mighei 100-200-μm matrix-rich fraction

# 7 Thermal Properties

The thermal properties of Bennu include parameters derived by observations taken in the near-, mid-, long- and far-IR. Ground-based and space-based telescopes have measured the reflected and thermal flux of Bennu over an IR wavelength range of 0.8 to 160 μm.

Thermal properties have been published in the journal articles

"Emery, J.P., Fernandez, Y.R., Kelley, M.S.P., Warden, K.T., Hergenrother, C., Lauretta, D.S., Drake, M.J., Campins, H., Ziffer, J. 2014. Thermal infrared observations and thermophysical characterization of OSIRIS-REx target asteroid (101955) Bennu. Icarus 234, 17-35."

and

"Müller, T.G., O'Rourke, L., Barucci, A.M., Pal, A., Kiss, C., Zeidler, P., Altieri, B., Gonzalez-Garcia, B.M., Kuppers, M., 2012. Physical properties of OSIRIS-REx target asteroid (101955) 1999 RQ36. Astronomy and Astrophysics 548, A36.".

## 7.1 Thermal Characteristics

Relevant Observations:

- 3.6-36 μm Spitzer Space Telescope Photometry and Spectroscopy: Josh Emery obtained director's Discretionary Time on Spitzer in May 2007. Spitzer observed Bennu during the time period 3 to 9 May 2007 and again in August 2012. Thermal spectra from 5.2 to 38 μm were measured with the Infrared Spectrograph (IRS) of opposite hemispheres of the body. Photometry at 16 and 22 μm was obtained with the IRS peak-up imaging (PUI) mode and at 3.6, 4.5, 5.8, and 8.0 μm with the Infrared Array Camera (IRAC). Because of greater sensitivity of the imaging modes, 10 equally distributed longitudes were targeted in order to search for rotational heterogeneities.
- 0-160 μm Herschel Space Observatory Photometry: Antonella Barucci used the Herschel Space Observatory on September 9, 2011 to measure the thermal flux at 70, 100 and 160 μm with the PACS instrument.
- 0.8-2.5 μm IRTF and Magellan Spectroscopy: Rick Binzel used the IRTF as part of the MIT-UH-IRTF Joint Campaign for NEO Spectral Reconnaissance. This near-Infrared spectral data (spanning from 0.82 to 2.5 μm) is consistent with Bennu being a dark carbonaceous object. Additional observations were obtained in August 2011 and May 2012 with the Magellan 6.5-m.

*Statement of problem of determining albedo*

Accurate knowledge of the visible reflectance of Bennu is important for spacecraft, instrument, and mission design. Visible reflectance, here, is described by the quantity "visible geometric albedo" ($p_v$), which is defined as the disk-integrated reflectance (in the astronomical V band at ~0.55 μm) of the object observed at 0° phase angle compared to a perfectly reflecting Lambert sphere at the same wavelength and geometry. Other quantifications of albedo and reflectance are used elsewhere in mission documentation.

Preliminary analysis determined an albedo of ~0.03. More recent analysis indicates that $p_v$ for Bennu is ~0.045. The present document briefly describes the analyses that lead to the higher geometric albedo and the uncertainties involved.

*General procedure for determining albedo*

Visible geometric albedo is a derived, rather than directly measured, quantity. Visible flux measurements contain the signature of both $p_v$ and the size. Therefore, an independent estimate of the size is necessary to derive $p_v$ from the visible flux. For a spherical body,

$$p_v = \frac{D_o^2}{D^2} 10^{-\frac{2}{5}H_v}, \qquad (1)$$

where D is the object diameter, $H_v$ is the visible absolute magnitude, and $D_o$ is a parameter set primarily by the brightness of the Sun $\left(D_o = 2 \cdot 1.496 \times 10^8 \frac{km}{AU} \cdot 10^{\frac{-m_{sun}}{5}}\right)$. A value of $D_o$=1329 is typically used.

An accurate determination of $H_v$ was determined by Hergenrother et al. (2013) from calibrated visible light curves and phase curve observations. Assuming a linear phase curve and allowing for a small opposition effect, they find $Hv = 20.56^{+0.05}_{-0.15}$. If the phase curve associated with the IAU H-G system were used, $H_v$ would be lower (brighter; $H_v$=19.97±0.26), because that system assumes a significant opposition effect. However, low-albedo asteroids, particularly those with $p_v$<0.05, generally do not exhibit significant opposition effects (e.g., Shevchenko et al. 2012). The linear phase curve with a small opposition effect component assumption is therefore justified, and the higher (fainter) value of $H_v$ is considered more reliable.

The remaining quantity required to estimate $p_v$ is the diameter of Bennu.

*Radar size inputs*

The size of Bennu was measured from radar imaging conducted in 1999 and 2005 from the Arecibo and Goldstone observatories. Shape models produced by Nolan et al. (2013) incorporated both the radar observations as well as lightcurve photometry. Bennu has a "spinning top" shape with an equatorial ridge similar to that of binary near-Earth asteroids. The long and intermediate axes are 565±10 m and 535±10m, respectively. The polar diameter is 508±52 m. Due to its "spinning top" shape, Bennu's mid-latitude dimensions are significantly smaller than its equatorial and polar dimensions resulting in a mean diameter of 492±20 m. The projected area at equatorial viewing aspects is 0.191±0.006 km$^2$, corresponding to a equivalent diameter of 493 m. Due to Bennu's pole orientation of -88° and inclination of 6°, all viewing circumstances are near-equatorial making an equivalent diameter of 493 m valid for most observations. All reported uncertainties are 1-sigma.

Solving equation (1) for a 3-sigma range of $H_v$ of 20.41-20.61 and projected area equivalent mean diameter of 432-552 m yields a visible geometric albedo $p_v$ = 0.043 $^{+0.022}_{-0.010}$ (3-sigma error).

*Thermal size determination*

Size of a body can also be determined from measurements of the thermal flux. Thermal flux measured at a given wavelength can be expressed as

$$F_\lambda = \frac{R^2}{\Delta^2} \varepsilon_\lambda \int B_\lambda(T) \cos\theta \, dA, \qquad (2)$$

where R is the object radius, Δ is the distance from the object to the observer, $\varepsilon_\lambda$ is the spectral emissivity, $B_\lambda$ is the Planck function at wavelength λ and temperature T, θ is the emission angle, and the integral is over the visible surface area of the object. For silicate-rich bodies (where mid-IR emissivities are ~0.9), the primary uncertainties in this expression are R and the temperature distribution on the body. The albedo is one factor in the temperature distribution, but its contribution is very small, particularly for low albedos. The temperature of a given surface facet scales roughly as $(1-A_B)^{1/4}$, where $A_B$, the Bond albedo, is related to $p_v$ by the phase integral (q): $A_B = p_v q$. For the linear phase coefficient of 0.040±0.003 found by Hergenrother et al. (2013), q=0.367±0.045. Taken together, even fairly significant changes in Bennu's albedo will not significantly affect the temperature distribution of the surface. The primary benefit of thermal observations in this case is therefore to estimate the size of Bennu.

The derived size does still depend on how the temperature distribution is modeled. Basically, the size will shift the overall spectral flux curve up or down, whereas the temperature will change the shape of the spectral flux curve as well as shifting it. These effects have to be unraveled to get a good estimate of the size. A common, simple thermal model is the Near-Earth Asteroid Thermal Model (NEATM) [a misnomer, because nothing in the model is specific to NEAs, but we seem to be stuck with the name now]. This model does not explicitly include heat conduction, but rather attempts to parameterize the effects in a fudge factor called the beaming parameter that is adjusted to artificially adjust surface temperatures to match the observed flux. The main failing of the model with respect to the observations of Bennu is that it assumes there is no flux whatsoever coming from the night side of the object. The Spitzer observations of Bennu were made at fairly high phase angle (~63°), so a significant portion of the night side was in the field of view, and in reality there is flux coming from that side. To compensate, the model increases the overall size. This overestimation of diameter by NEATM at large phase angles has been discussed in the literature.

When fitting the Spitzer data of Bennu with NEATM, the derived size (D~600m) is larger than the radar size and the resulting $p_v$ is ~0.03. However, these numbers are not reliable, for the reasons described in the previous paragraph.

Thermophysical models that explicitly include heat conduction and allow for non-zero temperatures anywhere on the body provide better estimates of physical properties. The only caveat is that their use requires significant ancillary data that is not available for most asteroids. Fortunately, Bennu is well-observed and therefore a good target for thermophysical modeling.

The plots below show the results of fitting one of the Spitzer flux spectra of Bennu. The modeling technique is to vary the size to find the best fit to the measured spectrum at a given thermal inertia. From this thermophysical modeling, the best fit thermal inertia is ~600, and that corresponds to a radius of ~500m (green pluses on bottom plot), which is smaller than the value found from NEATM. The smaller size in turn results in a higher albedo, from Eqn. 1 (also see the green triangles on the bottom plot in Figure 21). A factor of 2 in $\chi^2$ corresponds roughly to 1σ. The minimum $\chi^2$ in this case is ~1.3, so the 1σ range are the thermal inertia values that give $\chi^2 < 2.6$, roughly 350 to 750 J m$^{-2}$s$^{-1/2}$K$^{-1}$, corresponding to $p_v$~0.043±0.007. The relationship between thermal inertia and $p_v$ flattens out in the model solution such that a 3σ confidence limit in $\chi^2$ space still corresponds to a fairly small $p_v$ uncertainty ($p_v$~0.043±0.010, 3σ).

This value is also in agreement with the results of Müller et al. (2012), who published $p_v=0.043^{+0.015}_{-0.012}$.

*Conclusion/recommendation*

The mission should carry a visible geometric albedo $p_v$=0.045±0.015 (3-sigma uncertainty) for Bennu.

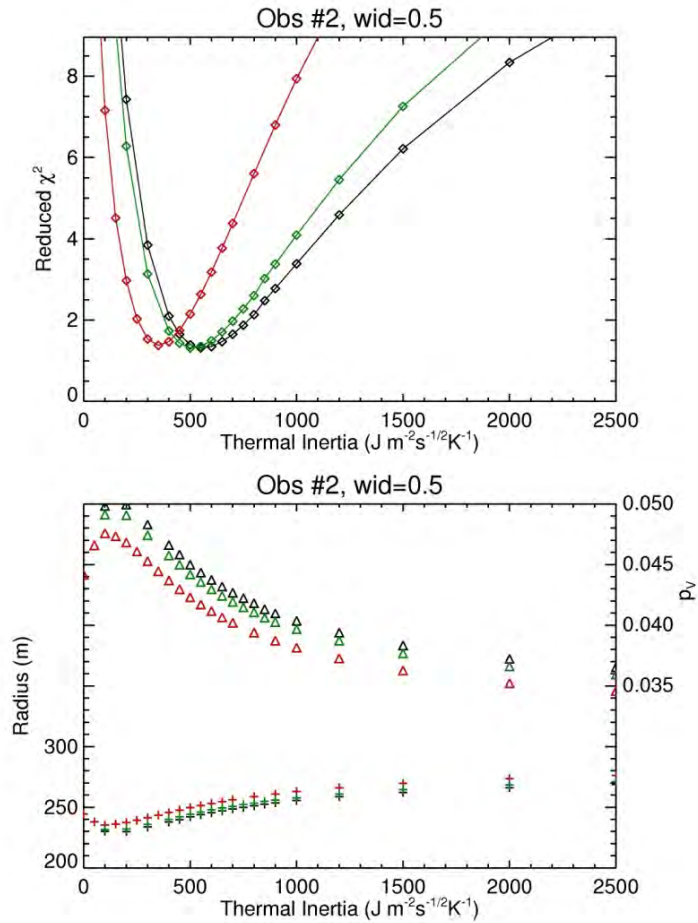

**Figure 21 - Results of thermophysical model fits to Spitzer IRS data of Bennu. Red points are for a smooth surface, Green are for a nominal roughness, and black are for the roughest possible surface. The top plot can be used to infer the best-fit thermal inertia, and the bottom plot illustrates the size and albedo that correspond to a given thermal inertia (or range).**

*Thermal inertia*

Figure 22 shows the thermophysical model (TPM) fit by Josh Emery to Spitzer IRS spectra and IRS/IRAC photometry. The thermal inertia derived from the model fits is 600 ± (-200/+100) J m$^{-2}$ s$^{-1/2}$ K$^{-1}$ (the lunar regolith is ~50 in these units). This range of values includes modeling different surface roughnesses. This moderately high thermal inertia suggests a somewhat blocky surface, perhaps similar to that of the similarly sized NEA Itokawa (
Figure 23). The inferred size of Bennu as derived from this model (D~635 ± 15 m) is in excellent agreement with radar observations. The current model assumes a moderate surface roughness (RMS slopes ~20°). The albedo derive from the model fits is ~0.028 ± 0.002. The Emery/Spitzer models also found a beaming parameter (η) of 1.68 ± 0.09 and subsolar temperature ($T_{ss}$) of 332 ± 5 K.

The thermal inertia found for Bennu is similar to those found for similar sized NEAs, including Itokawa (
Figure 23). These values confirm that Bennu is covered in regolith. Interestingly even the smallest asteroids (including one with an extremely rapid rotation period) with thermal inertia measurements appear to have regolith covering a significant fraction of their surfaces. Both the thermal inertia and phase function show Bennu to be similar to other NEAs and suggest a significant amount of regolith on its surface.

Additional confirmation of a very low albedo is provided by the near-IR spectra and phase function photometry. Modeling of the thermal tail observed in near-infrared spectra beyond 2 μm yields an albedo on the order of 0.03. Also, the slope of the phase function is a proxy for albedo. The relatively large phase function slope of 0.040 mag/deg is indicative of a low albedo (< 0.10).

The combination of orbital information, the radar-derived shape model and thermal infrared measurements has allowed Marco Delbo to produce a temperature model of the surface of Bennu. Input parameters for the model are:

> Shape Model – 1999RQ36.shape1.2008dec30.obj,
> Bolometric Bond's Albedo – 0.01,
> Ecliptic Longitude of Pole – 0.0°,
> Ecliptic Latitude of Pole - -90.0°,
> Sidereal Rotation Period – 4.2968 hours,
> Reference Rotation Phase – 0-350°, step size = 10°,
> Reference Epoch – JD 0.0,
> Emissivity – 0.9,
> Thermal Inertia – 50.0 – 600.0 J m-2 s-0.5 K-1,
> Mean Surface Slope – 0.0°.

Temperatures were calculated at perihelion for different times of the day (i.e. different rotational phases with a step of 10 degree)). The maximum temperature was then recorded for each surface element.

Figure 24 shows the MAXIMUM temperatures reached by each surface element of the Nolan's shape model of Bennu for 2 different value of the thermal inertia as function of longitude and latitude for three positions along the orbit, namely perihelion, aphelion and the ascending node. Figure 25 shows the MINIMUM temperatures. A thermal inertia of 50 J m$^{-2}$ s$^{-0.5}$ K$^{-1}$ is typical for lunar regolith and appropriate to represent the thermal behavior of region of the asteroid covered with fine regolith. A thermal inertia of 600 J m$^{-2}$ s$^{-0.5}$ K$^{-1}$ is the value derived by J. Emery for Bennu from disk-integrated Spitzer observations.

The main findings of the model are:

1. Lower values of the thermal inertia imply higher temperatures,
2. The polar regions are cooler than the equator no matter of the value of the thermal inertia,
3. The south pole is cooler than the north pole. Less temperature-altered organics are more likely to be found on the south pole rather than on the north pole.

The Herschel observations, methods of reduction and results are presented in Müller et al. (2012). The following text is from the Müller et al. (2012) abstract:

*"In September 2011, the Herschel Space Observatory performed an observation campaign with the PACS photometer observing the asteroid (101955) Bennu in the far infrared. The Herschel observations were analyzed, together with ESO VLT/VISIR and Spitzer/IRS data, by means of a thermophysical model in order to derive the physical properties of Bennu. We find the asteroid has an effective diameter in the range 480 to 511 m, a slightly elongated shape with a semi-major axis ratio of a/b = 1.04, a geometric albedo of 0.045 (+0.015, −0.012), and a retrograde rotation with a spin vector between –70 and –90° ecliptic latitude. The thermal emission at wavelengths below 12 μm-originating in the hot sub-solar region- shows that there may be large variations in roughness on the surface along the equatorial zone of Bennu, but further measurements are required for final proof. We determine that the asteroid has a disk-averaged thermal inertia of $\Gamma = 650$ J m$^{-2}$ s$^{-0.5}$ K$^{-1}$ with a 3-σ confidence range of 350 to 950 J m$^{-2}$ s$^{-0.5}$ K$^{-1}$, equivalent to what is observed for (25143) Itokawa and suggestive that Bennu has a similar surface texture and may also be a rubble-pile in nature. The low albedo indicates that Bennu very likely contains primitive volatile-rich material, consistent with its spectral type, and that it is an ideal target for the OSIRIS-REx sample return mission."*

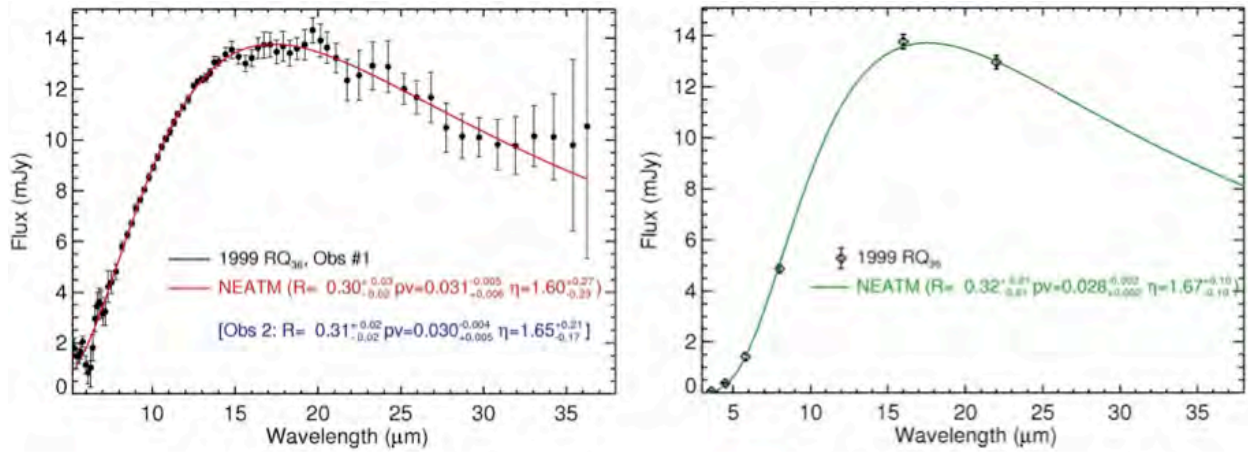

Figure 22 - IRS spectrum of Bennu binned by a factor of 8 and fit with a thermophysical model. The model fit suggests thermal inertia ~600 J m$^{-2}$s$^{-1/2}$K$^{-1}$ and diameter of ~610 m (giving $p_v$ ~0.03).

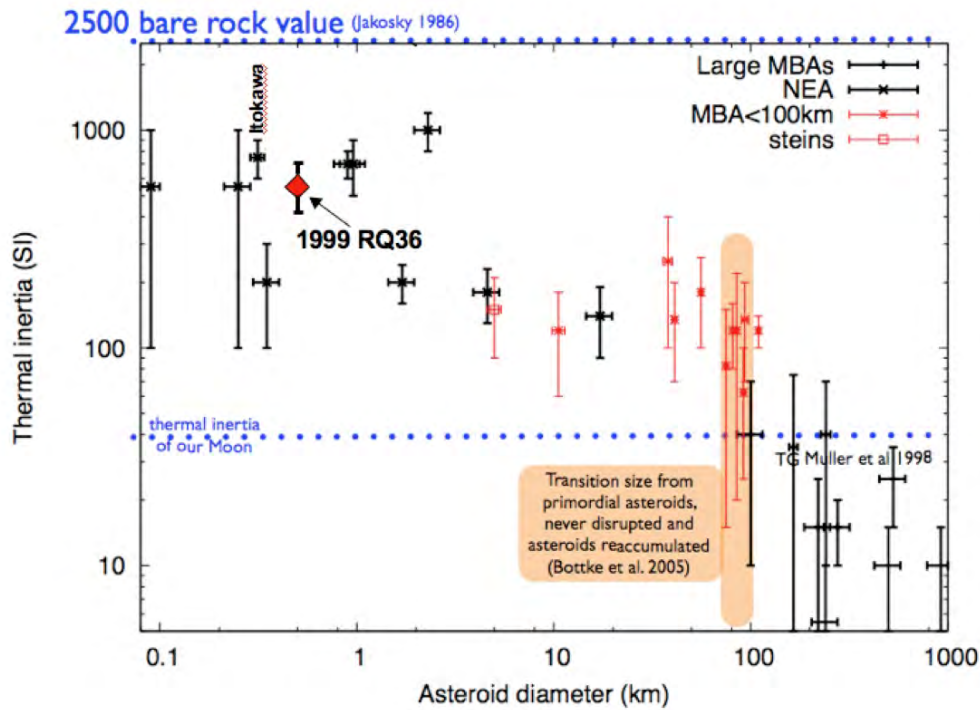

Figure 23 - Relationship between thermal inertia and asteroid diameter.

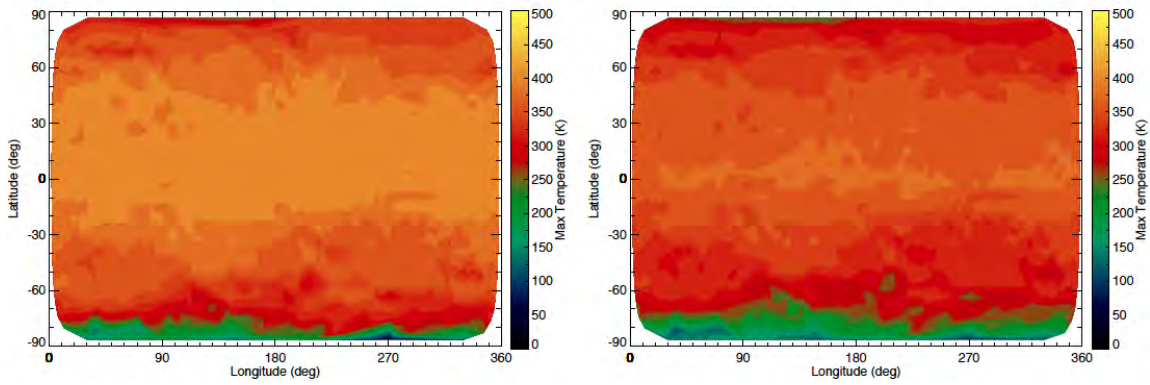

Figure 1: Perihelion, LEFT: $\Gamma=50\ J\ m^{-2}\ s^{-0.5}\ K^{-1}$, RIGHT: $\Gamma=600\ J\ m^{-2}\ s^{-0.5}\ K^{-1}$

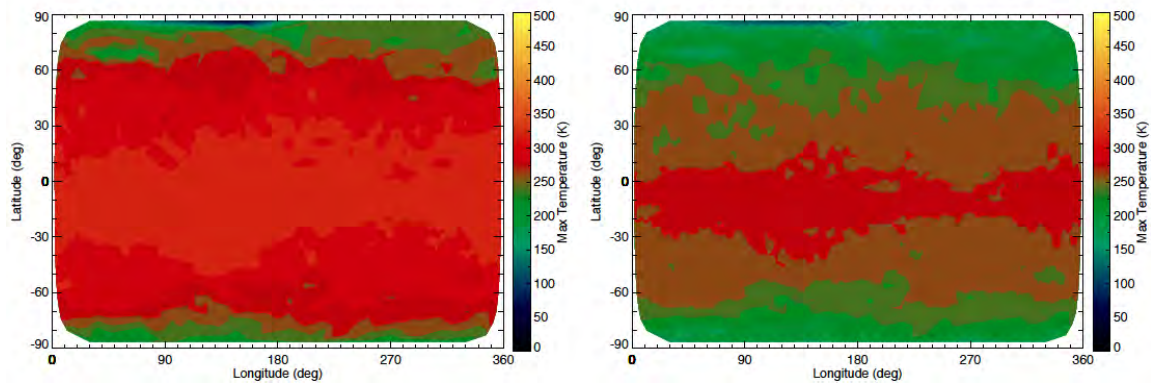

Figure 2: Aphelion, LEFT: $\Gamma=50\ J\ m^{-2}\ s^{-0.5}\ K^{-1}$, RIGHT: $\Gamma=600\ J\ m^{-2}\ s^{-0.5}\ K^{-1}$

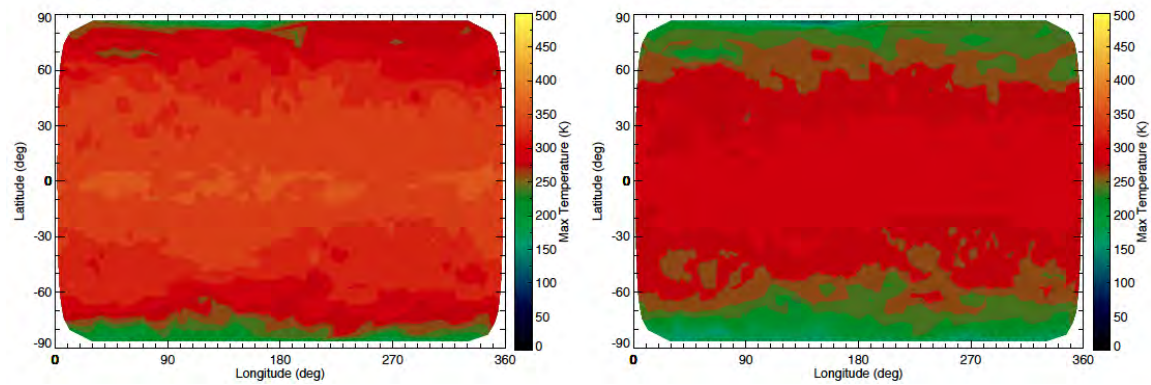

Figure 3: Ascending Node, LEFT: $\Gamma=50\ J\ m^{-2}\ s^{-0.5}\ K^{-1}$, RIGHT: $\Gamma=600\ J\ m^{-2}\ s^{-0.5}\ K^{-1}$

**Figure 24** - Maximum temperatures for thermal inertias of 50 (left) and 600 (right) J m$^{-2}$ s$^{-0.5}$ K$^{-1}$ at Perihelion, Aphelion and the point of Ascending Node.

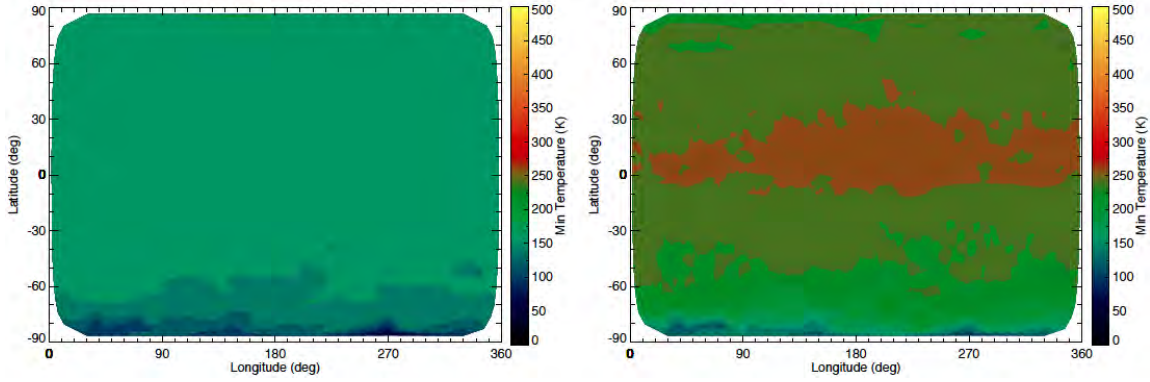

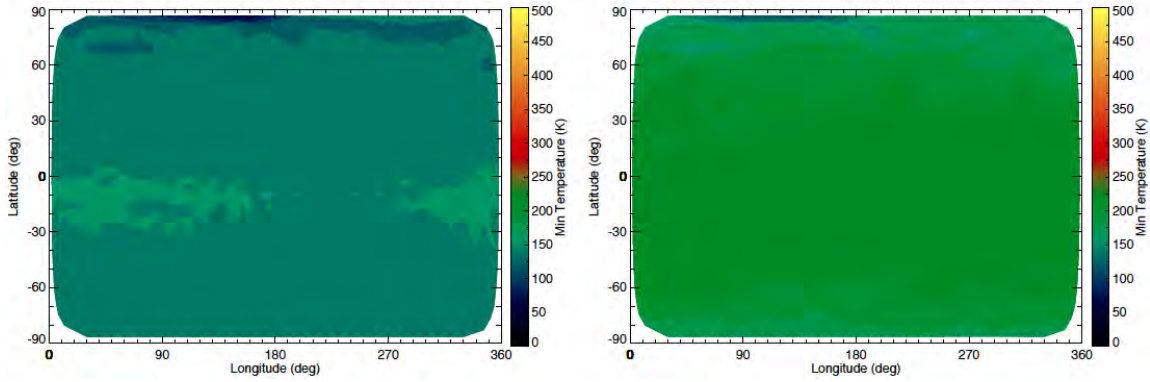

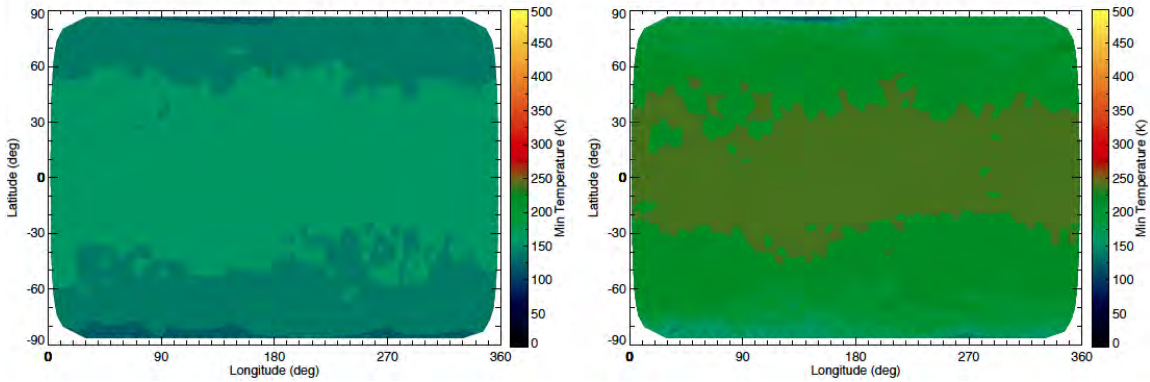

Figure 25 - Minimum temperatures for thermal inertias of 50 (left) and 600 (right) J m$^{-2}$ s$^{-0.5}$ K$^{-1}$ at Perihelion, Aphelion and the point of Ascending Node.

### 7.1.1 Geometric Albedo @ 0.54 μm

- Defined as the ratio of actual reflected light for a body at 0° phase angle to that of a Lambertian disk with the same cross-section. The geometric albedo is valid for the V-band (0.54 μm).
- Source: Josh Emery and Antonella Barucci – Emery et al. (2014)

$$P_v = 0.045 \pm 0.015 \text{ (3-sigma uncertainty)}$$

### 7.1.2 Geometric Albedo from 0.48 to 2.0 μm

- Defined as in 7.1.1 but for wavelengths from 0.48 to 2.0 μm
- Source: Carl Hergenrother

Same as parameter 6.2.1

### 7.1.3 Thermal Inertia

- Defined as the rate at which the temperature of a body approaches that of it's surrounding.
- Source: Josh Emery and Antonella Barucci – Emery et al. (2014)

$$\Gamma = 310 \pm 70 \text{ J m}^{-2} \text{ s}^{-0.5} \text{ K}^{-1} \text{ (1-sigma uncertainty)}$$
$$\text{(3-sigma uncertainty is} \pm 180 \text{ J m}^{-2} \text{ s}^{-0.5} \text{ K}^{-1}$$

### 7.1.4 Beaming Parameter

- Defined as the measure of the non-Lambertian, "beaming", scattering of airless surfaces. [*Note from J. Emery: For the large phase angle observations by Spitzer, NEATM is not an appropriate model, and gives incorrect physical/model parameters in order to match the flux. For example, the beaming parameter will change with smaller phase angles, so this beaming parameter would probably not work well for small phase angle thermal observations of Bennu.*]
- Source: Josh Emery and Antonella Barucci – Emery et al. (2014)

$$\eta = 1.55 \pm 0.03 \text{ (1-sigma uncertainty)}$$

### 7.1.5 Global Temperature Model

- Defined as a spatial model of the temperature of the surface of Bennu at different rotation phases and orbit location. This version of his model is for the orbital positions of perihelion, aphelion and ascending node, a bond albedo of 0.02, an IR emissivity of 0.9, thermal inertias of 50 and 600 J m$^{-2}$ s$^{-0.5}$ K$^{-1}$ and no surface roughness.
- Source: Marco Delbo

The data files for this model are not part of this public release of the OSIRIS-REx Bennu Design Reference Asteroid.

### 7.1.6 Bond Albedo

- Defined as is the fraction of power in the total electromagnetic radiation incident on an astronomical body that is scattered back out into space. It takes into account all wavelengths at all phase angle.
- Source: Josh Emery – Emery et al. (2014)

$$A = 0.0173 \pm 0.0027 \quad \text{(1-sigma uncertainty)}$$

# 8 Surface Analog Properties

*Surface Analog properties describe the characteristics of the surface analogs and how they react to external stimuli. Note that these are not characteristics of the regolith of (101955) Bennu but of terrestrial and meteorite material that are analogous to the surface material on Bennu.*

*Surface Analog Properties for Bennu have not been formally published and are assumptions and best guesses based on our understanding of meteorites and the surface of asteroids. This work has been internally reviewed by the OSIRIS-REx Science Team.*

## 8.1 Mechanical Characteristics

The surface of Bennu has not been imaged optically at a resolution that would confirm the existence of a regolith and provide information about the size distribution of such material. The assumption of a regolith, and in particular of a regolith with clasts that can be ingested by TAGSAM, is based entirely on the analyses of thermal inertia and radar polarization measurements.

In this report, consideration is given to the possibility that shape and slope models, the evolutionary and dynamical history of the asteroid, surface composition from spectroscopy, and comparisons with other asteroids, might provide additional clues about the regolith.

In addition to knowing that there is a regolith, we also need to understand its size distribution. First, there must be material of TAGSAM ingestible size. Second, this must not be mixed with clasts that are large enough to foul the TAGSAM mechanism, or sufficiently large to endanger the S/C itself. Knowledge of these is currently lacking from either observations (direct or indirect), or from modeling. In a recent report (Marshall, Dec 2011), the following classes of regolith size were identified for consideration in any development of a regolith model and what the model predicts about TAGSAM interaction:

INGEST RANGE: 0-2 cm. This is collectable by TAGSAM.

COMPLIANT RANGE: 2-30 cm. This might be capable of being pushed into the regolith to ensure full contact of the TAGSAM head with the regolith surface, and also does not offer significant resistance to any lateral sliding of the TAGSAM head. In the report just mentioned, the upper limit of this size fraction was derived by a calculation analysis of inertial and frictional resistance to the TAGSAM load of 53 N, as well as considerations about the size of the TAGSAM head.

HAZARD RANGE: >20 cm clasts may hinder surface compliance, snag the arm, or may be a S/C collision hazard.

Of the three basic sizes of concern, the relevance of the smallest (ingestible) is obvious (we might land there), and the relevance of the largest (hazardous) is also obvious (we won't land there). The relevance of an intermediate size (compliant) is not so obvious, but it could be a deciding factor in determining a site for TAG. This "unappreciated" size deserves further attention. We currently have no knowledge of any of these size fractions, but they are all equally important in assessing where to TAG. Fortunately, the OCAMS camera resolution will automatically assess the larger fractions during surveys for the ingestible size range.

Our knowledge of the regolith's physical characteristics that will have to be developed as soon as we arrive at Bennu includes the following:

1) The presence of ingestible material (< 2cm).
2) Enough ingestible material per unit area to satisfy the statistical level of confidence (currently 95%) that it will be encountered by TAGSAM.
3) An area of ingestible material equal to or greater in size than the long dimension of our landing ellipse.
4) No stones of hazard size in the landing ellipse, or an acceptably low statistically probability of encountering them (an inverse probability to the ingestible material).
5) An acceptable probability that there is low risk from the presence of nonhazardous, non-ingestible stones (compliant size as defined below).
6) No hazardous topographic slopes or geomorphic features of TBD steepness/size and TBD density of distribution within the landing ellipse.
7) No adverse geodetic gradients in the landing ellipse.

If we were visiting Itokawa, we would have a very good chance of acquiring regolith with TAGSAM, but this would only be possible because of the spatial sorting of clast sizes that has evolved in areas such as Muses C. Hence, the idea of "average" grain size is only meaningful in this context. Currently our expectations of spatial sorting by clast size on Bennu are undeveloped and even comparing Bennu with Itokawa might be considered misleading because of the very different asteroid histories. We do not know if there are equivalents to Muses C, given the different histories, shapes, and gravitational equipotential surfaces of the two bodies, nor if latitudinal drift of material to the equator actually causes size sorting and in which latitudinal direction.

We also have no direct or modeled insight into the compaction state of the regolith and the possibility that the surface might be a seismic lag deposit with vertical stratification (fines hidden below coarse surface cobbles). Vibrational compaction can provide very rigid granular structures, but how these behave in low gravity is not understood either intuitively, or from the results of experimentation or modeling.

So far, the discussion has considered only *static* properties of the regolith, but it is also important to understand the *dynamic* properties. These are specific to the reaction of the regolith to the impingement of the TAGSAM head, the subsequent fluidization with high-pressure gas, and the "pull-away" effects of dragging regolith with the collector. Understanding the relative roles of clast inertia, momentum, dilatancy, frictional vs. frictionless granular flow, intergranular cohesion, and other effects are key to the success of TAGSAM operations. Currently we have very limited experimental (microgravity) data in support of the dynamic properties. The virtual weightlessness of the material is the critical factor, and its effects are not particularly intuitive. Low-gravity flights have had encouraging results, but the results are also ambiguous owing to the effects of negative g's during experiment initiation. These flights also dealt with small stones of ingestible size, but not the larger compliant clasts. Fortunately, the OSIRIS team now has members dedicated to modeling these dynamic properties. Additionally, a possible method of simulating Bennu in the lab using an inverted force system of fluid-suspended materials has been suggested in the Marshall (Dec 2011) report.

If we were pressing TAGSAM slowly into the regolith, the weight of the stones would be most important – inertia would have a secondary roll. But we are expecting to push stones into compliance with the head in a relatively short time (1-s, TBD). In this case, the inertia of the stones is the most important parameter and this is more difficult to simulate experimentally than weight (or lack thereof).

*Thermal Inertia Measurements*

Because thermal inertia is the key source of information regarding the regolith, the following section incorporates the bulk of the supplied material.

*Synopsis*: Measurements of thermal emission from Bennu indicate a regolith dominated by grains with diameters larger than a few hundred microns and smaller than a few centimeters. Two groups using independent measurements have both derived a thermal inertia of ~600 J m$^{-2}$ K$^{-1}$ s$^{-1/2}$. This is more than an order of magnitude larger than the thermal inertia of the Moon, indicating a somewhat coarser regolith. The average grain size of the lunar regolith is ~60 μm. The derived thermal inertia of Bennu is, on the other hand, much lower than bedrock, indicating that the regolith grains are smaller than the diurnal thermal skin depth, which is a few cm on Bennu. More precise estimates of grain size would require new laboratory measurements of thermal conduction as a function of grain size taken under vacuum, detailed modeling of heat transport across grain contacts, or both.

*Theory*: Qualitatively, the relationship between thermal inertia and grain size is commonly observed – unconsolidated sediment (e.g., sand on a beach) heats up and cools down quickly, whereas adjacent gravel, asphalt, or sidewalks remain cool well into the day and warm well after sunset. The basic reason for this relationship is the interruption of thermal conduction pathways by grain boundaries. Other factors besides simply grain size – e.g., contact area and inter-grain heat transport (dominated by gas-phase conduction on bodies with atmospheres and by radiative transport in vacuum) – can heavily influence the bulk conductivity, resulting in a relationship between conductivity and grain size that is complicated. The grain size/conductivity relationship under vacuum, in particular, is not very well quantified. However, the qualitative relationship certainly holds and can be used to place broad but useful constraints on the grain size and texture of regolith.

In practice, thermal emission data as measured by orbiting spacecraft and instruments on the ground are regularly used on Earth and Mars for distinguishing between bedrock vs dust or consolidated vs unconsolidated sediments. Mapping of quantitative thermal inertia values to discrete grain-size ranges is possible due to laboratory measurements of thermal conductivity dependencies under appropriate environmental conditions for Earth and Mars. Remote measurements of the bulk thermal inertia of the Moon combined with in situ and laboratory measurements of lunar soils and the upper bound imposed by the thermal inertia of silicate bedrock provide the only concrete tie points for linking derived thermal inertias of airless bodies to an average grain size. The thermal inertia of the Moon, which has a mean regolith grain size of ~60 **μ**m, is ~50 J m$^{-2}$ K$^{-1}$ s$^{-1/2}$, whereas that of silicate bedrock, defined to be slabs or rocks larger than the thermal skin depth (which itself depends on spin-rate as well as other parameters), is ~2500 J m$^{-2}$ K$^{-1}$ s$^{-1/2}$. Thermal emission measurements of a variety of large, airless bodies (asteroids, icy moons) generally all result in estimates of low, lunar-like thermal inertias, likely indicating similarly well-developed regoliths. Several small near-Earth asteroids (NEAs), on the other hand, have thermal inertias significantly in excess of the lunar value, but none so far as high as bedrock. Only two of these NEAs have been imaged up-close: Eros and Itokawa. The thermal inertias of Eros and Itokawa derived from ground-based thermal emission measurements are ~150 and 750 J m$^{-2}$ K$^{-1}$ s$^{-1/2}$, respectively. Images of Eros show a fairly well developed regolith, possibly with a larger rock fraction than the Moon, but detailed grain-size distributions for the regolith itself are not possible from the orbital images. Images of Itokawa show a much blockier (i.e., composed of larger rocks) surface than the Moon or Eros, but again, detailed size distributions for grains smaller than a few cm are not possible. Eros and Itokawa therefore only provide qualitative comparisons for linking thermal inertia to grain size.

*Observations*: Thermal emission measurements of Bennu have been collected by two different space telescopes – Spitzer and Herschel. The Spitzer measurements were conducted during the period 4 – 9 May 2007. They consist of spectra covering 5.2 to 38 μm taken of opposite hemispheres (integration time ~1/2 a rotation period) and photometric measurements at 3.6, 4.5, 5.8, 8.0, 16, and 22 μm taken at 10 different longitudes (integration times very short compared to rotation period). The spectra are relatively low signal-to noise (S/N), but fully cover a broad range that includes the peak in emission – a sensitive indicator of temperature. The photometric measurements are higher S/N, but more sparsely cover the spectral range. Furthermore, the 16 and 22 μm data were taken several days after the 3.6 to 8.0 μm data (two different instruments), but are rephased to produce a lower resolution spectral flux curve at each of the 10 longitudes.

The Herschel measurements were conducted on Sept 9, 2011. They consist of photometric measurements at 70, 100, and 160 μm. Supporting measurements at shorter wavelengths were made during the same month using the VLT ground-based facility.

*Analysis*: Thermal fluxes are interpreted in terms of physical properties through the use of models of surface energy balance. In order to constrain thermal inertia, the model must explicitly include subsurface heat conduction. The telescopic measurements are all full-disk – the asteroid is not spatially resolved. Asteroid thermal models therefore solve for temperatures across the surface of the asteroid, calculate the thermal flux from each surface facet, and then integrate the fluxes over the entire visible surface at each wavelength to calculate a model spectrum. The temperature distribution is altered (by varying key input parameters such as thermal inertia) to find the best fit to the measured spectral fluxes. The resulting thermal inertia estimates are therefore somewhat model dependent. The biggest model uncertainties are generally shape, spin-pole orientation, and surface roughness. For Bennu, the last of these is the most severe.

Thermal inertia provides some indication of the regolith texture, including grain size. However, no good quantitative relationship between thermal inertia (or conductivity) and grain size has been established for use under vacuum conditions. Laboratory measurements under vacuum have not been sufficiently systematic or well-documented, and theoretical calculations are hindered by the complexity of grain-to-grain contacts. The most reliable connections to be drawn, therefore, rely on the comparisons to the Moon and silicate bedrock mentioned above.

*Results*: Thermal models of the Spitzer data alone by Emery and models that include the Spitzer, Herschel, and VLT data by Barucci/Müller both find a thermal inertia of ~600 J m$^{-2}$ K$^{-1}$ s$^{-1/2}$ for Bennu. The formal value presented by Emery has been 550±150 J m$^{-2}$ K$^{-1}$ s$^{-1/2}$. This uncertainty includes the full range of physically possible surface roughness (smooth to completely covered in hemispherical roughness elements). The formal value presented by Barucci/Müller is 600 ± 50 Barucci/Müller. This assumes a nominal surface roughness that is compatible with observed surfaces. The two reported uncertainties might therefore be thought of as something like a 2 to 3-σ Value and a 1 to 2-σ value, respectively. In any case, the thermal inertia values agree quite well.

The thermal inertia of Bennu is more than an order of magnitude larger than that of the Moon. This implies that the average grain size on Bennu is larger than that of the lunar regolith. Alternatively, the grains themselves could be similar size, but the grain-to-grain contacts larger than on the Moon (e.g., by cementation or weaker grain strength). On the other hand, the thermal inertia of Bennu is a factor of several smaller than that of bedrock. This suggests that the average grain size is smaller than the diurnal thermal skin depth. The diurnal skin depth (ls) is given by

$$l_s = \sqrt{\frac{k}{\rho c_p \omega}}, \text{ and } \Gamma = \sqrt{k c_p \rho}, \text{ giving } l_s = \frac{\Gamma}{c_p \rho \sqrt{\omega}},$$

where k is thermal conductivity, cp is heat capacity, $\rho$ is grain density, $\omega$ is rotation rate, and $\Gamma$ is thermal inertia. For the known rotation period of Bennu (~4.29 hr), reasonable values for $c_p$ and $\rho$, and the derived thermal inertia, ls for Bennu is 2 to 4 cm. The average grain size on Bennu must therefore be smaller than 2 to 4 cm. The regolith grain size on Bennu is therefore expected to fall somewhere between a few hundred microns and a few cm. Given that the thermal inertia of Bennu is between that of the two end members, it is reasonable to expect that the mean grain size is on the order of a few mm.

The Spitzer photometric measurements sample the rotation period at 10 evenly spaced intervals. It is important to recognize that although the observations strictly integrate flux over the entire visible hemisphere, the majority of the contribution comes from a restricted region around the hottest longitude. The 10 intervals therefore really do end up sampling the surface spatially to some extent. Nevertheless, those data show no evidence for variation in temperature (i.e. thermal inertia) across the surface; Bennu appears to be rotationally homogenous in terms of its thermal properties, and therefore regolith texture/grain size.

*Calibration/Ground-truth*: This is covered above in the Theory section. The only reliable ground-truth is the Moon and bedrock values of thermal inertia. Results for Eros and Itokawa support the broad inferences on regolith grain size from those two end members. Note that for Itokawa, the slower rotation period and higher thermal inertia result in a larger skin depth. Therefore, similar arguments as used for Bennu would allow for a larger upper limit to the grain size on Itokawa, consistent with imaging results.

*Limitations of technique*: The limitations and uncertainties relate mainly to inferring grain size from the thermal measurements. Nevertheless, the upper limit in grain size is considered robust. There is no physical mechanism that could lower the thermal inertia from the bedrock value without decreasing the grain size. On the other hand, cementation or increased surface area at the grain-to-grain contacts could increase thermal inertia for a fine-grained regolith. The lower limit may therefore be less robust. The thermal modeling is fairly straightforward, and the pertinent parameters, with the exception of surface roughness, are well known from other measurements (visible lightcurves, radar). The uncertainties listed above capture the observations and modeling uncertainties well and are completely swamped by the lack of precision possible in converting to grain size.

*Potential improvements*: Improvements in the constraints placed on grain size from thermal measurements would rely on gaining a better quantitative grasp of the relationship between grain size and thermal conductivity in vacuum conditions. There are potentially two approaches to this problem: laboratory measurements and numerical modeling. Laboratory measurements would likely be the most useful here. With the proper setup, appropriate analogs could be used to make the direct measurements that would enable a tie point to be established in terms of grain size. Numerical modeling holds some promise, but would be substantially complicated by the fact that the nature (i.e. surface area) of grain-to-grain contacts in the regolith of Bennu is not known and is likely to remain unconstrained until samples are returned. Measurements of heat capacities of potential analogs would help to improve precision on the calculation of $l_s$, but this improvement would be minor compared to actual measurements of conductivities in vacuum.

Observationally, continued temporal coverage would continue to refine understanding of the thermal properties of the surface. Observations at different phase angles, particularly corresponding to apparitions that would see the "morning" side of the asteroid (the current observations see the "afternoon" side) would proved better constraints on the surface temperature distribution. Observations as Bennu traverses different parts of its orbit would potentially constrain the thermal properties to a greater depth, corresponding to the annual thermal wave. But neither of these would really improve nor refine the estimate of grain size if the laboratory and/or numerical modeling work are not undertaken.

---

*Further to the above discussion are comments by Marco Delbo:*

I do not fully agree with the interpretation of the value for thermal inertia = 600 J m-2 s-0.5 K-1 as a regolith dominated by grains with diameters larger than few hundred microns and 'smaller than a few cm'.

In my opinion, the problem is, as noted by John, that a value of thermal inertia of 2500 J m-2 s-0.5 K-1 might not be appropriate for the rocks on Bennu larger than the skin depth, and here is why: recent lab measurements by Opeil et al. (2010) show lower-than-expected thermal conductivities of the bulk of meteorites. In particular, the Cold Bokkeveld (CM2) meteorites has a bulk (or solid) thermal inertia of 644 J m-2 s-0.5 K-1 and thus very similar to Bennu.

Note that the best spectral analogs for Bennu are the CM, CR, or CI chondrites, though none are a perfect match (Lauretta et al.) Cold Bokkeveld is a CM2.

To the best of my knowledge, Opeil et al. have not done any measurements of CR, or CI chondrites.

---

*Patrick Michel also added some comments:*

Bennu is a small body whose density and albedo are both very low. This puts it closer to comets in any continuum of objects that is bounded by comets on one side and S-type asteroids like Itokawa on the other. Bennu will be one of the dimmest objects we have ever visited, while Itokawa was one of the brightest, if not the brightest, especially for that size class. Unfortunately, we have not imaged yet any object like Bennu in details and so rely only on educated guesses as to how to describe its regolith. Because the regolith properties on Itokawa have often been assumed as a reference to define the possible regolith properties on Bennu, a discussion follows regarding the validity of this assumption. First let's recall some facts:

1. Bennu is a B-type asteroid with a shape and likely a composition (based on the spectrum and albedo) that are greatly different from Itokawa.

2. Bennu has a very low bulk density (about half that of Itokawa).

Then, let's derive some possible implications of these two facts:

1. Bennu, based at least on its shape, has experienced a different dynamical and collisional history from that of Itokawa. As a consequence, the evolution of the regolith may have been subjected to similar but also different processes (such as YORP spin-up that may explain its shape, while Itokawa's shape does not seem to have been sculpted by this process) and/or environments. This may have lead to different regolith properties.

2. Bennu has a density that is so low that it is possibly macroporous AND microporous. The microporosity of Bennu's material may lead to a different behavior when subjected to physical processes such as impacts, compared with the behavior of a nonporous material. In effect, pore crushing (compaction) is involved in impacts on microporous material, which causes additional energy dissipation (compared to the same process on non-porous objects) and drastically influences the impact outcome (size and velocity distributions of fragments). Therefore, if regolith forms at least partially due to the impact process, we can expect different properties from those produced from a nonmicroporous material, such as Itokawa's one, although we don't have yet a good understanding of what these differences might be. The same holds true regarding thermal fragmentation, if it also pays a role in regolith production. Nevertheless, one may argue that it is interesting that despite the difference in composition, bulk density etc .. Itokawa and Bennu have the same thermal inertia, suggesting domination by grains at the millimeter-scale, and that Hartley 2 also looks strangely similar to Itokawa (without activity). Maybe this tells us something about the behavior of different materials submitted to the same physical processes in micro-g conditions. This similarity may suggest that those material differences have minor effect on their response to those processes. However regarding thermal inertia, the similarity in thermal inertia may not really tell us that the regolith properties are similar in the two cases. Recall that the thermal inertia measurements from the ground are disk integrated (over the whole body) and there is no crystal clear understanding about what it means regarding the actual distribution and properties of the regolith.

Experts on thermal aspects indicate (and may confirm) that the solution is not necessarily unique, and therefore, two similar thermal inertias may mean two different sorts of regolith (only the disk-integrated skin depth, for what it means, is similar).

For instance, it may mean that either we have a deep layer of fine grain regolith, or a shallow layer of large grain regolith. As for Hartley 2, being a comet with apparent activity (jets), it is also difficult to compare, because this body is actually submitted to processes that are different than those experienced by Bennu. So, we may speculate that the apparent similarity of Itokawa and Hartley 2, despite completely different compositions, may not be observed if both bodies were submitted to the same processes. For instance, under the same process, does the rock-type on Bennu weather in the same way as a metamorphic rock on Itokawa? The answer may not be trivial. Although one hypothesis is that Itokawa and Bennu may respond in a similar way to similar processes, another is that, at least with respect to weathering (not break-up, re-lithification) their weathering products could be different based on the original texture, size, and modal mineralogy, and therefore the regolith formed that way may lead to different products.

The story of disk integrated asteroid thermal inertia vs grain size deserves a lot of attention."

---

*And finally, Ben Clark and John Marshall have questioned* the assertion that composition will necessarily dominate the type of regolith that evolves –physics may trump all other considerations and lead to more or less the same material comminution products within the impact gardened layer.

Something else not considered in sufficient depth is how an asteroid retains the regolith as well as how it develops it. It is clear that there is a trend towards fewer fine fractions as airless bodies get smaller. What is the mechanism of loss –increased centrifugal effects, selective ejection of fines by physical or electrostatic forces…?

---

The RDWG is continuing to evaluate the interpretation of thermal inertia data for Bennu because this is the primary indicator of the size distribution available for TAGSAM collection. Interpretation of the thermal inertia as indicating mm size material is still a contentious issue. It has seen numerous email exchanges, too numerous to be documented here. However, the stream of ideas noted above gives an idea of the issues and our current interpretations of the available data. The current concern focuses on why Itokawa and Bennu have very similar thermal inertia values, but the regolith is three orders of magnitude larger on Itokawa. The cause of this discrepancy, and especially the role of composition (about which we have heard no definitive concepts) needs to be debated in future RDWG sessions.

Additionally there are bodies of data that provide ground truth for the thermal inertia data. These are thermal inertia measurements for objects such as Mars for which we have direct grain size measurements and observations. These data, however, do not currently exist in our OSIRIS documentation trail.

**Figure 23** illustrates the relationship between asteroid diameter and thermal. It is still awaiting more analysis of the data for Bennu (notably absent from this compilation). Itokawa and JU3 have similar thermal inertia values. The implication of this figure is that small asteroids do not retain the finest fraction of the regolith.

*Radar Polarization Ratio*

The radar polarization ratio (0.18) suggests a smooth surface of fine-grained material in the size range ingestible by TAGSAM (this is virtually the only statement regarding radar polarization in the CSR and needs to be expanded). Similarly to the thermal inertia measurements, the data need to be grounded against other observations. In particular, it is not documented if radar polarization measurements have been made of other airless bodies as point sources for which we have direct confirmation of grain size from observations. Do such data exist? Also, similarly to the thermal inertia measurements, the radar data have significant degrees of uncertainty that have not been bounded with high levels of confidence at this time. The methods of deriving the data for both thermal inertia and radar polarization are not documented in an OSIRIS paper trail. It is anecdotal that radar data possibly indicate latitudinal sorting of material by size, but no documentation of this is evident. Radar polarization assessment is incomplete at this time; input to this DRA document has not been provided.

*Shape and Slope Models*

These were much more extensively examined at the Tucson STM than in the CSR. The rather good correspondence between rotational axis and the asteroid shape strongly suggests a dependence of the latter on the former, viz., that the equatorial bulge and the generally equant shape and sphericity of the object reflect a balance between gravitational and centrifugal forces. In other words, the body is deformable and not a solid rock object. It is not definitively determined if this reflects whole-body deformation (consistent with the spinning of a rubble pile), or the migration of surficial material to the equatorial region.

Surface slope maps that have been produced from modeling indicate that the general "down" gradient is towards the equator (see `Figure 5` in Section 2). These models depict gravitational potential relative to topographic slope and are therefore hard to translate directly into geomorphic processes that we are familiar with on Earth. Nevertheless, these models indicate an immature surface (otherwise the slopes would all be zero) that may still be evolving into a smoother asteroid surface, or that alternatively may have frozen in time. In the latter case, the shape of Bennu and the surface slopes would not reflect the present rotational state of the asteroid, but rather, a previous more dynamic situation (spin-up). Do we know which scenario favors clast size sorting by latitude -- cataclysmic spin-up or long-term stable equatorial creep, or indeed, if either supports size sorting? From our terrestrial experience, we know that clasts are sorted by size in many gravitational situations, screes for example, but they are also not sorted in debris flows and landslides that give rise to diamictites. Debris flows have been observed on asteroid surfaces –they are not unlike the fanglomerates of desert pediments, and coincidentally very similar to the surface of Itokawa.

Because we do not understand how "average" clast size (from thermal inertia or radar polarization) translates into spatially (geographically) separated size distributions, it is important to develop models that we can use to translate shape and slope models into sorting models. Sorting is likely to be key to the success of OSIRIS in finding ingestible materials. Plans for such modeling are not determined for Phase B at this time.

Shape and slope information can be used, along with other measurement constraints on the regolith of Bennu, to aid in modeling the hypothetical evolution of the asteroid's surface. However, it is difficult to make strong statements about the presence of regolith on the asteroid from computations alone; our understanding of material in this environment is quite speculative. It is noted that shape and slope models tell us nothing about the likely size distribution of clasts within the regolith.

In an accompanying document, D. Scheeres provides an extensive discussion of surface slopes and the asteroidal geoid. There is some reference to regolith movement but apparently such modeling tells nothing about latitudinal sorting of material (we want sorting to concentrate fines), nor the size of material preferentially transported by topographic and geodetic gradients, nor the instigating mechanisms for transport (impact, centrifugal stress threshold). The most important point made by Dan's modeling is the suggestion of surface mobility of material.

*Evolutionary and Dynamical History*

Because Bennu is >> 200 m in diameter, it has a much slower spin rate than smaller objects and therefore is less likely to have centrifugally ejected all its regolith into space. This, however, does not argue for the existence of a regolith, but merely the fact that if there was a regolith at one time, it would not have been lost by this process.

If we know, or can surmise, the history of Bennu from its perambulations through the Solar System, can we deduce a likely impact or fragmentation history and what this might say about the likelihood of fine-grained regolith on the asteroid? If Bennu is a fragment of a larger asteroid, how much of the smaller impact comminution products would likely be produced and (importantly) retained by the body.

However, according to B. Bottke, we probably do not know enough about the dynamical history of Bennu to constrain the nature of the regolith at this time. To do so would require an integrated model that includes as input the likely formation mechanism of the body (i.e., was Bennu likely to accrete loose ejected material after its formation in a possible family-forming event), the subsequent dynamical evolution of the body via Yarkovsky and resonances, the collisional evolution of the body coupled to its dynamical evolution, a determination of how main belt and small particle collisions produce regolith on small bodies, and then how YORP has influenced the body's spin rate and removal of any regolith produced all along its evolution.

We are not close to generating such an estimate, and even if we were, it would be extremely model-dependant. Any such model would likely be superceded by direct observations, given our likely uncertainties, as well as what happens via YORP spin up, which probably dominates many of the other effects (as suggested by Dan Scheeres' work).

*Composition and Spectroscopy*

Could we gain insight into the properties of the regolith through our knowledge of the composition of the asteroid? Composition affects radar reflectivity, thermal inertia, clast friability, intergranular "stickiness", and other properties. Input from the CMWG (H. Connolly) indicates that nothing is predicted about the regolith size distribution from compositional knowledge. Despite the fact that composition is likely to affect thermal inertia, our prime source of size information, there is no resolution on the efficacy of thermal inertia data interpretation with respect to composition (nor indeed, with respect to other factors). This is an ongoing debate within the RDWG.

*Asteroid Analogs*

Some of the key issues regarding asteroid comparisons have already been addressed in the above text. Recent discussions indicate that this could be a very fruitful, if somewhat contentious, area to investigate. The histories of imaged asteroids vary greatly. So too do their compositions. Although general relationships between asteroid size and regolith character can be established, it is not certain where we place Bennu in the scheme of things. What controls regolith development and size distribution – chemistry, physics, evolutionary events? It is currently being debated in the RDWG as to whether Itokawa can be considered a legitimate analog for Bennu in terms of regolith size distribution.

Hayabusa revealed that there was much more regolith present on Itokawa than predicted by theoretical considerations. If we were visiting Itokawa, this would be good news. But basically, it means that the predictions were wrong and perhaps only fortuitously in the right direction. What were the Hayabusa predictions based on, and can we learn from their mistakes?

---

The Carbonaceous Meteorite Working Group (CMWG) produced a report released on May 4, 2012 that investigated the properties of meteorite analogs to Bennu. The relevant portions of that report are excerpted here:

*Solid material density and microporosity*

The possible microporosity and density of the solid material composing Bennu may be estimated on the basis of density and porosity measurements of its most likely meteorite analogs: CM and CI. However, note that meteorites may have been subjected to compaction processes, possibly during the extraction of their parent bodies (numerical simulations of impact on porous materials tend to show that the ejected material is generally more compact than its original state within the parent body), and also that our meteorite collection may be biased towards the stronger component of materials in space that can survive atmospheric entry. So, the values provided by those meteorites may be considered as possible upper limit in terms of the bulk density, compared to the actual density of the solid material composing Bennu. A number of recent measurements of the density and porosity of various types of carbonaceous chondrites, including the CM and CI groups have been reported (see `Table 5`) (Macke et al. 2011; Ibrahim and Hildebrand 2012). These results build on the work of Britt and Consolmagno (2003) that summarizes a broad array of data stretching back decades. Without going into the details of the measurements, the following subsections summarize for these two groups the relevant information.

CM Meteorites

Bulk densities are rather low, although not as low as Orgueil (see below for CI). They average 2.20 g/cc, with a range from 1.88 to 2.47 g/cc. Grain densities average 2.92 g/cc, with a low of 2.74 g/cc and a maximum value of 3.26 g/cc. CMs are quite porous, averaging 24.7% and ranging from 15.0% to 36.7%.

CI Meteorites

CI meteorites are highly porous, extremely friable, and quite rare, with only nine meteorites of the type known. Because of their extreme friability, collections managers have been reluctant to grant access for whole-rock study even using the relatively benign methods employed in this study, and no new report is provided for CIs by Macke et al. (2011). However, Consolmagno and Britt (1998) report physical properties of Orgueil from the Vatican collection. Its grain density (2.43 g/cc) and bulk density (1.58 g/cc) place it among the least dense of chondrites, and give it a very high porosity of 35%. It should be noted that porosity of the extremely primitive and friable ungrouped CC Tagish Lake, determined by methods comparable to Macke et al.'s one, was comparably high at an average of 40% over multiple stones (Hildebrand et al. 2006). Such high porosities are therefore neither unreasonable nor unheard of.

Summary

If we assume CM-like material for the solid material of Bennu, the bulk density may be about 2.2 g/cc (and porosity about 25%); if we assume CI-material, then the bulk density may be lower (1.6 g/cc for Orgueil) and microporosity higher (35%). So, if one wants to stick to the range of measured values for CI and CM, we may consider the measured range accounting for both types, i.e. within 1.6 – 2.5 g/cc. However, because meteorites are likely to be more compacted or less fragile than pristine material, we may consider a range of bulk densities including lower values, e.g. within 1.0 – 2.0 g/cc. The microporosity of this material would then be from 20 to 40%, maybe higher if pristine material is not as compacted (microporous materials on Earth can have a much greater fraction of microporosity: for instance, without claiming this is a good analog, a bloc of solid pumice can have a microporosity as high as 80-90%).

*Macroporosity*

The question of macroporosity is even more difficult. What we may say is that if the bulk density of Bennu is 0.98 g/cc, and if we assume that the bulk density of its solid component is 2 g/cc (upper range above, already including microporosity), its macroporosity is 51%; say 50%. Note that usually, the exercise is done using the grain density of meteorites, instead of their bulk density, and the derived porosity for the whole asteroid is higher as it includes both micro and macroporosity. Here only macroporosity is considered, having already been addressed in the previous section.

However, what does 50% macroporosity mean? No one knows so the following is speculation. It can correspond to 50% of void space distributed uniformly in large voids separating a large number of small monolithic (microporous) blocs. It can also correspond to <50% of void space distributed in small voids separating a small number of big monolithic (microporous) blocs, added with a thick layer of loose regolith, to reach the 50% level. This regolith may still be composed of the same kind of solid material, but instead of being cohesive, as large monolithic blocks, it may be composed of structurally loose smaller components (cohesionless, but possibly with a Mohr-Coulomb/Drucker-Prager shear strength behavior), maybe down to grain size. Its level of compaction (or how much packed it is) depends on its dynamical history (and various processes such as thermal cycling) and on how much we leave for large voids in the macroporosity fraction, which unfortunately cannot be constrained at this stage except on a purely arbitrary basis. In other words, there are different combinations of large void fraction and regolith (and its packing) fraction that can lead to the same whole macroporosity. Unfortunately at this stage, there is no obvious rationale that can give a plausible estimate of the level of compaction expected for the regolith material on Bennu. This goes back to the more general discussion about the regolith properties expected on Bennu, which is not the scope of this report.

Note that in the close future, we plan to look at the possible regolith formation and evolution on Bennu, based on its possible dynamical history, the history of Earth approach distances (to check whether they may have helped freshening the surface) etc … But this is an ambitious project, requiring simulations of regolith evolution under various effects and the investigation of a large parameter space that cannot be done on a time short enough to provide information at the current stage.

Table 5 – Taken directly from Table 2 in Macke et al. (2011)

Physical property averages for carbonaceous chondrite falls by petrographic type.

| Petrographic type | No. meteorites | Bulk density (g cm$^{-3}$) Average | Bulk density Max. Min. | Grain density (g cm$^{-3}$) Average | Grain density Max. Min. | Porosity (%) Average | Porosity Max. Min. | Magnetic susceptibility (log χ) Average | Magnetic susceptibility Max. Min. |
|---|---|---|---|---|---|---|---|---|---|
| 1 | 1 | 1.57 | 1.57 / 1.57 | 2.42 | 2.42 / 2.42 | 34.9 | 34.9 / 34.9 | 4.49 | 4.49 / 4.49 |
| 2 | 13 | 2.26 ± 0.09 | 3.05 / 1.88 | 2.93 ± 0.05 | 3.37 / 2.74 | 23.1 ± 2.2 | 36.7 / 9.5 | 4.03 ± 0.16 | 5.11 / 3.30 |
| 3 | 10 | 2.90 ± 0.08 | 3.22 / 2.42 | 3.63 ± 0.03 | 3.74 / 3.50 | 21.0 ± 2.7 | 34.2 / 8.3 | 4.39 ± 0.12 | 4.81 / 3.65 |
| 4 | 4 | 3.04 ± 0.19 | 3.23 / 2.85 | 3.58 ± 0.02 | 3.60 / 3.56 | 15.0 ± 5.8 | 20.8 / 9.2 | 4.55 ± 0.12 | 4.67 / 4.43 |

Note: "±" denotes 1 SD among the meteorites represented.

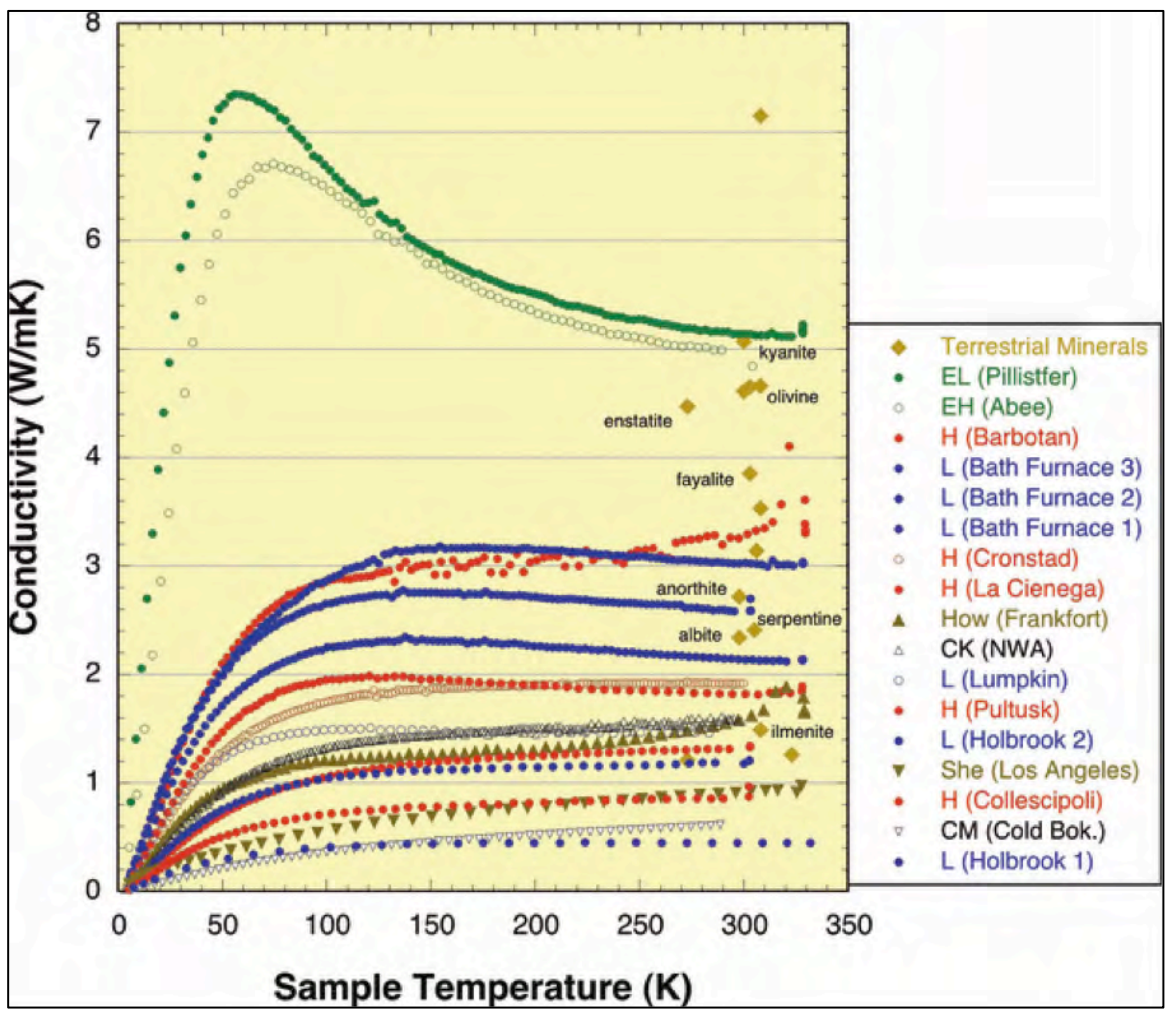

Figure 26 - Taken directly from Figure 3 in Opeil et al. (2012). Original Opeil et al. (2012) caption: "*Thermal conductivity of meteorites as a function of temperature. New results (this paper) are indicated with solid symbols; results previously reported in Opeil et al. (2010) are open symbols. Different classes of meteorites are indicated by the different colors, and the individual samples are identified, to the right, in the order of their conductivity at 300 K, from highest to lowest conductivities.*"

*Strength of Materials*

Limited laboratory data are available on the compressive and tensile strengths of meteoritic materials. Tsuchiyama and colleagues (Tsuchiyama et al. 2009) have obtained tensile strengths of CM, CI and Tagish Lake (TL) chondrites. The tensile strengths range from ~0.3 to 30 MPa with a general trend of lower strength from CM>CI>TL. Higher petrographic grade ordinary chondrites have typical tensile strengths of ~30-40 MPa and compressive strengths ranging from ~60 to 450 MPa (Popova et al. 2011). The "average" tensile strength of Murchison is ~2 MPa, and the averages for CIs and TL are slightly lower. The compressive strength of Murchison is reported to be ~50 MPa (Tsuchiyama et al., 2009).

Material strengths have also been inferred from the break-up of bolides with well-constrained trajectories (Popova et al. 2011). Popova et al. estimate the strength of Tagish Lake between 0.3 and 1.3 MPa based on the height of earliest fragmentation and major fragmentation, respectively. More importantly, they note that the bulk strengths on atmospheric entry are low and describe the bolides as "weak and crumbly" objects compared to the measured tensile strengths of recovered fragments.

*Thermal Properties*

Thermal conductivities of chondritic meteorites range from ~0.5 to 4.5 W/m/K and are nearly constant with temperatures >100K as shown in `Figure 26` (Szurgot 2011; Opeil et al. 2012). Opeil et al. (2012) note that the thermal conductivity of meteorites are, in general, lower by up to a factor of 3-10 than that of a collection of the pure minerals from which they are made. They also note that a linear relationship exists between thermal conductivity and the inverse of porosity and suggest that microcracks and porosity provide a barrier to the transmission of thermal energy in these objects.

### 8.1.1 Tensile Strength

- Defined as the maximum stress that a material can withstand while being stretched or pulled before necking, which is when the specimen's cross-section starts to significantly contract.
- Source: Lindsay Keller and the Carbonaceous Meteorite Working Group

   Tensile Strength = 0.3 to 3.0 MPa

### 8.1.2 Bulk Material Density

- Defined as the bulk density of the surface regolith material.
- Source: Lindsay Keller and the Carbonaceous Meteorite Working Group

   Bulk Material Density = 1.0 to 2.0 g cm$^{-3}$

### 8.1.3 Bulk Material Porosity

- Defined as the porosity (or fraction of void spaces) of the surface regolith material.
- Source: Lindsay Keller and the Carbonaceous Meteorite Working Group

Bulk Material Porosity = 20 to 40%

### 8.1.4 Cohesion

- Defined as the component of shear strength of a rock or soil that is independent of interparticle friction.
- Source: Regolith Development Working Group

Cohesion = 0.1 pa to 1 mpa

### 8.1.5 Friction Angle

- Defined as the angle between normal and resultant force that is attained when failure occurs in response to a shear force. The range is within the values for many terrestrial geologic materials.
- Source: Lindsay Keller and the Carbonaceous Meteorite Working Group

Friction Angle = 15° to 47°

### 8.1.6 Young's Modulus

- Defined as a measure of the stiffness of an elastic material. It is the ratio of the stress over strain.
- Source: Lindsay Keller and the Carbonaceous Meteorite Working Group

Young's Modulus (E) = 100 kpa to 700 mpa

### 8.1.7 Surface Grain Density

- Defined as the density of the surface grain material of Bennu. This parameter is derived from Arecibo radar observations presented in Nolan et al. (2013). Using the Magri et al. (2001) linearized "Eros-calibrated" method, we obtain a grain density $\sigma_{OC,\ BENNU} / \sigma_{OC,\ Eros} = 0.12 / 0.25 = 1/2$ that of Eros. Magri et al. assume Eros to have L-chondrite composition, with a grain density of 3.75 g cm$^{-3}$ and a porosity of about 50%, giving a bulk density of about 1.8 gm cm$^{-3}$ (for Eros), and thus grain / bulk densities of 1.8 / 0.9 g cm$_{-3}$ for Bennu. These densities are not, a priori, unreasonable estimates for Bennu. Chesley et al. (2012), obtain an estimate of the bulk density of the entire asteroid to be ~ 1 from a measurement of the Yarkovsky effect on the orbit of Bennu, suggesting a fairly uniform porosity for the entire asteroid, though the uncertainties are large enough to prevent drawing any firm conclusions.
- Source: Steve Chesley

$$d \text{ (surface grain)} = 1.8 \text{ g cm}^{-3}$$

### 8.1.8 Surface Bulk Density

- Defined as the mass of all the surface material divided by its volume of Bennu. This parameter is derived from Arecibo radar observations presented in Nolan et al. (2013). See the description of 2.2.4 for additional information.
- Source: Steve Chesley

$$d \text{ (surface bulk)} = 0.9 \text{ g cm}^{-3}$$

### 8.1.9 Worst Case Surface Map

- Defined as a synthetic map of the surface of Bennu with properties that represent the worst case for TAGSAM sampleability. When defining a "worst-case" surface for Muses-C, there are several considerations, two of which are end-members: the worst case for large rocks that could endanger the mechanical safety of the arm or the sampler, and the worst-case for small rocks that jeopardize the sampling success. The map covers an area of 18 m x 18 m with a grid spacing of 5 mm. The size of the map is representative of the TAGSAM error ellipse.
- Source: Beau Bierhaus

see Appendix A

## 8.2 Electrical Characteristics

The electrical conductivities (ECs) of carbonaceous chondrites have been measured at low temperatures (<300K). From the data in Brecher et al. (1975) (see Table 6 and Figure 27), the anhydrous, high petrographic grade chondrites typically show the highest electrical conductivities, the CMs are intermediate, and the one CI chondrite measured had an EC similar to terrestrial serpentinites.

**Table 6 – Adapted from Table 1 in Brecher et al. (1975)**

Electrical conductivity in meteorites and terrestrial rocks

| Sample | Type | $\sigma(300°K)$ $(ohm-cm)^{-1}$ | Low-temperature $\sigma_0$ $(ohm-cm)^{-1}$ |
|---|---|---|---|
| *Carbonaceous chondrites* | | | |
| Orgueil | C-1 | $3-8 \times 10^{-11}$ | – |
| Cold Bokkeveld | C-2 | $4-7.5 \times 10^{-11}$ | – |
| Mighei | C-2 | $0.4-4 \times 10^{-9}$ | – |
| | | $3 \times 10^{-9}$ | – |
| | | $2 \times 10^{-8}$ | $2.8 \times 10^{-8}$ |
| Murchison | C-2 | $5 \times 10^{-9}$ | $3.8 \times 10^{-9}$ |
| Lancè | C-30 | $0.2-2 \times 10^{-5}$ | – |
| | | $1.4-3.5 \times 10^{-5}$ | – |
| | | $2 \times 10^{-5}$ | – |
| | | $1.2 \times 10^{-5}$ | $2.2 \times 10^{-5}$ |
| Allende | C-3V | $0.2-5 \times 10^{-6}$ | – |
| | | $4-5.5 \times 10^{-7}$ | – |
| | | $2.7 \times 10^{-7}$ | $4.8 \times 10^{-7}$ |
| | | $2.7 \times 10^{-7}$ | – |
| Ornans | C-4,50 | $1-4 \times 10^{-11}$ | – |
| *Other meteorites* | | | |
| Kelly | LL-4 | $2 \times 10^{-9}$ | – |
| Arriba | L-5 | $1-8 \times 10^{-6}$ | – |
| Leedey | L-6 | $4 \times 10^{-7}$ | – |
| Plainview | H-5 | $1.6-11 \times 10^{-4}$ | – |
| Richardson | H-5 | $1.8-6 \times 10^{-7}$ | – |
| Rose City | H-6 | $7-15 \times 10^{-2}$ | – |
| Pinto Mts. | (L) | $7 \times 10^{-9}$ | – |
| Elenovka | (L) | $8 \times 10^{-6}$ | – |
| Norton Cty. | Ach. | $1.2 \times 10^{-8}$ | – |
| Goalpara | Ach. | $1-4 \times 10^{-3}$ | – |
| | | $8.5 \times 10^{-4}$ | – |
| *Terrestrial rocks* | | | |
| Serpentine | – | $1-10 \times 10^{-10}$ | – |
| | | $4 \times 10^{-9}$ | $3 \times 10^{-9}$ |
| Serpentine marble | – | $1-20 \times 10^{-11}$ | – |
| | | $5 \times 10^{-10}$ | $1.2 \times 10^{-10}$ |
| Olivine | – | $1 \times 10^{-12}$ | – |

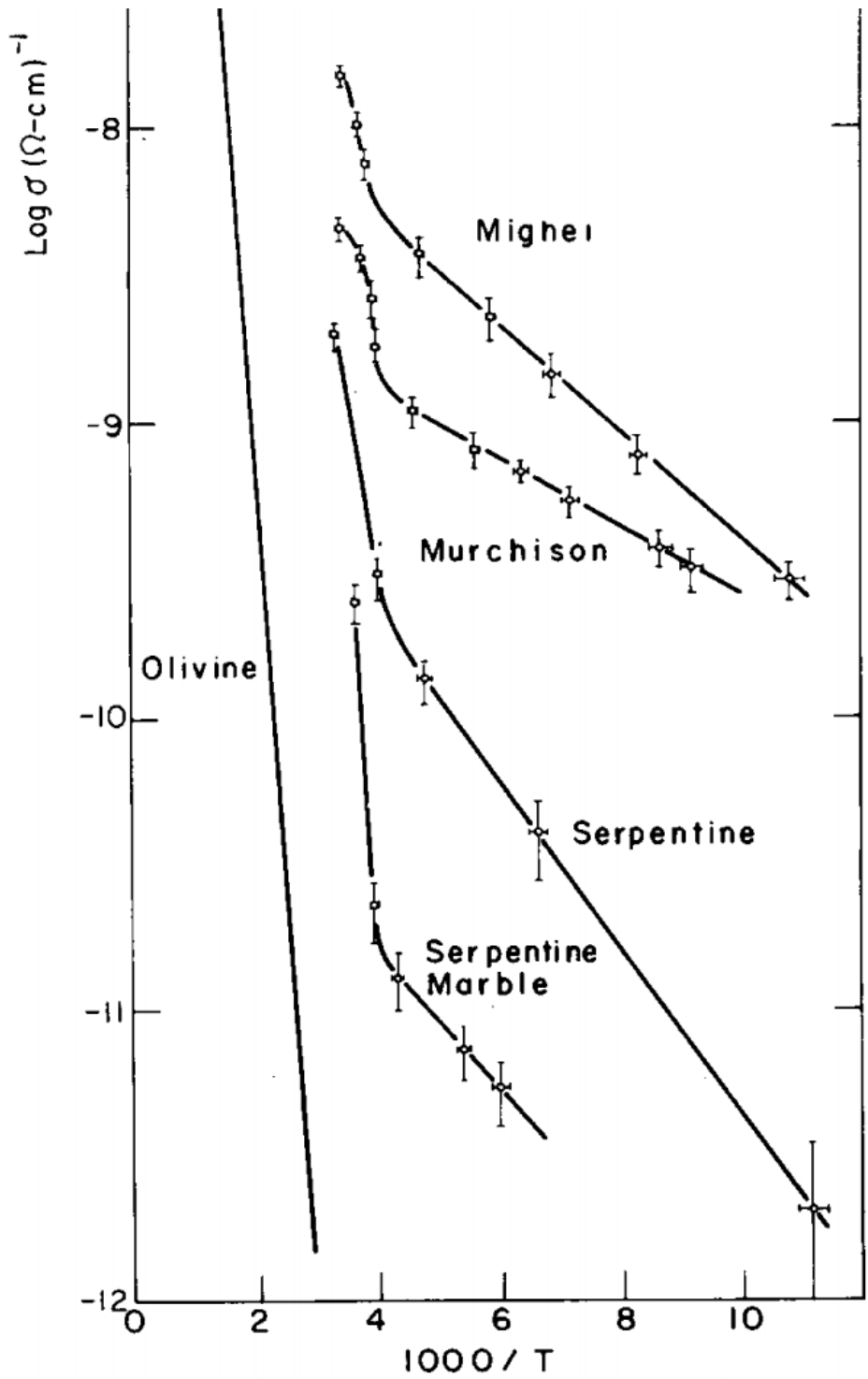

Figure 27 - Figure adapted from Figure 2 of Brecher et al. (1975, Earth and Planetary Science Letters 28, 37-45) showing the electrical conductivities of two CM chondrites and two terrestrial serpentinites as a function of temperature.

# 9 Environmental Properties

*Environmental properties are specifically parameters that model the near-asteroid environment. This includes natural satellites, dust and gas released from the surface of Bennu and electrical properties.*

## 9.1 Satellites

*Some of the parameters presented in this section are based on a preliminary density and mass. Though Chesley et al. (2014) has produced a better density and mass estimate, the preliminary one is used in the determination of 9.1.1 Radius of Hill Sphere, 9.1.2 Radius of Stability Sphere for 1-m Satellites, 9.1.3 Radius of Stability Sphere for 10-cm Satellites and 9.1.4 Radius of Stability Sphere for 1-cm Satellites. Updated maps are being produced.*

The Hill Sphere due to gravity and tides alone is computed from the approximate formula

$$R_H = (\frac{\mu}{3\mu_S})^{1/3} q$$

where $\mu_S = 1.327 \times 10^{11}$ km$^3$/s$^2$ is the solar gravitational parameter and $q = 1.345 \times 10^8$ km is the asteroid orbit perihelion. For the nominal asteroid density and uncertainty this equals

$$R_H = 29.5 \pm 1.5 \text{ km}$$

The true region where an object may be trapped in a long-term stable orbit will actually depend on the object's orbit plane orientation. For a circular orbit in a direct orbit relative to the asteroid's heliocentric orbit plane the limiting semi-major axis can equal one-half of the Hill radius and will be formally unstable for inclinations between ~ 45° → 135° due to Kozai resonances with the sun (although these orbits can be stable over relatively long periods of time still). Retrograde orbits can actually be much farther than the Hill radius and still be stable and, technically, in orbit about the asteroid (albeit traveling quite slow relative to the asteroid). Given this range of uncertainty, thus, adoption of the Hill radius is usually conservative. Once solar radiation pressure effects are taken into account, the range of semi-major axis for stability generally shrinks. An approximate formula for the maximum semi-major axis for an object to be trapped in orbit is

$$a_M = \frac{\sqrt{3}}{4}\sqrt{\frac{\mu B}{\Phi_S}d}$$

where $\mu$ is the asteroid gravitational parameter, B is the body's mass to area ratio in kg/m$^2$, $\Phi_S$ ~ $1 \times 10^8$ km$^3$ kg/s$^2$/m$^2$ is the solar radiation constant and d is the asteroid-sun distance. We note that the smallest value of $a_M$ occurs at perihelion, and we will use this as our limiting value. This limit does not take into account the same effects as the Hill Sphere, and thus can predict stable semi-major axes that may not be realistic. An example of this is below, with the 1 meter sized objects. It is possible to account for both effects simultaneously, but the computations become more complex and probably aren't needed at this juncture. If we model a body of density ρ in units of kg/m$^3$ as a sphere with radius R in meters, its mass to area ratio is

$$B = \frac{4}{3}\rho R$$

In the following we assume a bulk density of $2 \times 10^3$ kg/m$^3$ for our bodies. Then we can create a table (see Table 7) with the limiting values of semi-major axis for different sized bodies. This provides another measure of the possible distance bodies could be from the asteroid. These are semi-major axes, thus the orbit can also have an associated eccentricity. For the smaller particles this eccentricity is quite limited, however for the larger range of the 1-meter object it can become relatively large before it impacts with the asteroid. We note that twice this semi-major axis is beyond the Hill Radius computed above, indicating that gravity and SRP effects are competing at a similar level for this size of body. A more detailed study of this case is warranted, but is put off for the future.

*Most Satellite Properties have not been formally published but have been internally reviewed by the OSIRIS-REx Science Team. Lower size limits of any existing satellites have been published in*

"Nolan, M.C., Magri, M., Howell, E.S., Benner, L.A.M., Giorgini, J.D., Hergenrother, C.W., Hudson, R.S., Lauretta, D.S., Margot, J-.L., 2013. Shape model and surface properties of the OSIRIS-REx target asteroid (101955) Bennu from radar and lightcurve observations. Icarus 226, 629-640.".

**Table 7 - Limiting semi-major axes for bodies of different radius, computes assuming a bulk density of 2 gm/cm$^{-3}$.**

| $r$ | $B$ | $a_m$ |
| --- | --- | --- |
| m | kg m$^{-2}$ | km |
| 0.01 | 27 | 2 |
| 0.10 | 270 | 6.3 |
| 1.00 | 2700 | 20 |

### 9.1.1 Radius of Hill Sphere

- Defined as the radius of the gravitational sphere of influence of Bennu. Within this sphere Bennu is the dominant gravitational source.
- Source: Dan Scheeres

$$R_H = 29.5 \pm 1.5 \text{ km}$$

### 9.1.2 Radius of Stability Sphere for 1-m Satellites

- Defined as the volume of space about Bennu where 1-m satellites are found on stable orbits.
- Source: Dan Scheeres

$$R_{1\text{-}m} = 20 \text{ km}$$

### 9.1.3 Radius of Stability Sphere for 10-cm Satellites

- Defined as the volume of space about Bennu where 10-cm satellites are found on stable orbits.
- Source: Dan Scheeres

$$R_{10\text{-}cm} = 12 \text{ km}$$

### 9.1.4 Radius of Stability Sphere for 1-cm Satellites

- Defined as the volume of space about Bennu where 1-cm satellites are found on stable orbits.
- Source: Dan Scheeres

$$R_{1\text{-}cm} = 4 \text{ km}$$

### 9.1.5 Lower Size Limit of Ground-based Radar Detection (slow rotating satellite)

- Defined as the smallest size satellite of Bennu detectable by the Arecibo radar observations taken in 1999 and 2005. Any satellites larger than the lower size limit would have been detected within 300 km of Bennu. The detection size limit is dependent on the rotation rate of the satellite and scales with the rotation period to the -1/3 power. The 2-sigma detection limit for a satellite with the radar albedo of Bennu and a slow tidally-locked rotation period (order of many hours) in the highest SNR radar data is 2 m.
- Source: Mike Nolan – Nolan et al. (2013)

$$D_{min}(24 \text{ hour rotation period}) = 2 \text{ m}$$

### 9.1.6 Lower Size Limit of Ground-based Radar Detection (rapidly rotating satellite)

- Defined as the smallest size satellite of Bennu detectable by the Arecibo radar observations taken in 1999 and 2005. Any satellites larger than the lower size limit would have been detected within 300 km of Bennu. The 2-sigma detection limit for a satellite with the radar albedo of Bennu and a very rapid rotation period (0.01 hours) in the highest SNR radar data is 20 m. Such a fast rotation period has been observed for asteroids with similar diameters (Hergenrother and Whiteley 2011).
- Source: Mike Nolan – Nolan et al. (2013)

$$D_{min}(1 \text{ min rapid rotator}) = 15 \text{ m}$$

## 9.2 Near-Asteroid Environment

Upper limits for *dust concentration, dust mass* and *dust production rates* are based on analysis of Spitzer IR data taken in mid-2007. Note, that no observations, including Spitzer, have detected dust in the vicinity of Bennu. As a result, these measurements are upper limits. Rates and concentration of dust can be greater than these upper limits if Bennu experiences a recent bout of activity. These upper limits define the limit of detection of our best observations.

*Summary*

The coma flux limit is estimated with two different methods: 1) measure the surface brightness at 7 to 10 pixels from the asteroid and assume a 1/distance surface brightness coma profile, 2) sum the point spread function (PSF) subtracted image inside of 7 pixels (within 681 km of Bennu). The 3-sigma flux upper-limits for the coma in a 7-pixel radius aperture are 0.58 mJy at ~16 microns (actual bandwidth is 13.3 to 18.7 microns and henceforth denoted as Blue) and 0.72 mJy at ~22 microns (actual bandwidth is 18.5 to 26.0 microns and henceforth denoted as Red).

These limits are then used to compute mass upper-limits for various grain sizes. See the table below. Now that this work has been done, it is easy to compute these numbers for other size distributions, grain sizes or compositions, if the OSIRIS-REx team needs them. This report includes results for solid amorphous carbon grains, and solid amorphous olivine grains with a 50/50 Fe/Mg ratio.

*Methods and Analysis*

Each of the peakup images were scaled onto a half-pixel grid, centered on the object, then each pixel's distance from the center was computed to get the radial profile of the image. Peakup images refer to images made by the Spitzer infrared spectrograph at 16 and 22 microns. Originally intended to center objects on the slit, these small-scale images have scientific utility. Next, all 11 profiles are combined together to get a super profile. Then, a PSF computed with STinyTim (a standard model for the PSF of Spitzer) is treated in the same manner and this is fit to the super profile. A scale factor and a background term are fit to the profile at 1 to 5 pixels for Blue, and 1 to 7 pixels for Red. These fits are plotted in `Figure 28`. The inset contains the red and blue residuals. The residuals are pretty clean except for the core (where there are fewer pixels to average over). Generating a set of STinyTim PSFs with different subpixel offsets, and generating a PSF super profile could improve this, but it shouldn't change the results.

Residuals in Blue at 5 to 10 pixels   =   $0.024 \pm 0.009$ MJy/sr/pixel

Residuals in Red at 7 to 10 pixels   =   $0.006 \pm 0.013$ MJy/sr/pixel

The Blue residuals are 2.7-sigma above the background, but the color is completely wrong for dust. A 10,000 K blackbody has a Blue/Red ratio of 1.9. If the background term in the PSF fitting took the signature of dust out of the inner of the radial profile, one would expect the outer radial profile would be less than zero, because the surface brightness of a coma should decrease with distance, but the background term is the same for all radii.

The background fit is significantly correlated with the brightness of the point source. Therefore, we cannot include this error source in the surface brightness errors.

Fit a PSF to the Red and Blue radial profiles out to 7 pixels, then check the residuals at 7 to 10 pixels, using 1.1 pixel smoothed PSFs.

Mean surface brightness:

Blue at 7 to 10 pixels (MJy/sr)   =   $-0.017 \pm 0.010$

Red at 7 to 10 pixels (MJy/sr)   =   $-0.019 \pm 0.013$

Total flux:

Blue inside 7 pixels (mJy)   =   $0.16 \pm 0.14$

Red inside 7 pixels (mJy)   =   $0.02 \pm 0.15$

Now, assume a 3-sigma coma, what is the flux upper-limit?

Image scale is 1.835 "/pixel

|  | Blue | Red |  |
|---|---|---|---|
| 3-sig (mJy) | 0.48 | 0.72 | from Surface Brightness and $\rho^{-1}$ profile |
|  | 0.58 | 0.47 | from flux |

For mass-loss rates, assume a nominal comet coma profile and use a velocity of 10 m/s. Mass-loss, Q, scales directly with v: $Q = 2 M v / (\pi R)$, M is mass (g), v is velocity (0.01 km/s), and R is aperture radius (km).

Assuming these are solid amorphous carbon grains, what are the mass upper-limits? Use 0.58 mJy for Blue and 0.72 mJy for Red. The composition determines the radiative equilibrium temperature and thermal emission efficiencies (see Harker et al. 2002, ApJ, 580, 579-597 for details). Note that these are solid grains: 3.3 g/cm$^3$ for silicates, 2.5 g/cm$^3$ for amorphous carbon.

The following tables have several rows of monodisperse distributions, then the bottom row in each table shows upper limits for a wide range of grains that follows a size-frequency distribution of dn/da=a$^{-3.5}$, which has been found to generally fit comet comae very well.

For solid amorphous carbon grains:

```
                                  Blue
            ----------------------------------
    size        F(a)     N(a)      mass       Q(a)
    (um)       (mJy)               (g)        (g/s)
    ---------  --------  --------  --------   --------
    0.01       5.54e-23  1.05e+22  1.10e+05   1.46e-01
    0.10       5.90e-20  9.84e+18  1.03e+05   1.38e-01
    1.00       2.57e-17  2.25e+16  2.36e+05   3.15e-01
    10.00      3.83e-15  1.52e+14  1.59e+06   2.12e+00
    100.00     3.48e-13  1.67e+12  1.75e+07   2.33e+01
    1000.00    3.37e-11  1.72e+10  1.80e+08   2.41e+02
    0.1-1000*  3.28e-20  1.77e+19  3.67e+06   4.90e+00

                                  Red
            ----------------------------------
    size        F(a)     N(a)      mass       Q(a)
    (um)       (mJy)               (g)        (g/s)
    ---------  --------  --------  --------   --------
    0.01       2.34e-23  3.07e+22  3.22e+05   4.30e-01
    0.10       2.46e-20  2.92e+19  3.06e+05   4.09e-01
    1.00       1.17e-17  6.15e+16  6.44e+05   8.60e-01
    10.00      4.28e-15  1.68e+14  1.76e+06   2.35e+00
    100.00     3.42e-13  2.10e+12  2.20e+07   2.94e+01
    1000.00    3.16e-11  2.28e+10  2.38e+08   3.18e+02
    0.1-1000*  2.74e-20  2.63e+19  5.45e+06   7.28e+00
```

* dn/da = a$^{-3.5}$

where F(a) is Flux, N(a) is Number of particles and Q(a) is Production rate.

For solid amorphous olivine grains with a 50/50 Fe/Mg ratio:

```
                            Blue
        ----------------------------------------
 size       F(a)      N(a)      mass      Q(a)
 (um)      (mJy)                (g)       (g/s)
---------  --------  --------  --------  --------
 0.01      7.50e-23  7.74e+21  1.07e+05  1.43e-01
 0.10      1.19e-19  4.87e+18  6.74e+04  8.99e-02
 1.00      5.08e-17  1.14e+16  1.58e+05  2.11e-01
 10.00     4.31e-15  1.35e+14  1.86e+06  2.48e+00
 100.00    4.22e-13  1.37e+12  1.90e+07  2.54e+01
 1000.00   4.21e-11  1.38e+10  1.90e+08  2.54e+02
 0.1-1000* 3.93e-20  1.47e+19  4.03e+06  5.39e+00

                            Red
        ----------------------------------------
 size       F(a)      N(a)      mass      Q(a)
 (um)      (mJy)                (g)       (g/s)
---------  --------  --------  --------  --------
 0.01      4.30e-23  1.68e+22  2.32e+05  3.09e-01
 0.10      6.23e-20  1.16e+19  1.60e+05  2.13e-01
 1.00      3.43e-17  2.10e+16  2.91e+05  3.88e-01
 10.00     5.11e-15  1.41e+14  1.95e+06  2.60e+00
 100.00    3.94e-13  1.83e+12  2.52e+07  3.37e+01
 1000.00   3.62e-11  1.99e+10  2.75e+08  3.67e+02
 0.1-1000* 3.95e-20  1.82e+19  4.98e+06  6.66e+00
```

* dn/da = $a^{-3.5}$, all grains have the same velocity

where F(a) is Flux, N(a) is Number of particles and Q(a) is Production rate.

Now, instead we use an ejection velocity of 0 m/s, and let solar radiation pressure remove grains from the surface. This model may be more realistic for an asteroid than the above comet-based model (i.e., little or no outgassing to lift grains). I compute mass-loss rates using my (fully 3D) dust dynamical model, the orbit of the asteroid, Spitzer's viewing geometry, and again two compositions (amorphous carbon and 50/50 Fe/Mg amorphous olivine). The compositions affect the radiation pressure efficiencies, which are computed with Mie-theory. The model is described in Kelley 2006 (PhD thesis, Univ. of Minnesota), and Kelley et al. 2008 (Icarus, 193, 572-587). Note that the model does not account for the mass of the asteroid, i.e., it doesn't consider how the grains escape from the asteroid, just that they leave the potential well with v ~ 0 m/s. In short, mass-loss rate upper limits are of order magnitude 1 g/s, with larger values for large grains (>=100 micron)

For amorphous carbon:

```
                      Blue
          -------------------------
size      mass      mean v    Q(a)
(um)      (g)       m/s       (g/s)
--------- --------  --------  --------
0.01      1.10e+05  2.02e+02  2.95e+00
0.10      1.03e+05  2.60e+02  3.58e+00
1.00      2.36e+05  7.94e+01  2.50e+00
10.00     1.59e+06  2.17e+01  4.61e+00
100.00    1.75e+07  6.65e+00  1.55e+01
1000.00   1.80e+08  2.04e+00  4.90e+01

                      Red
          -------------------------
size      mass      mean v    Q(a)
(um)      (g)       m/s       (g/s)
--------- --------  --------  --------
0.01      3.22e+05  2.02e+02  8.67e+00
0.10      3.06e+05  2.60e+02  1.06e+01
1.00      6.44e+05  7.94e+01  6.83e+00
10.00     1.76e+06  2.17e+01  5.11e+00
100.00    2.20e+07  6.65e+00  1.96e+01
1000.00   2.38e+08  2.04e+00  6.48e+01
```

where V is velocity and Q(a) is Production rate.

For 50/50 Fe/Mg amorphous olivine:

```
                         Blue
              --------------------------
   size       mass     mean v   Q(a)
   (um)       (g)      m/s      (g/s)
   ---------  -------- -------- --------
   0.01       1.07e+05 7.16e+01 1.02e+00
   0.10       6.74e+04 1.49e+02 1.34e+00
   1.00       1.58e+05 6.73e+01 1.42e+00
   10.00      1.86e+06 1.88e+01 4.67e+00
   100.00     1.90e+07 5.76e+00 1.46e+01
   1000.00    1.90e+08 1.75e+00 4.45e+01

                         Red
              --------------------------
   size       mass     mean v   Q(a)
   (um)       (g)      m/s      (g/s)
   ---------  -------- -------- --------
   0.01       2.32e+05 7.16e+01 2.21e+00
   0.10       1.60e+05 1.49e+02 3.18e+00
   1.00       2.91e+05 6.73e+01 2.61e+00
   10.00      1.95e+06 1.88e+01 4.89e+00
   100.00     2.52e+07 5.76e+00 1.94e+01
   1000.00    2.75e+08 1.75e+00 6.43e+01
```

where v is velocity and Q(a) is Production rate.

The mass-loss rates for the smallest grains are higher than the previous tables based on 10 m/s ejection velocity. This is because these particles are easily accelerated by radiation pressure to speeds higher than 10 m/s. The mass-loss rates of the large particles are reduced, because radiation pressure weakly pushes them. I suggest that the second tables are a more realistic estimate of the mass-loss rate upper limit.

Near-Asteroid Environment Properties have been published in

"Emery, J.P., Fernandez, Y.R., Kelley, M.S.P., Warden, K.T., Hergenrother, C., Lauretta, D.S., Drake, M.J., Campins, H., Ziffer, J. 2014. Thermal infrared observations and thermophysical characterization of OSIRIS-REx target asteroid (101955) Bennu. Icarus 234, 17-35.".

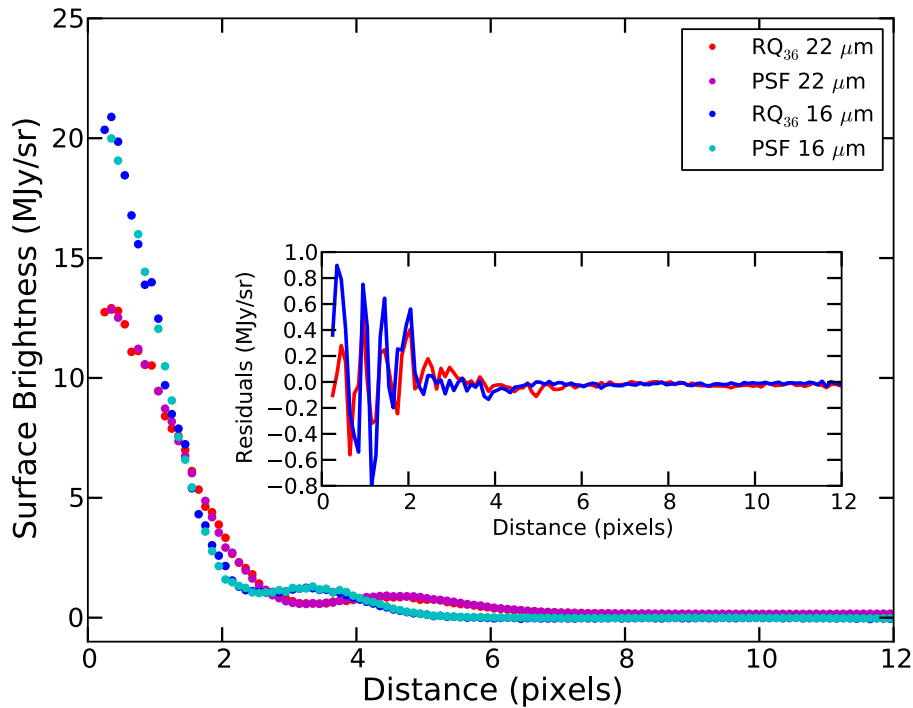

**Figure 28 - Comparison of Bennu PSF profiles at 16 and 22 microns compared with modeled PSFs.**

### 9.2.1 Dust Concentration Upper Limit

- Defined as the upper limit for the concentration of dust within 700 km of Bennu based on the lack of any detectable dust in the May 2007 Spitzer 16 and 22-µm images.
- Source: Mike Kelley (U. of Maryland, non-OSIRIS-REx member) and Josh Emery

$$C_{dust} = 1.5e+19 \text{ particles}$$

### 9.2.2 Dust Mass Upper Limit

- Defined as the upper limit for the mass of dust within 4750 km of Bennu based on the lack of any detectable dust in the May 2007 Spitzer 16 and 22-μm images.
- Source: Mike Kelley (U. of Maryland, non-OSIRIS-REx member) and Josh Emery – Emery et al. (2013)

$$M_{dust} = 10^6 \text{ g}$$

### 9.2.3 Dust Production Upper Limit

- Defined as the upper limit for the production rate of dust from the surface of Bennu based on the lack of any detectable dust in the May 2007 Spitzer 16 and 22-μm images.
- Source: Mike Kelley (U. of Maryland, non-OSIRIS-REx member) and Josh Emery

$$Q_{dust} = 1 \text{ g s}^{-1}$$

Krugly, Yu.N., Belskaya, I.N., Shevchenko, V.G., Chiorny, V.G., Velichko, F.P., Mottola, S., Erikson, A., Hahn, G., Nathues, A., Neukum, G., Gaftonyuk, N.M., Dotto, E., 2002. The near-Earth objects follow-up program IV. CCD photometry in 1996-1999. Icarus 158, 294-304.

Lauretta, D.S., Bartels, A.E., Barucci, M.A., Bierhaus, E.B., Binzel, R.P., Bottke, W.F., Campins, H., Chesley, S.R., Clark, B.C., Clark, B.E., Cloutis, E.A., Connolly, H.C., Crombie, M.K., Delbó, M., Dworkin, J.P., Emery, J.P., Glavin, D.P., Hamilton, V.E., Hergenrother, C.W., Johnson, C.L., Keller, L.P., Michel, P., Nolan, M.C., Sandford, S.A., Scheeres, D.J., Simon, A.A., Sutter, B.M., Vokrouhlický, Walsh, K.J. 2014. The OSIRIS-REx target asteroid 101955 Bennu: constraints on its physical, geological, and dynamical nature from astronomical observations. Meteorit. Planet. Sci., in press.

Lester, T.P., McCall, M.L., Tatum, J.B. 1979. Theory of planetary photometry. J. Royal Astro. Soc. Canada 73, 233-257.

Li et al. 2004. Photometric analysis of Eros from NEAR data. Icarus 172, 415-431.

Li et al. 2009. Photometric analysis of the nucleus of Comet 81P/Wild2 from Stardust images. Icarus 204, 209-226.

Macke, R.J, Consolmagno, G.J., Britt, D.T. 2011. Density, porosity, and magnetic susceptibility of carbonaceous chondrites. Meteorit. Planet. Sci. 46, 1842-1862.

Magri, C., Consolmagno, G.J., Ostro, S.J., Benner, L.A.M., Beeney, B.R., 2001. Radar constraints on asteroid regolith compositions using 433 Eros as ground truth. Meteorit. Planet. Sci. 36, 1697–1709.

Magri, C., Ostro, S.J., Scheeres, D.J., Nolan, M.C., Giorgini, J.D., Benner, L.A.M., Margot, J.-L., 2007. Radar observations and a physical model of asteroid 1580 Betulia. Icarus 186, 152-1777.

Miller, J.K., Konopliv, A.S., Antreasian, P.G., Bordi, J.J., Chesley, S., Helfrich, C.E., Owen, W.M., Wang, T.C., Williams, B.G., Yeomans, D.K., and Scheeres, D.J., 2002. Determination of shape, gravity and rotational state of Asteroid 433 Eros, Icarus 155, 3-17.

Milliken, R.E., Mustard, J.F., 2006. Estimating absolute $H_2O$ content of low-albedo materials using reflectance spectroscopy. Lunar Planet. Sci., 37, p. 1954 (abstract).

Moroz, L., Pieters, C.M., 1991. Reflectance spectra of some fractions of Migei and Murchison CM chondrites in the range of 0.3–2.6 μm. Lunar Planet. Sci., 22, pp. 923–924 (abstract).

Morris, R.V., Lawson, C.A., Gibson, E.K., Lauer, H.V., Nace, G.A., Stewart, C. 1985. Spectral and other physicochemical properties of submicron powders of hematite (alpha-Fe2O3), maghemite (gamma-Fe2O3), magnetite (Fe3O4), goethite (alpha-FeOOH), and lepidocrocite (gamma-FeOOH). J. Geophys. Res. 90, 3126-3144.

Moyer, T. D. (1971). Mathematical formulation of the Double-Precision Orbit Determination Program (DPODP). Technical Report JPL-TR-32-1527.

Muinonen, K., Belskaya, I.N., Cellino, A., Delbó, M., Levasseur-Regourd, A-.C., Penttilä, A., Tedesco, E.F., 2010. A three-parameter magnitude phase function for asteroids. Icarus 209, 542-555.

Muller, T.G., O'Rourke, L., Barucci, A.M., Pal, A., Kiss, C., Zeidler, P., Altieri, B., Gonzalez-Garcia, B.M., Kuppers, M., 2012. Physical properties of OSIRIS-REx target asteroid (101955) 1999 RQ36. Astronomy and Astrophysics 548, A36.
93

**III. Appendices**

**Appendix A: An Analysis of the Muses-C Rock Distribution and Surface Roughness, and Application to Conditions at Asteroid Bennu**

Special Report for OSIRIS-REx:

# An Analysis of the Muses-C Rock Distribution and Surface Roughness, and Application to Conditions at Asteroid Bennu

Authors:
Beau Bierhaus, Lockheed Martin
Olivier Barnouin, Applied Physics Laboratory
Bob Gaskell, Planetary Science Institute

Document: Version 5

## 1. Introduction
This memorandum describes the development plan for addressing the following statement of task:

"A digital topography model, consistent with a format that LM can ingest into their analysis software, that is representative of our best estimate of the boulder/rock distribution of the smoothest portion of the Itokawa MUSES-C region down to 2cm in diameter. If reasonable estimates cannot be made to the 2cm size, a larger size will be provided with a rationale for using the larger size as a limit."

*Data Types for a Muses-C Model*
There are two primary data types contributing to our understanding of the rock distribution in the Muses-C region. The bulk of this document addresses these data types:

- **Section 2:** measurements of the rock size-frequency distribution (SFD)
- **Section 3:** topographic information from LIDAR and stereo-photoclinometry
- **Section 4:** Muses-C in the context of other solar system bodies, and proposed synthetic surface that is representative of Muses-C.

The following sections describe our understanding of Muses-C through the perspective of each data set.

## 2. Measured rock size-frequency distribution (SFD)

The measurements (done by Bierhaus) were made using the APL Small Body MappingTool (SBMT). The three images, listed in Table 1, are the three highest-resolution images acquired by Itokawa on its descent to Muses-C. The outline of the three images and their relative location appear in Figure 1; a mosaic of the three images and the measured rocks appear in Figure 2. The area measured is 63.4 m$^2$. The measurements in the images and described here were made to ensure a completely sampled population of ≥20 cm size rocks – a threshold relevant to TAGSAM and spacecraft safety – and are estimated to be complete down to 10 cm. While in principle these images are of sufficient quality to enable measurements at a 5 pixel completeness limit criteria (which, if applied to the lowest resolution image results in a dimensioned completeness limit of 3.75 cm), the labor required to achieve measurements at a 5 pixel completeness limit is beyond the scope of this exercise. Thus we divide our description of the particle SFD into two regimes: (i) the measured SFD, complete (within error bars) at sizes ≥ 10 cm, and (ii) an extrapolated SFD for sizes between 10 cm and 2 cm.

Table 1: Hayabusa images of Itokawa's Muses-C region used to determine the particle SFD of the region.

| Image Name | Pixel scale [cm/pix] |
|---|---|
| st_2563511720_v.fit | 0.75 |
| st_2563537820_v.fit | 0.63 |
| st_2563607030_v.fit | 0.58 |

The measurements were made using the SBMT "ellipse tool", for which a user identifies three points: the two end points of the long axis of a feature, and then one side of the small axis. From those three points, the SBMT fits an ellipse. The sizes reported here are for the long axis only.

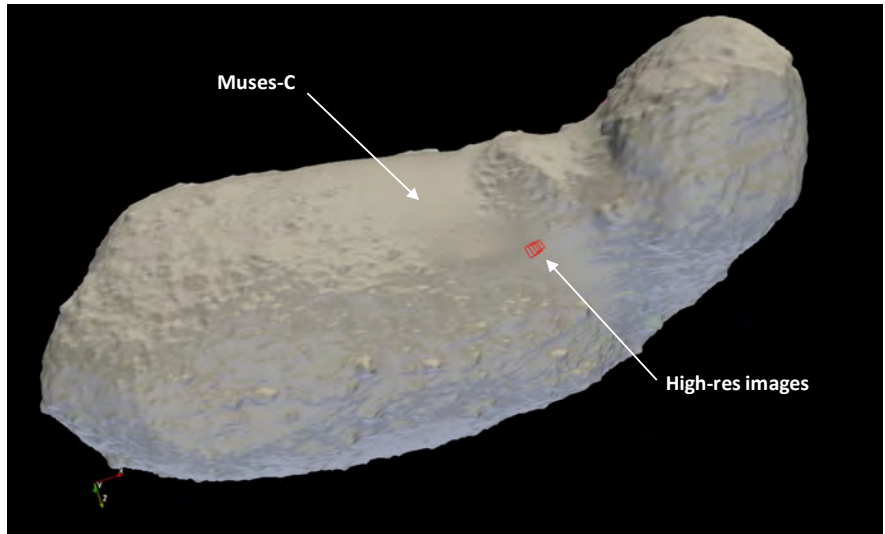

**Figure 1: A shape model (made by Bob Gaskell) of Itokawa, oriented to show Muses-C and the footprint of the three high-resolutions images in Table 1.**

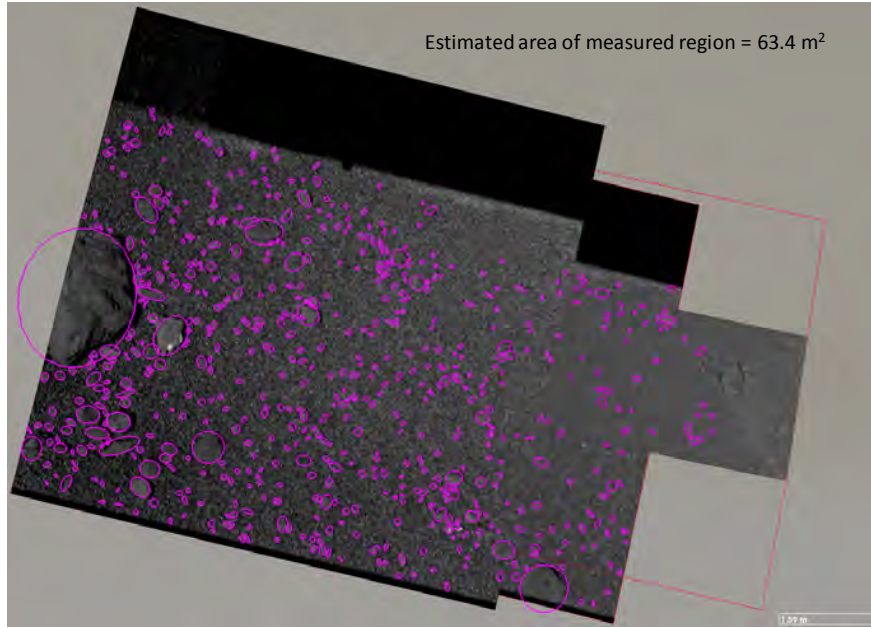

**Figure 2: The three high-resolution images outlined in red in Figure 1, and listed in Table 1, seen as displayed in the APL SBMT. The purple ellipses are the rock measurements.**

A differential SFD is described by:

$$dN = kD^{-b}dD$$

where *dN* is the number of boulders per unit area in the diameter range between *D* and *D + dD*, and *k* and *b* are constants of the power-law. The cumulative distribution is the integral of the differential distribution, or:

$$N_c = \int dN = \int kD^{-b}dD = \frac{k}{b+1}D^{-b+1} = cD^{-a}$$

where *a = b-1* and *c = k/(b+1)*. When displayed in log-log plots, such as in Figure 3, the power-law exponents are equivalent to a linear slope, and thus *b* and *c* are often referred to as the SFD "slopes", and we use that nomenclature here. In the clean world of mathematics, the differential slope is always one unit "steeper" than the cumulative slope; the dirty reality of data and using a weighted fit often means that the best-fit values for the differential and cumulative slopes does not exactly follow this relationship. The parameters *k*, *b*, *a*, and *c* are fit to the data using a non-linear least-squares method, minimizing the sum of squared residuals. Poisson counting statistics define the measurement error, thus the error bars are calculated using $\sqrt{N}$. Table 2 lists the best fit parameters for both the differential and cumulative SFDs, while Figure 3 is the

---

Special Report for OSIRIS-REx　　　　　　　　　　　　　　　　　　　　　　　　Page 3

differential and cumulative plots, respectively, of the measured particle SFD and the best fit power-law to the data.

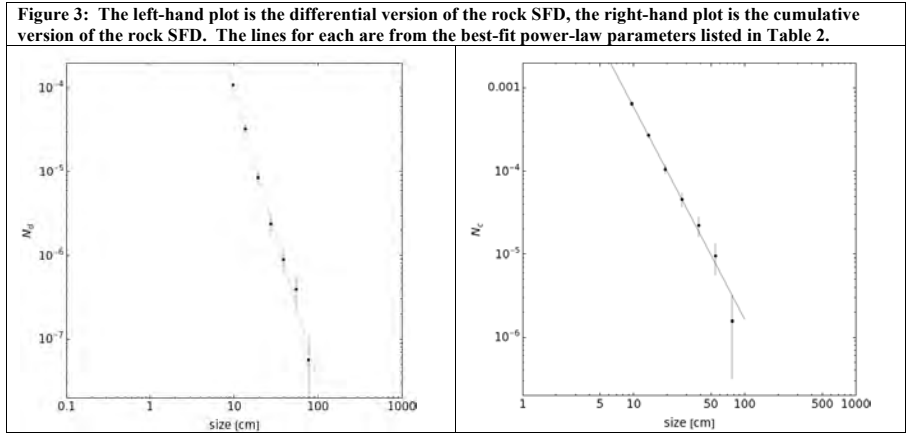

Figure 3: The left-hand plot is the differential version of the rock SFD, the right-hand plot is the cumulative version of the rock SFD. The lines for each are from the best-fit power-law parameters listed in Table 2.

| Table 2: Best-fit values to the differential and cumulative size-frequency distributions. Note these are for length scales in cm. | |
|---|---|
| Differential Fit (k, b) | Cumulative Fit (c, a) |
| k = 0.27 ± 0.02 | c = 0.21 ± 0.01 |
| b = 3.46 ± 0.03 | a = 2.55 ± 0.02 |

*2.1 Discussion of the Slope Values and Corresponding Rock Populations*

The cumulative slope $a = 2.55 \pm 0.02$ presented here is remarkably consistent with a cumulative slope value of $a = 2.56$ that Bashar Rizk derived during OSIRIS-REx Phase A. Remarkable because: (i) it is rare for two independent measurements of an SFD to match so closely, and (ii) he examined a very small subset of a single image, focusing on particle sizes that are < 20 pixels across. The smallest diameters shown here overlap with the largest diameters from his work, lending confidence that there is a consistent SFD for sizes between 100 cm and several cm.

*Rocks > 20 cm*. Rocks that are > 20 cm in size have special consideration for TAGSAM safety, see the special report prepared by Ben Clark for the Sample Site Selection Working Group. The special report includes two rock measurements that provide roughly consistent estimates for > 20 cm rock densities of ~0.44 rocks/m$^2$. Using the best fit cumulative power-law above (i.e. $N_c = 0.21\ D^{-2.55}$ for $D = 20$ cm) gives an estimate of 1 rock/m$^2$. One purely historical grounds, the comparison to within a factor of 2 for these kinds of measurements is acceptable. A discussion between Bierhaus and Clark concluded that the principle reason for the discrepancy is the analytical method: Bierhaus examined the source data at full resolution, while Clark used a figure from a science paper. The more conservative approach is to use the higher-density estimate.

*Rocks < 10 cm*. Rocks < 10 cm were not completely sampled by the measurements, and thus we estimate their populations by application of the best-fit power law. Using the differential format of the SFD, we estimate the number of rocks between 1 and 2 cm, material that is within the size range of ingestible material for TAGSAM. Using the following expression for the differential SFD

$$\Delta N = kD^{-b}\Delta D$$

with $k = 0.27$, $b = 3.46$, $D = 1.3$ cm (the weighted average size between 1 and 2 cm, given an approximately -3 power-law slope), and $\Delta D = 1$ cm, $\Delta N = 1089$ rocks/m$^2$ for sizes between 1 and 2 cm.

Is this a reasonable value?

This result is sensitive to the value selected for D (the mid-point of the diameter range of interest); larger values of $D$ lead to lower values for $\Delta N$. We provide a lower bound for the rocks of these sizes by assuming D = 2 cm, which leads to $\Delta N = 245$ rocks/m$^2$.

And, in fact, the nominal value is likely somewhat high. To explain why, we use a third means of plotting the SFD: the relative plot, or R-plot. An R-plot is the differential data normalized by a -3 differential slope; thus a -3 differential slope appears as a horizontal line. The value of an R-plot is that the vertical height of data in an R-plot is easily understood in terms of spatial density – a higher R-value corresponds to a higher spatial density, and vice versa. Decades of analysis has shown that there is an upper-limit to the maximum spatial density of impact craters on planetary surfaces; at some point the surface is "saturated" with craters, such that (on average) the formation of a new crater necessarily obliterates a pre-existing crater of similar size. The saturation value in an R-format plot, for many planetary surfaces, is approximately R = 0.2; the value can vary slightly for various circumstances, but is generally the same. Boulders suffer the same areal constraint that craters do; only so many boulders of a given size can be exposed on a surface before they start to cover one-another up.

Figure 4 plots the R-version of the Muses-C data, along with the nominal saturation limit of R = 0.2 (horizontal dashed line), and the estimated data point for 1-2 cm rocks (blue square). The estimated data point is R ~ 0.3, above the saturation line. This is a small enough increase above the nominal saturation value that this value could exist, higher values are progressively more unlikely.

Thus, a reasonable estimated bound on the population 1-2 cm surface rocks is between 245 and 1089 rocks/m$^2$.

Assuming a density of 1.5 g/cm$^3$, a 1 cm diameter particle has a mass of 0.79 g, and a 2 cm diameter particle has a mass of 6.28 g. If the collection consisted of particles only between 1 and 2 cm diameter, TAGSAM would need to ingest between 29 and 190 particles to meet the 150 g sample mass requirement. These numbers compare favorably to the estimated number of particles of these sizes resting on the surface (245 to 1089), and do not take into account access to sub-surface particles, which substantially increases the amount of sampleable material.

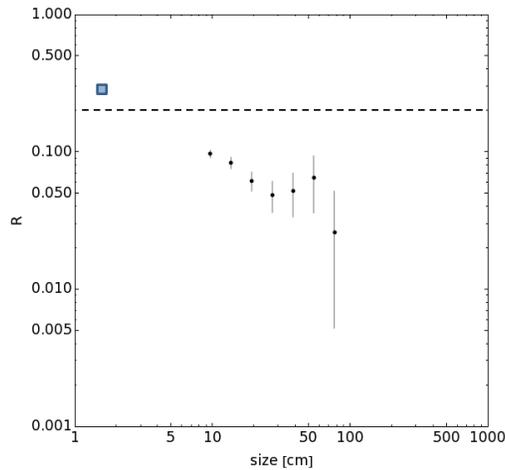

**Figure 4: A relative-format plot of the Muses-C rock data. The horizontal dashed line is a saturation value of 0.2, while the blue square is the theoretical point for rocks between 1-2 cm. See text for discussion.**

### *2.2 Relevance of These Data to Muses-C As a Whole, and Comparison to Bennu*

The three images measured represent a modest fraction of the totality of the Muses-C area, and they also are from a region of Muses C near a transition to other terrains. Examination of the full resolution images, and lower-resolution images that provide context for the high-res images, shows rough terrain to the left and right of the region measured in Figure 2 – indeed, it is obvious from Figure 2 that there is a gradient in the spatial density of rocks, from higher to lower spatial density as one moves from left to right.

Figure 5 shows a version of the Itokawa shape model, slightly zoomed in and reoriented relative to Figure 1, that is shaded to emphasize the difference in surface roughness between Muses-C and the rest of Itokawa. The high-resolution images (and thus the touch-down location of Itokawa) occur near a boundary between Muses-C and rougher terrain. It is possible, although currently unquantified, that the measured region is not a true average of the Muses C region. At least qualitatively, the true center of Muses-C seen in Figure 5 appears more smooth then the boundary region in which the high-resolution images appear. This suggests that the measurements above are a "worst-case" rock population for Muses-C.

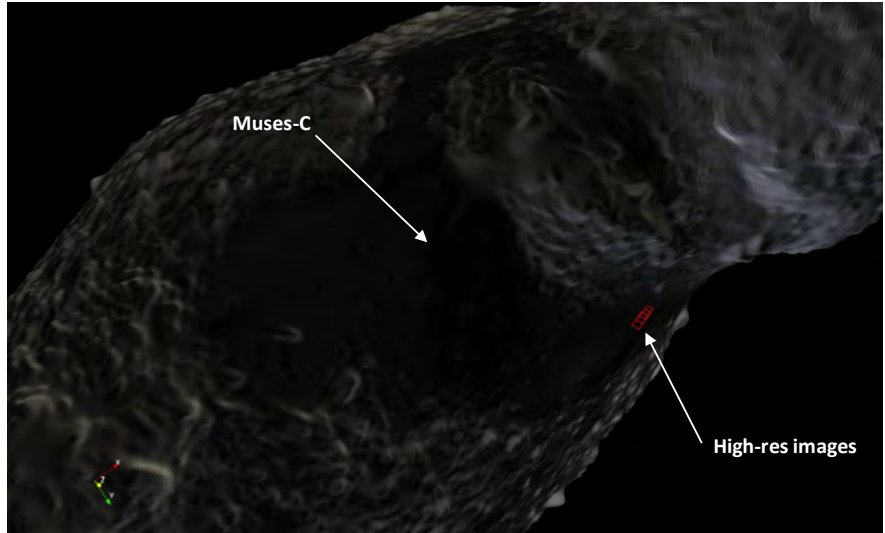

**Figure 5: A version of the shape model, shaded to emphasize the difference in roughness between the Muses-C region and the rest of the asteroid.**

## 2.3 Comparison with Bennu

The radar shape model for Bennu includes one definitive detection of a boulder within the resolution limits of the radar data. We estimate the rock population on Bennu by the assuming the following:

- the boulder is 7 m diameter, and that there is only one
- Bennu is a sphere with a radius of 250 m
- a cumulative power-law exponent of -3

With those assumptions, we solve for the coefficient $c$ of the cumulative power law:

$$c = \frac{N}{A}D^{-a}$$

Using N = 1, A (the area of the asteroid) = 1.96 x $10^5$ m$^2$, D = 7 m, and $a$ = 3, then $c$ = 1.48 x $10^{-8}$. The corresponding hypothetical, cumulative power-law for rocks on Bennu is:

$$N_c = 1.48 \times 10^{-8} D^{-3}$$

Figure 6 plots the cumulative Muses-C measurements (black data points, as in Figure 3) as well as the hypothetical Bennu SFD, the red line. The hypothetical SFD is an order-of-magnitude estimate, and may be incorrect by even more than an order of magnitude. Nevertheless, Figure 5 demonstrates that, within the limited constraints of the available data we have for both Muses-C and Bennu, that Muses-C has a higher-density rock population – and thus Muses-C can be considered a legitimate upper-bound for the rock population expected on Bennu.

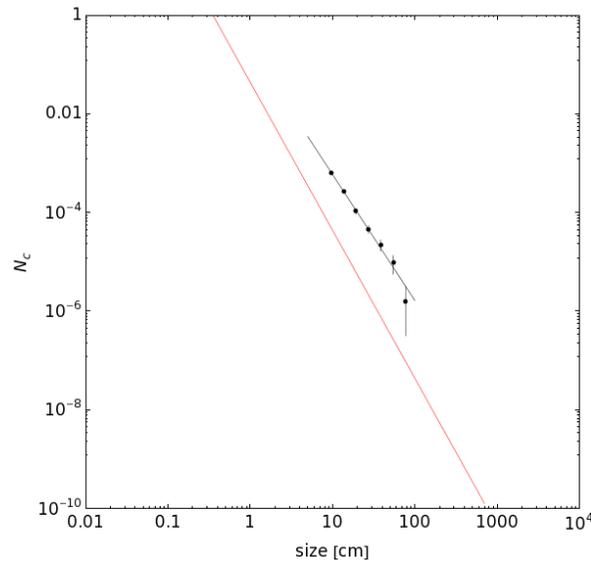

**Figure 6: A log-log plot of both the Itokawa Muses-C cumulative rock measurements (black points and best-fit SFD represented by the black line, see Figure 2), and a hypothetical cumulative SFD for Bennu (red line), assuming a cumulative slope of -3, and a single 7 m boulder is the largest boulder on the asteroid. Even with a slightly steeper slope (-3 vs. -2.55), the hypothetical rock population on Bennu is less dense than on Itokawa.**

## 3. Topographic Information for Muses-C

Topographic information from the Muses-C region is available from two independent data sources:
1. *LIDAR data*. The LIDAR data are a direct measurement of the range between the spacecraft and the surface, so given accurate knowledge of the spacecraft-asteroid relative geometry, the LIDAR data are a direct measurement of surface topography. In practice, navigation uncertainties, and the "single-track" design of the LIDAR (i.e. it samples 1D swaths over the surface) introduce uncertainties and limitations in the extent to which the data apply over broader surface regions.
2. *Stereophotoclinomtry (SPC)*. SPC uses camera images, and information on lighting angles, to translate pixel intensity values to slope values – and therefore heights. Heights are thus a derived quantity, rather than a direct measurement. However, the benefit to SPC is that one can estimate height and surface roughness values for broad surface areas, given that images of sufficient utility exist.

The combination of these two data sources are complimentary: the Hayabusa LIDAR data are direct measurements but limited in areal extent, while the SPC data are indirect measurements but provide constraints over a broader surface area. The following sections summarize our understanding of Muses-C topography from each of these two data sets.

*3.1 Obtaining meaningful topographic data from the Hayabusa LIDAR*
Estimates of the Hayabusa spacecraft position relative to Itokawa were, upon arrival, insufficiently accurate to generate useful topographic profiles with the Hayabusa laser altimeter (LIDAR). To obtain useful data for analyzing the surface properties of Itokawa, a new algorithm was developed to better locate the spacecraft relative to the asteroid (Barnouin-Jha et al., 2008). Housekeeping data were used, combining a LIDAR range every 2 min with an x-y pixel measurement obtained by the Wide Angle Camera (WAC) of the illuminated centroid of Itokawa. With the accurately known pointing of the LIDAR and the WAC, and a shape model of Itokawa, we determine the location of the spacecraft and hence the LIDAR footprint to within 10m. In more recent efforts [Barnouin-Jha et al, 2013], the location of individual LIDAR shots was further improved to an RMS of 6m, by using least-squares to minimize the difference between an Itokawa shape model and the topography measured by the Hayabusa LIDAR.

**Investigating the topography from HAYABUSA LIDAR.** The improvements in the Hayabusa LIDAR has permitted directly measuring topographic features on Itokawa. An example in the Muses-C regio is shown in Figure 7 (from Barnouin-Jha et al. 2008). The results show that within the ranging precision of ~0.5 m (shown by the small bumps in the data in the profiles) that this region is smooth. (Neither transect crosses the region shown in Figure 2.) The precision of the HAYABUSA lidar is insufficient to provide direct measurements of topographic values at scales relative to TAGSAM. Nevertheless, roughness assessments can be used to extrapolate what the topography might be at smaller scales.

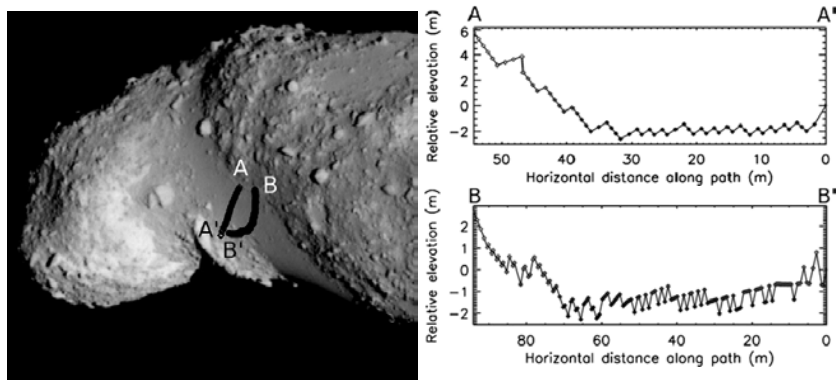

**Figure 7: Taken from Barnouin et al. (2008). The top panel is an image of the Muses-C region with two lidar transects. The middle panel is the corresponding lidar data for the A-A' transect. The bottom panel is the lidar data for the B-B' transect. All the data show elevation relative to geoid (see Barnouin-Jha et al., 2008 for a definition).**

**Computing Surface Roughness**: The improved LIDAR footprint location enabled a quantitative assessment of variations in surface roughness of Itokawa, which can be compared to data for the surface roughness of other asteroids and planets Eros. Roughness is characterized as in [Shepard et al. (2001)] using individual LIDAR profiles to determine the standard deviation σ (square root of Allan variance) of height differences versus baseline $L$. The standard deviation is $\sigma = \langle [e(s)-e(s+L)]^2 \rangle^{0.5}$, where $e$ is elevation along a LIDAR profile at a distance $s$. The value of $e$ is computed relative to the local geoid, and then de-trended to remove the regional slope. The results for this roughness across the surface of Itokawa are shown for several chosen areas. In most planetary settings, surface roughness is usually self-affine. This means that the standard deviation of height differences, measured for all pairs of samples in a track separated by a baseline L, obeys

$$\sigma = C_o \left( \frac{L}{L_o} \right)^H$$

where $L_0$ is always 1 m, and $C_o$ is a normalizing constant. The quantity $H$ is called the Hurst exponent and is related to fractal dimension γ by the relation γ = 2 − H. Examples of the standard deviation heights or roughness is shown in Figure 8 for regions on Eros, and on the Moon.

A preliminary analysis [Barnouin-Jha et al., 2012] of a few regions on Itokawa indicates that this asteroid much rougher than either Eros or the Moon (Figure 9). The surface appears not to have much of a self-affine nature, comparable to complex surfaces shown in Figure 8 on the Moon. The complex character of Itokawa's topography has been attributed to its rubble pile structure, which does not have sufficient strength to sustain any topography at long baselines, and has significant rough rubble present at shorter baselines.

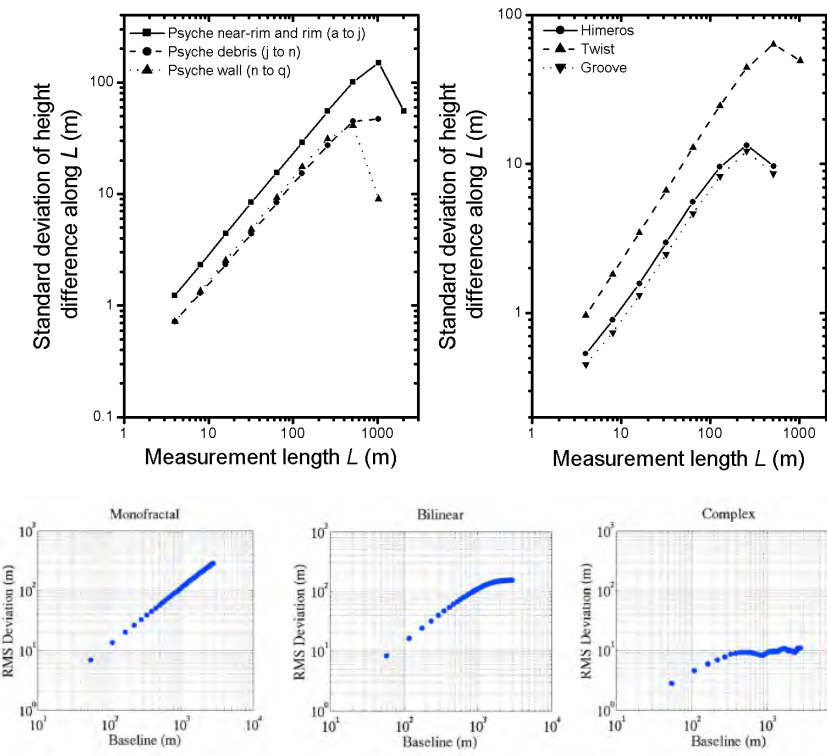

**Figure 8. Surface roughness on Eros (Top; from Cheng et al., 2002) and the Moon (Bottom; from Rosenburg et al., 2010). Younger surface on the Moon such as the lunar Mare tend to be complex and less rough than the lunar highlands. Eros is significantly rougher than the Moon, at comparable baseline scales.**

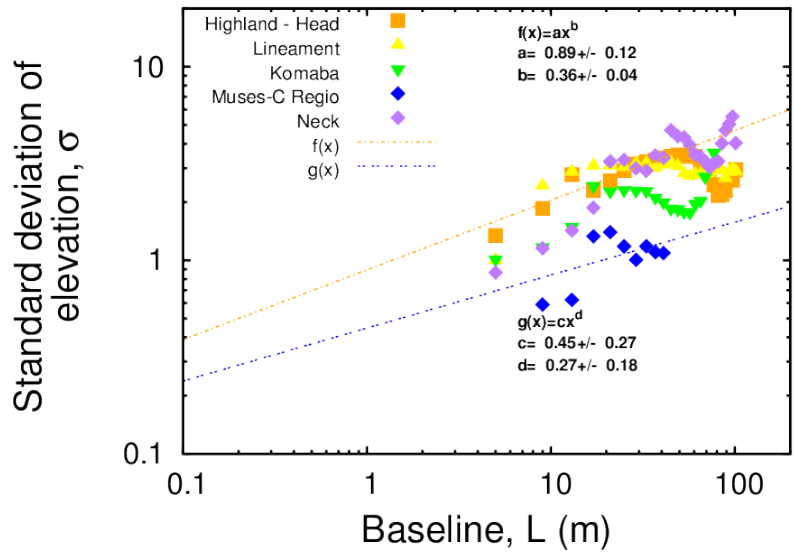

**Figure 9. Standard deviations of the height for various baselines length scales L on Itokawa from Barnouin-Jha et al. (2008). In all cases, the pre-exiting long wavelength slopes were removed. Fits show a best guess at how the topography might behave at short baseline lengths.**

For the purpose of generating a synthetic surface that is comparable to the TAGSAM, we can make a best guess of the surface characteristic at TAGSAM by fitting a power law function to the data obtained for both the highlands and Muses-C regio of Itokawa, and then extrapolating the results to smaller baselines assuming that the generally self-affine nature of planetary surfaces is maintained. Those fits are also shown on Figure 9.

### 3.2 Investigating the topography using SPC from HAYABUSA Images

Bob Gaskell created a shape model of the Muses-C region, combining image data with resolutions between 20 cm and 50 m, (Figure 10). The surface roughness of Itokawa in the smoothest portions of the Muses-C region using topography derived from stereo-photoclinometry (SPC) show a self-affine function ($H = 0.88$; $\sigma_o = Co = 0.412$), which compares much more favorably with Eros and even the Moon (see Table 3). The observed discrepancy between the results from the Hayabusa LIDAR data and the SPC data may be because of some uncertainties associated with location of the Hayabusa LIDAR. Location errors of up to 10 m could mean that some of the LIDAR tracks analyzed by Barnouin-Jha et al. (2008) that were thought to be in the Muses-C region might not have been sampling uniquely this region. Further, the shape model generated by SPC has a tendency to smooth short wavelength topography.

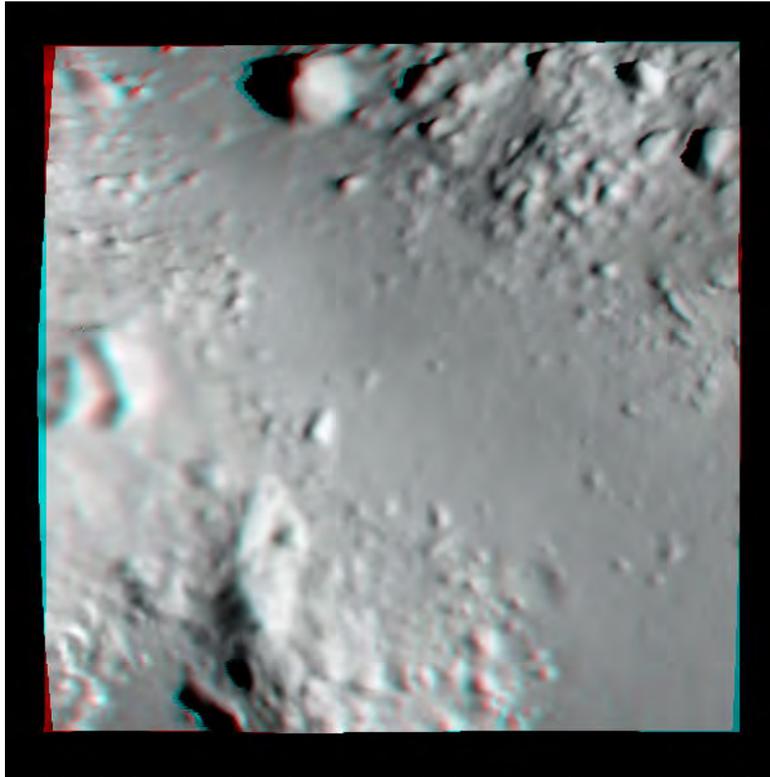
MUSESC Lt= 9.82S Ln=320.36W Rd= 0.122 Sz=0.120 km
**Figure 10: The Muses-C region from Gaskell's shape model.**

## 4. Context with other Solar System Bodies, and a Model for Muses-C

Table 3 lists the Hurst parameters for several solar system surfaces; Figure 11 plots these idealized fits relative to one another, including the two Muses-C best fit Hurst values, one from the LIDAR and one from SPC. The "clean" linear fits do not capture the variation seen in the real data, but they provide a convenient means to compare the data. Examination of Table 3, and of the lunar data in Figure 8, demonstrate that surfaces appearing broadly similar "to the eye", or surfaces that are similarly dominated by impact processes, can express different Hurst parameters. Indeed, Shepard et al. (2001) suggest that Hurst values cannot be uniquely diagnostic of geologic processes, they are simply a measure of surface roughness.

Broadly speaking, there are three categories of Hurst values (Shepard et al. 2001):
1. surfaces with H < 0.5 – These are surfaces that are "rough" at small length scales, and which transition to smoother topography at larger scales. At very small scales an example is sandpaper. At the scale of the grains the surface is rough, but at the scale of the width and length of the paper the surface topography is small. A slightly larger scale version, with more direct application to Muses-C, is a gravel driveway. The surface is rough at the scale of the particles that comprise the gravel, but at the scale of road width, the topography is smooth.
2. Surfaces with H > 0.5 – These are surfaces that preserve roughness as length scale increases. At the limit of H = 1, the surface is self-similar at all scales, and will be equally rough regardless of length scale.
3. Surfaces with H = 0.5 – These are surfaces are termed "Brownian" surfaces, since Brownian motion will produce a surface of this type. The surface can vary randomly between smooth and rough across all scales.

Shepard et al. (2001) suggest that natural surfaces tend to have values near 0.5, though that may be an artifact of limited data sets. The Rosenburg et al. (2011) analysis of lunar surface roughness (see Figure 8), using the extensive (essentially global), consistent, and relatively high-quality LRO laser altimeter data, derive Hurst exponents of > 0.76 across the surface, with values for some parts of the Moon exceeding 0.95 for length scales between a few meters and a few kilometers.

When defining a "worst-case" surface for Muses-C, there are several considerations, two of which are end-members: the worst case for large rocks that could endanger the mechanical safety of the arm or the sampler, and the worst-case for small rocks that jeopardize the sampling success. The first case would have a Hurst exponent > 0.5, the second case would have a Hurst exponent < 0.5. Interestingly, the two estimates for the Muses-C surface roughness fall into both categories. The LIDAR-derived Hurst exponent is 0.27, whereas the SPC-derived exponent is 0.88. RADAR observations (circular polarization ratios of the returns) indicate that Bennu is likely to be smoother than Itokawa, and may be comparable to 433 Eros at RADAR wavelength scales of 7-20 cm. In addition, measurements of the Muses-C rock SFD predict a higher spatial density than what is expected from the single resolved boulder in the radar shape model (see Section 3.3). Because a primary use of the worst-case Muses-C surface is to evaluate spacecraft safety rather than sampleability, we recommend using a value that resides in the H > 0.5 space.

Gaskell's shape model software has the ability to generate rock populations described by a power law, and to overlay that population on a surface which has some inherent roughness. Using this technique, Gaskell created several synthetic asteroid surfaces, with the measured rock population described in Section 2 as an input. These original attempts were able to match the rock population, but the derived Hurst parameters were inconsistent with the measured parameters from other surfaces. Varying the inherent roughness, in conjunction with the rock population, ultimately led to a surface that could match both the Muses-C rock measurements and fall within the range of Hurst parameters measured on Itokawa.

*4.1 The synthetic surface of Bennu*
Recall the two parameters for a power-law are a constant coefficient and the power-law exponent. For rock populations, the coefficient relates to the density of objects, with a larger value corresponding to a greater density. The exponent controls the relative proportion of large objects to small objects. We recommend a worst-case surface that preserves the exponent, but increases the coefficient. The rationale is as follows:

*Preserving the exponent*. A steeper exponent, i.e. < -2.55, will have more smaller particles relative to large particles. This means that the fractional area covered by large objects will decrease, and thus not be representative of a worst-case in regards to spacecraft safety. A more shallow exponent, i.e. > -2.55, will increase the fractional area of large objects. While this may initially appear to meet the intent for defining a worst-case surface, the nature of a power-law means that this method could easily push worst-case to unnecessarily unrealistic topographies, with the potential for expensive and unneeded modifications to the spacecraft design. Further, existing data of boulders on planetary surfaces (see, e.g. Bart and Melosh 2010), and on asteroids (see, e.g. Chapman et al. 2002 or Michikami et al. 2008) indicate that rocks and boulders typically follow exponents that are near this value or steeper, and thus shallower slopes are not well justified by existing data.

*Increasing the coefficient.* Because we cannot be sure that the direct measurements of the Muses-C rock population reflect a worst-case environment within an area that, more broadly may ultimately be sampleable, we require a means to increase the rock density. As described above, changing the power-law exponent achieves this goal with a potentially excessive result. In contrast, an increase of the overall rock density preserves the relative proportion of large rocks to small rocks, while increasing the total number across all sizes. We recommend a factor of 2 increase in density, which is sufficiently large to reflect the uncertainty of this analysis, but not so large that the surface character departs from the Muses-C observations.

Thus, using the best-fit cumulative slope of -2.55 and a coefficient of 0.4, Gaskell generated the surface in Figure 11. The dimension of the entire synthetic surface is 18 m x 18 m, with 5 mm grid spacing. The Hurst parameters for several scales are listed in Table 3 (and included in Figure 14). At several scales the surface has a Hurst exponent near 0.6, which meets the > 0.5 criteria. The surface was constructed with CreatorP, a program that has the capability of generating a 6.88 x 1018 vector asteroid, or any piece of the asteroid's surface at that resolution (about a nanometer for Bennu). The vectors are labeled by indices corresponding to grids on the faces of a cube, with each grid having a maximum of 260 cells (230 +1 x 230 +1 vector labels). The grid points are assigned a "level" with the highest level belonging to the corner vectors of the cube, and the next highest to the center of each face. New, lower-level, grid points are constructed at the mid point of four neighboring higher level points, arranged in a square aligned with the grid or rotated 45°.

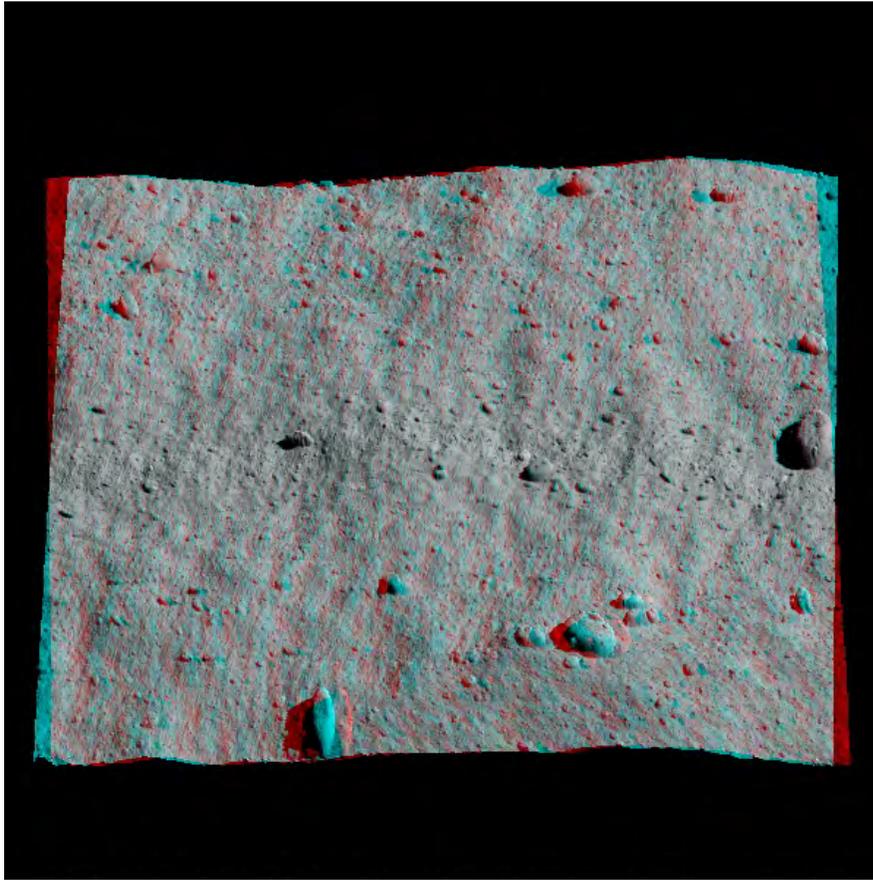

**Figure 11. Anaglyph of the complete 17.5 m x 17.5 m synthetic surface.**

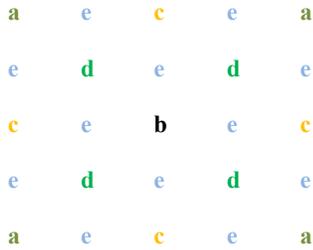

**Figure 12. Successively lower levels in a grid (a>b>c>d>e).  Edge points have "parents" on another cube face.**

Surfaces are built up by the successive application of various processes, with each one designed to proceed to progressively lower levels. The current surface used only a rock generator and a fractal surface generator. The latter fills in lower level grid points by adding a random height to the average of the parents' heights, where the amplitude of the random distribution is scale dependent. Roughness of the rocks is also accomplished through this "stochastic interpolation". Rocks are generated from large to small, filling in cells at a certain level with rocks with a diameter distribution commensurate with the cell size. Rocks have a randomly selected ellipticity, size, height and albedo. The algorithms are designed so that a surface patch can be constructed that is completely consistent with the construction of the entire body at that resolution, and with overlapping patches at different resolutions. Two other processes are currently included in CreatorP. Cratering proceeds much like addition of rocks, except that the order in which craters appear is randomly determined, so that large craters can obliterate small ones, or small craters can appear inside larger ones. The final algorithm mimics the migration of regolith to low gravitational/rotational potential areas, a process seen to be at work on Itokawa.

CreatorP is not capable of generating a surface of the desired size and resolution. Instead, 121 overlapping surfaces were created, each 3 m x 3 m at 5 mm resolution. These were then put together using a program called BIGMAP to produce the final product. One of these maps, from the center of the large 18 m x 18 m map in Figure 11, is shown in Figure 13.

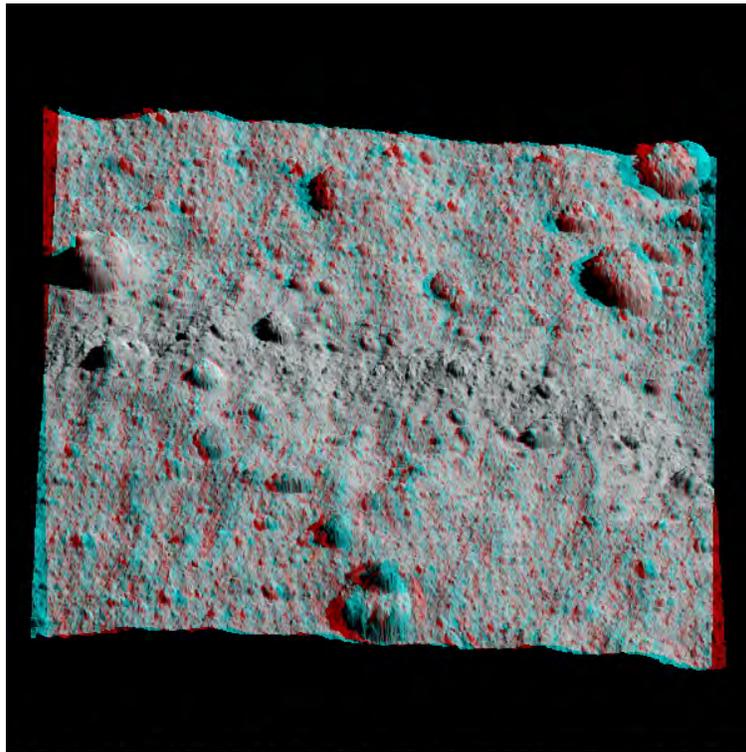

BD0606 Lt=39.02S Ln=329.01W Rd= 0.251 Sz=0.003 km
**Figure 13. Anaglyph of a 3 m x 3 m constituent map at 5 mm resolution.**

CreatorP has a heritage going back to simulations of the Martian surface (Gaskell, 1993) that were used to study rover behavior in a variety of terrains to determine inadequacies in flight software, and to study the landing characteristics of air bag systems. Earlier Creator comet and asteroid generators were used in simulations in preparation for the NEAR, Deep Impact and DAWN missions. The "P" in CreatorP indicates that the software is now portable, producing the same results on any platform given the same starting conditions.

The synthetic data set is provided as a digital elevation model, and is identified by the filename MUSESCWORSTCASE.MAP[1]. The use of a 18 m x 18 m surface allows analysis over an area that is more representative of the TAG uncertainty ellipse, and also provides a range of cases for tools that are limited to analyzing smaller areas.

> **Added by DRA editor: the file MUSESCWORSTCASE.MAP can be Found on ODOCS at / OSIRIS-REx 04.0 Science and Technology / Design Reference Asteroid / RQ36_Surface_Properties**

Table 3: Surface roughness parameters from a variety of solar system surfaces.

| Region | Hurst Exponent (H) | σ | Reference |
|---|---|---|---|
| Eros – Himeros | 0.81 | 0.18 | Cheng et al. (2002) |
| Eros – Himeros | 0.90 | 0.41 | Gaskell SPC data |
| Eros – Twist | 0.89 | 0.30 | Cheng et al. (2002) |
| Eros – Groove | 0.82 | 0.14 | Cheng et al. (2002) |
| Eros – Psyche rim | 0.91 | 0.35 | Cheng et al. (2002) |
| Eros – Psyche wall | 0.87 | 0.23 | Cheng et al. (2002) |
| Eros – Shoemaker | 0.91 | 0.33 | Gaskell SPC data |
| Earth – a'a flow | 0.55 | 0.46 | Shepard and Campbell (1998) |
| Mars Pathfinder site | ~0.5 | NA | Shepard and Campbell (1998) |
| lunar regolith | 0.5-0.7 | 0.02-0.07 | Helfenstein and Shepard (1999) |
| Itokawa – "rough" | 0.36 | 0.89 | Barnouin-Jha et al. (2008), lidar |
| Itokawa – Muses-C | 0.27 | 0.45 | Barnouin-Jha et al. (2008), lidar |
| Itokawa – Muses-C | 0.88 | 0.41 | Gaskell SPC data |
| RQ36_a | 0.72 | 0.13 | Gaskell RQ36 model, circa 2012 |
| RQ36_b | 0.56 | 0.10 | Gaskell RQ36 model, circa 2012 |
| Bennu, 32 cm detrend | 0.57 | 0.08 | Gaskell Bennu model, June 2013 |
| Bennu, 100 cm detrend | 0.57 | 0.09 | Gaskell Bennu model, June 2013 |
| Bennu, 2.5 m detrend | 0.61 | 0.11 | Gaskell Bennu model, June 2013 |
| Bennu, 5 m detrend | 0.60 | 0.11 | Gaskell Bennu model, June 2013 |

---

[1] To ensure that you have the correct file, the MD5 checksum for MUSESCWORSTCASE.MAP is d8d4312ca12a9e646ee728e5bbb3883a

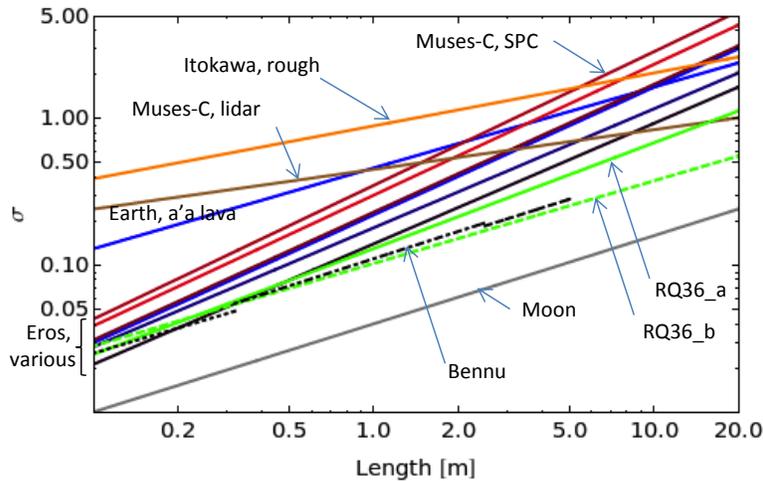

**Figure 14. Idealized fits of the Hurst parameters for various asteroids, the Moon and the Earth. The Hurst parameters are listed in Table 3.**

## 5. References

Barnouin-Jha et al. (2008). Small-scale topography of 25143 Itokawa from the Hayabusa laser altimeter, *Icarus*, **198**, pp. 108-124.

Bart and Melosh (2010). Distributions of boulders ejected from lunar craters, *Icarus*, **209**, pp. 337-357.

Chapman et al. (2002). Impact History of Eros: Craters and Boulders, *Icarus*, **155**, pp. 104-118.

Cheng et al. (2002). Small-Scale Topography of 433 Eros from Laser Altimetry and Imaging, *Icarus*, **155**, pp. 51-74.

Gaskell, R.W. (1993). Martian Surface Simulations, *Journal of Geophysical Research*, **96**, E6, pp. 11099-11103.

Michikami et al. (2008). Size-frequency statistics of boulders on global surface of asteroid Itokawa, *Earth Planets Space*, **60**, pp. 13-20.

Rosenburg et al. (2011). Global surface slopes and roughness of the Moon from the Lunar Orbiter Laser Altimeter, *Journal of Geophysical Research*.

Shepard, M. K. et al. *Journal of Geophysical Research*, 106:32777–32796, 2001.

## 6. File Format

The following describes the MAPFILE.MAP structure as of 02 July 2013.

RECORD LENGTH:  72

FIRST RECORD:

BYTE 01-06:  CHARACTER*6       'UNUSED'
BYTE 07-10:  MSB REAL*4        SCALE   Pixel scale of map in km
BYTE 11-12:  LSB UNSIGNED SHORT   QSZ     Map size = 2*QSZ+1 X 2*QSZ+1 pixels
BYTE 13-15:  BYTE              * See Below
BYTE 16-19:  MSB REAL*4        V(1)
BYTE 20-23:  MSB REAL*4        V(2)    V = body fixed lanamark vector (km)
BYTE 24-27:  MSB REAL*4        V(3)
BYTE 28-31:  MSB REAL*4        UX(1)
BYTE 32-35:  MSB REAL*4        UX(2)   Ux = maplet ref plane x unit vector
BYTE 36-39:  MSB REAL*4        UX(3)
BYTE 40-43:  MSB REAL*4        UY(1)
BYTE 44-47:  MSB REAL*4        UY(2)   Uy = maplet ref plane y unit vector
BYTE 48-51:  MSB REAL*4        UY(3)
BYTE 52-55:  MSB REAL*4        UZ(1)
BYTE 56-59:  MSB REAL*4        UZ(2)   Uz = maplet plane z (normal) vector
BYTE 60-63:  MSB REAL*4        UZ(3)
BYTE 64-67:  MSB REAL*4        HSCALE = max(abs(ht))/30000
BYTE 68-71:  MSB REAL*4        * See Below
BYTE 72   :  UNUSED BYTE       CHAR(0)

SUBSEQUENT RECORDS ARE 24 3 BYTE ENTRIES PER RECORD, padded to 24 entries for last record.

ENTRY:

BYTE 01-02:  INTEGER*2         NINT(HT/HSCALE)) (-30000,30000)
BYTE 03   :  BYTE              NINT(RELATIVE ALBEDO X 200) (0,199) **

* These two entries are respectively relative and absolute magnitude of position uncertainty.  Subsequent versions of the file format may replace these with gravity direction (13-15) and magnitude (68-71).

** Albedo entry is limited to (1,199). A value of 0 indicated no data.